\documentclass[longauth]{aa}  

\usepackage{graphicx}
\usepackage{txfonts}
\usepackage{lipsum}
\usepackage{subcaption}         
\usepackage{lscape}             
\usepackage{placeins}           
\usepackage{dcolumn}
\usepackage{array}
\usepackage{etoolbox}   
\usepackage{orcidlink}
\usepackage{natbib}
\usepackage{xcolor}
\usepackage{hyperref}
\hypersetup{
    colorlinks=true,
    linkcolor=blue,  
    citecolor=blue,
    }
\usepackage[version=4]{mhchem}
\usepackage{savesym}
\usepackage{multirow}
\savesymbol{tablenum}
\usepackage{siunitx}
\restoresymbol{SIX}{tablenum}
\usepackage{gensymb}

\begin{document}
   \title{The Multi-phase Biconical Outflow in the local IR-Luminous Merger IRASF01364-1042}
   \author{Y. Song\inst{1,2,3}\thanks{\email{yiqing.song@eso.org}}\and 
          V. U \inst{4,5} \and 
          J. Kader \inst{5} \and 
          M. Bianchin \inst{5,6,7} \fnmsep\thanks{IAU Gruber Fellow} \and 
          J. Agostino \inst{8} \and 
          L. Barcos-Mu\~{n}oz \inst{9,10} \and 
          N. Torres-Alb\`{a} \inst{10} \and 
          A. Medling \inst{8} \and 
          C. Ricci \inst{11, 12} \and 
          L. Armus \inst{4} \and 
          T. Lai  \inst{4} \and 
          L. Ghodsi \inst{13} \and 
          A. Evans \inst{9,10} \and
          H. Inami \inst{14} \and 
          S. Stierwalt \inst{15} \and
          T. Bohn \inst{14} \and
          K. L. Emig \inst{9} \and 
          V. Buiten \inst{16} \and
          G. Donnelly \inst{8} \and
          E. Treister \inst{17} \and 
          A. Jimenez-Gallardo \inst{1} \and
          T. B\"{o}ker \inst{18} \and 
          D. Kakkad \inst{19} \and  
          S. Linden \inst{20} \and  
          P. van der Werf \inst{16} \and 
          V. Charmandaris \inst{21,22,23} \and 
          R. Remigio \inst{5} \and 
          L. Lenki\'{c} \inst{4} \and
          C. Finlez \inst{1,2} \and 
          M. Sanchez-Garc\'{i}a \inst{22,23} \and
          C. Lofaro \inst{22,23} \and
          A. Saravia \inst{9,10} \and
          I. Yoon \inst{9} \and
          G. C. Privon \inst{9,10,24} \and 
          T. Gao \inst{25,26} \and 
          H. Aziz \inst{5} \and 
          R. McGurk \inst{27} \and
          T. D\'{i}az-Santos \inst{22,23}\and 
          D. Kunneriath \inst{9} \and 
          David B. Sanders \inst{28}
        }

   \institute{European Southern Observatory, Alonso de Córdova, 3107, Vitacura, Santiago, 763-0355, Chile
            \and Joint ALMA Observatory, Alonso de C\'{o}rdova, 3107, Vitacura, Santiago, 763-0355, Chile 
            \and Max-Planck-Institut f\"{u}r Radioastronomie, Auf dem H\"{u}gel 69, 53121, Bonn, Germany 
            \and IPAC, California Institute of Technology, 1200 E. California Boulevard, Pasadena, CA 91125, USA
            \and  Department of Physics and Astronomy, 4129 Frederick Reines Hall, University of California, Irvine, CA 92697, USA
            \and Instituto de Astrof\' isica de Canarias, Calle V\' ia L\'actea, s/n, E-38205, La Laguna, Tenerife, Spain
            \and Departamento de Astrof\' isica, Universidad de La Laguna, E-38206, La Laguna, Tenerife, Spain
            \and Ritter Astrophysical Research Center and Department of Physics and Astronomy, University of Toledo, Toledo, OH
43606, USA
            \and National Radio Astronomy Observatory, 520 Edgemont Road, Charlottesville, VA 22903, USA
            \and Department of Astronomy, University of Virginia, 530 McCormick Road, Charlottesville, VA 22903, USA
            \and Department of Astronomy, University of Geneva, Chemin Pegasi 51, 1290, Versoix, Switzerland
            \and Instituto de Estudios Astrof\'isicos, Facultad de Ingenier\'ia y Ciencias, Universidad Diego Portales, Av. Ej\'ercito Libertador 441, Santiago, Chile
            \and Department of Physics \& Astronomy, University of British Columbia, 6224 Agricultural Road, Vancouver, BC, V6T 1Z1, Canada
            \and Hiroshima Astrophysical Science Center, Hiroshima University, 1-3-1 Kagamiyama, Higashi-Hiroshima, Hiroshima 739-8526, Japan
            \and Occidental College, Physics Department, 1600 Campus Road, Los Angeles, CA 90042, USA
            \and Leiden Observatory, Leiden University, PO Box 9513, 2300 RA Leiden, The Netherlands
            \and Instituto de Alta Investigaci{\'{o}}n, Universidad de Tarapac{\'{a}}, Casilla 7D, Arica, Chile
            \and European Space Agency, c/o STScI, 3700 San Martin Drive, Baltimore, MD 21218, USA
            \and Centre for Astrophysics Research, University of Hertfordshire, College Lane, Hatfield, AL10 9AB, UK
            \and Steward Observatory, University of Arizona, 933 North Cherry Avenue, Tucson, AZ 85721, USA
            \and School of Sciences, European University Cyprus, Diogenes Street, Engomi, 1516 Nicosia, Cyprus
            \and Institute of Astrophysics, Foundation for Research and Technology-Hellas (FORTH), Heraklion, 70013, Greece
            \and Department of Physics, University of Crete, Heraklion, 71003, Greece
            \and Department of Astronomy, University of Florida, P.O. Box 112055, Gainesville, FL 32611, USA
            \and Research School of Astronomy and Astrophysics, Australian National University, Weston Creek, ACT 2611, Australia
            \and ARC Centre of Excellence for All Sky Astrophysics in 3 Dimensions (ASTRO 3D); Australia
            \and W. M. Keck Observatory, Kamuela, HI, USA
            \and Institute for Astronomy, University of Hawaii, 2680 Woodlawn Drive, Honolulu, HI 96822, USA
            }

  \abstract 
   {}
   {We investigate the spatially-resolved ISM properties of the local ($z = 0.048$), IR-luminous ($L_{\rm
    IR} = 10^{11.87}$\,L$_\odot$), late-stage galaxy merger IRAS F01364-1042, combining multi-wavelength IFU observations from
    \textit{JWST/MIRI-MRS}, ALMA and Keck/KCWI. The system exhibits excess emission from mid-IR \ce{H2} lines relative to dust, as identified in a previous \textit{Spitzer/IRS} survey of local IR-luminous galaxies (LIRGs). The bright \ce{H2} emission was speculated to arise from widespread shocks associated with a powerful starburst/AGN-driven outflow.} 
   {Using our multi-wavelength datasets, we assess and compare the resolved morphology and kinematics of multi-phase gas tracers on sub-kpc scales. We construct emission line maps of several key tracers of the ionized (e.g., [Ne\,II]\,12.8$\mu$m, [O\,III]$\lambda5007$), warm molecular (e.g., \ce{H2}\,0-0\,S(3)), and cold molecular gas (e.g., CO (J$=2-1$)), and perform detailed decomposition of spectra extracted in resolved regions across the areas of emission, with a focus on the central $\,\sim 3\,$kpc of the system covered by all our datasets.}
   {We confirm the presence of a multi-phase galactic biconical outflow along the minor axis of a highly inclined rotating disk, based on the multi-phase gas morphology and prominent line broadening (FWHM $> 500$\,km\,s$^{-1}$) in the ionized gas tracers detected out to $\sim$\,5\,kpc from the nucleus. The molecular gas tracers are detected out to $\sim $2\,kpc, and exhibit a distinct ``X''-shaped morphology and different velocity fields from the ionized gas, which we attribute to the higher concentration of molecular gas near the disk. We adopt a simple model to interpret the observed multiphase gas kinematics and infer an outflow velocity of $\sim\,$500 - 600\,km\,s$^{-1}$  $\sim\,$350\,km\,s$^{-1}$, and $\sim\,$200 - 300\,km\,s$^{-1}$, in the ionized, warm and cold molecular phase, respectively, with corresponding mass outflow rates of $\sim 0.3 - 2.3$, $\sim 31$, and $\sim 38 - 240$\,M$_\odot\,$yr$^{-1}$. The cold molecular phase dominates both the total mass outflow rate and the associated kinetic energy ($\sim\,2 - 8 \times 10^{42}$\,erg\,s$^{-1}$). Via  \textit{JWST/MIRI-MRS} detection of the [Ne\,V]\,14.3$\mu$m line, we identify, for the first time, a dust-obscured AGN in IRAS\,F01364-1042. The low inferred AGN bolometric luminosity ($1.2 - 1.8 \times 10^{43}$\,erg\,s$^{-1}$) suggests that the nuclear starburst alone, with a star formation rate of $\sim 40 - 60$\,M$_\odot$\,yr$^{-1}$, can account for the energy required to drive the outflow, though a more active AGN phase in the recent past may have also played a role.}
   {Our work showcases the necessity of multi-wavelength observations in interpreting the gas dynamics in merger-driven dusty starbursts, and the capability of \textit{JWST/MIRI-MRS} in uncovering obscured low-luminosity AGN that may be common in these systems.}
   {}
\maketitle
\nolinenumbers 
\section{Introduction}\label{sec:intro}
\begin{figure*}[t!]
    \centering
    \includegraphics[scale=0.255]{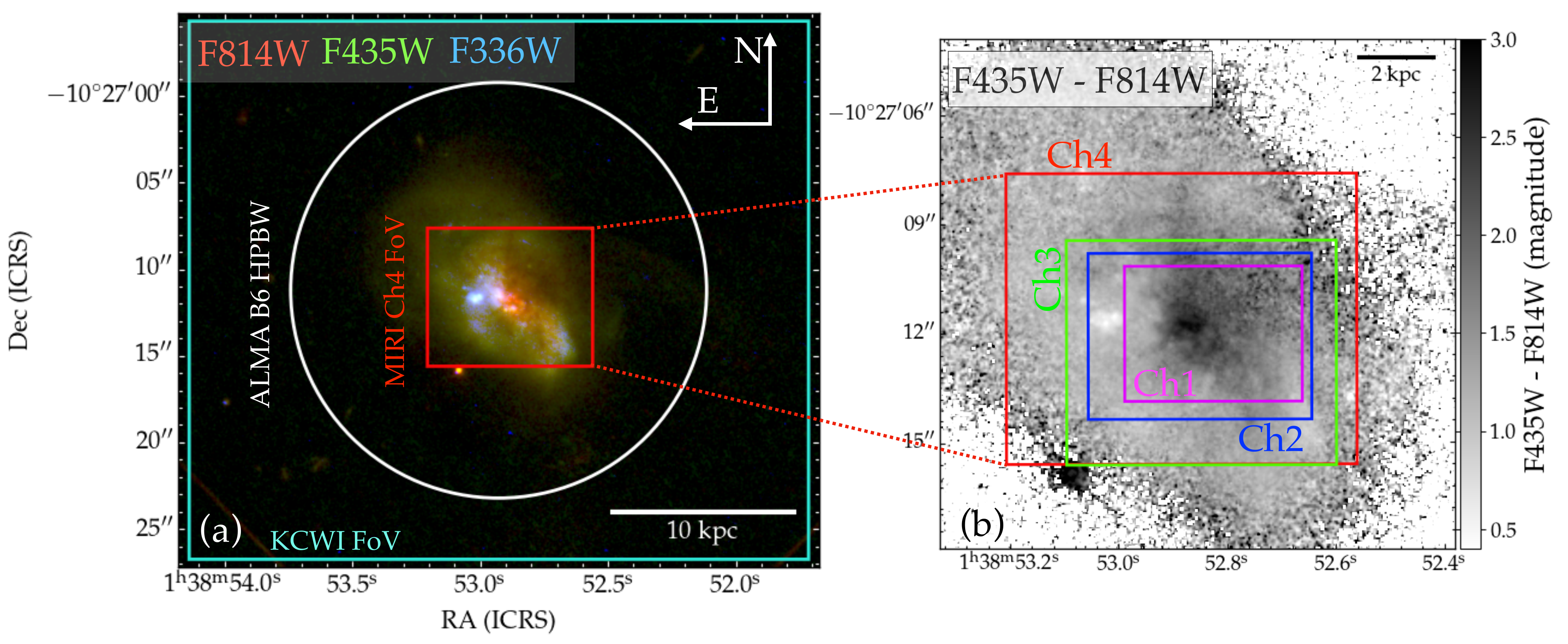}
    \caption{ \textit{HST} images of IRAS F01364-1042.  (a): False color image with F336W, F435W and F814W filters, which are represented in blue, green and red, respectively. The Half-Power-Beam-Width (HPBW) of the ALMA Band 6 primary beam, and the fields-of-view (FoVs) of the calibrated data cubes from Keck/KCWI and the longest wavelength channel of \textit{JWST/MIRI-MRS} (i.e., ch4) are outlined in white, cyan and red, respectively. (b) Zoom-in of the \textit{HST} F435W - F814W map in magnitude, where a darker grey shade indicates a redder (F435W - F814W) color, and thus higher dust extinction. Colored boxes outline the FoVs of the \textit{JWST/MIRI-MRS} ch1-ch4 cubes. Prominent, non-uniform dust structures can be seen extending from the nucleus out to $\sim$\,5\,kpc into the galaxy. The large FoVs of the Keck/KCWI and ALMA datasets allow characterization of the ISM across the entire system, while \textit{JWST/MIRI-MRS} probes the immediate vicinity of the dust-obscured nucleus. \label{fig:fig1}}
\end{figure*}

\indent Gas-rich galaxy mergers represent a crucial transformative phase in the
formation and evolution of massive galaxies. Simulations have predicted that
nuclear gas inflows due to strong gravitational torque during a merger would
trigger both a powerful nuclear starburst and accretion onto the central
supermassive black hole (SMBH) \citep[e.g.,][]{matteo05,springel05,hopkins06}.
During this phase, rapid, simultaneous growth of the galaxy's stellar content and the SMBH takes place behind thick layers of dust and gas \citep[e.g.,][]{blecha18}, until the energy
released by the accreting SMBH (i.e., Active Galactic Nuclei, or AGN) disperses the obscuring materials,
revealing the bright and powerful central engine as a quasar \citep{springel05b,
hopkins05}. This clearing process, so called ``AGN feedback", in the form of winds and outflows,
is thought to be responsible for shutting down galaxy star formation (SF) and further
SMBH growth, which eventually leads to the formation of massive quiescent galaxies \citep[e.g.,][]{springel05b,khalatyan08,hopkins13}. Throughout this transformation, the SMBH and its host galaxy would evolve in tandem \citep[e.g.,][]{silk98, kauffmann00, wyithe03}, which may explain the observed cut-off at the bright end of the local galaxy luminosity distribution
\citep[e.g.,][]{bower06}, and the widely observed tight relations 
between the masses of SMBHs and various properties of their host galaxies
\citep[i.e., stellar velocity dispersion, bulge mass, and luminosity; e.g.,
][]{magorrian98,ferrarese00,mcconnell13}. \\
\indent Luminous and Ultra-luminous Infra-Red Galaxies (LIRGs:
$L_{\rm IR [8 - 1000\mu m]} > 10^{11-12} L_\odot$; ULIRGs: $L_{\rm IR [8 -
1000\mu m]} > 10^{12} L_\odot$) in the local Universe ($z < 0.1$) provide the
best laboratories to study in detail the fueling and feedback processes that are
associated with the above evolutionary scenario. Predominantly triggered by
interactions/mergers between gas-rich spirals \citep[see review by
][]{sanders96}, these systems are known to host extremely compact ($r \lesssim
100$\,pc) nuclear starbursts that reach above
1000\,$M_\odot$\,yr$^{-1}$\,kpc$^{-2}$ in star formation rate (SFR) surface density
\citep[$\Sigma_{\rm SFR}$; e.g.,][]{condon91, bm17, koala-goals, larson20,
song22}. In many cases, these starbursts are accompanied by powerful, but heavily-obscured AGN ($N_{\rm H} > 10^{23}$
cm$^{-2}$) that are detected only in the radio, mid-IR and hard X-rays
\citep[e.g.,][]{iwasawa11, stierwalt13, petric11, vardoulaki15, inami18,
torres18,ricci21}, and can even reach quasar luminosity
\citep[][]{boksenberg77,sanders88,rupke13}. These extreme nuclear activities also
drive powerful outflows that have been observed ubiquitously in local
U/LIRGs across different ISM phases and physical scales \citep[e.g.,][]{heckman90, rupke05, veilleux05,
veilleux13, spoon13, cicone14,hill14, cazzoli16, bm18, ps18, koala-goals, lutz20, fluetsch21,
perna21, lamperti22}, manifesting signs of rapid galactic scale transformation as predicted by
theoretical works. However, many studies associate these outflows with the nuclear starburst rather than AGN activity \citep[e.g.][]{rupke05-agn, lamperti22}, the role of which remains challenging to quantify as their signatures are often attenuated by the heavy nuclear dust extinction. \\
\indent The supreme spatial resolution and sensitivity of the \textit{James Webb Space Telescope} \citep[JWST;][]{jwst} have enabled a new, detailed view of the multi-phase ISM
in local U/LIRGs, providing low-extinction and spatially-resolved diagnostics on the properties of the interstellar dust, ionized, hot ($T \sim 1000 - 3000$\,K) and warm ($ 100 \lesssim T \lesssim 1000$\,K) molecular gas around their obscured nuclei on sub-kpc scales, as has been successfully
demonstrated with the Early Science Release program 1328 \citep[PI: L. Armus, A.
Evans; e.g.,][]{lai22,u22,lai23,armus23,rich23,bianchin24,buiten24,bohn24}. To
this end, we have obtained new \textit{JWST/MIRI-MRS} observations in Cycle-1 (1717;
PI: V. U) to study the ISM properties in seven local U/LIRGs from the Great Observatories All-sky LIRG
Survey \citep[GOALS;][]{armus09} selected based on their extraordinarily bright Near- and Mid-IR (NIR and MIR) \ce{H2} emission. These galaxies were first highlighted by \cite{stierwalt13, stierwalt14} based on their high \ce{H2}/PAH luminosity ratios measured with 
\textit{Spitzer/IRS} on $>$\,kpc scales, which were attributed to enhanced
rotational \ce{H2} emission from either AGN heating or
shocked warm molecular gas, rather than weak PAH emission that traces SF. Observations with Keck/OSIRIS similarly revealed enhanced NIR ro-vibrational \ce{H2} relative to Br$\gamma$ emission in the immediate vicinity of these galactic nuclei at spatial resolutions of $\lesssim \,$100\,pc \citep[e.g.][]{medling14,koala-goals}, pointing to nuclear activity as a key driver behind the enhanced \ce{H2} emission. These new \textit{JWST/MIRI-MRS} observations, 
further complemented by ancillary multi-wavelength IFU datasets from Keck,
VLT and ALMA, allow comprehensive multiphase diagnostics
on the nuclear power source and putative outflows in these bright \ce{H2}-emitters at matched, sub-kpc scales. Several case studies utilizing these datasets have been presented by \cite{u22}, \cite{bianchin24} and \cite{kader26}, who identified powerful multi-phase outflows driven by AGN photo-ionization and/or radio jets in the Seyfert galaxies NGC\,7469 and VV\,340a. \\ 
\indent In this work we focus on a starburst-dominated galaxy from the sample,
the late-stage galaxy merger and LIRG \citep[$L_{\rm IR} =
10^{11.87}$\,L$_\odot$;][]{armus09}, IRASF01364$-$1042 (hereafter as
IRASF01364), shown in Figure \ref{fig:fig1}. As noted by \cite{stierwalt14}, IRASF01364 shows distinctly high $L_{\ce{H2}}/L_{\rm PAH}$ and
$L_{\ce{H2}}/L_{\rm IR}$ ratios among galaxies in the GOALS sample, making it a
promising target to search for shocked molecular outflows at higher spatial resolution. While its compactness and high luminosity surface density in the radio suggest
an AGN presence \citep{vardoulaki,bm17,song22}, observations in the mid-IR (MIR), X-ray and
optical all indicate starburst activity as the dominant source of its bolometric luminosity
\citep{stierwalt13,veilleux95,yuan10,iwasawa11,ds17}. Interpretation
of these results is complicated by the relatively low spatial/spectral
resolution of the datasets, as well as high dust content in the source, as evidenced by its high IR/UV
ratio \citep[$\sim 250$;][]{howell10} and deep 9.7~$\mu$m silicate absorption feature
\citep[$s_{\rm 9.7} = -1.27$;][]{stierwalt13}. High-resolution ground-based
NIR IFU observations reveal a 200\,pc-radius rotating nuclear gas disk in
ro-vibrational \ce{H2} and Pa$\alpha$ emission, with the former showing
turbulent motion beyond the nuclear disk that potential traces a nuclear molecular outflow
\citep{medling14, medling15, koala-goals}.\\
\indent In this work we report the first confirmation and characterization of
a kpc-scale multi-phase biconical outflow and a dust-obscured AGN in IRASF01364,
combining new sensitive IFU observations from \textit{JWST/MIRI-MRS},
Keck/KCWI and ALMA. The article is organized as follows: In
Section \ref{sec:data} we present the observations utilized in this work and the data reduction procedures. We describe the analyses performed on the multi-wavelength datasets in Section
\ref{sec:analysis} and present the key results in Section \ref{sec:result} where we identify the kinematic and morphological signature of the outflow, as well as 
characterize the spatially-resolved properties of the dust and multi-phase gas in the system. We combine these results to infer the energetics of the newly-confirmed outflow and its origin in Section \ref{sec:discussion}, in the context of similar galactic outflows reported in the literature. We summarize the main findings in Section \ref{sec:summary}.\\
\indent Throughout this article, we adopt cosmological parameters $H_0 =
70$\,km\,s$^{-1}$Mpc$^{-1}$, $\Omega_{\rm vaccum} = 0.72$ and $\Omega_{\rm matter}
= 0.28$ based on the 5\,yr Wilkinson Microwave Anisotropy Probe result
\citep{hinshaw09} and the three-attractor model \citep{mould00}. IRASF01364 has a redshift of $z = 0.04819$ determined from optical stellar absorption features (see Section \ref{sec:kcwi_data}), corresponding to a
luminosity distance of $D_{\rm L} = 214\,$Mpc, with a scale of
945\,pc/$''$.
\begin{figure*}
    \centering
    \includegraphics[width=\linewidth]{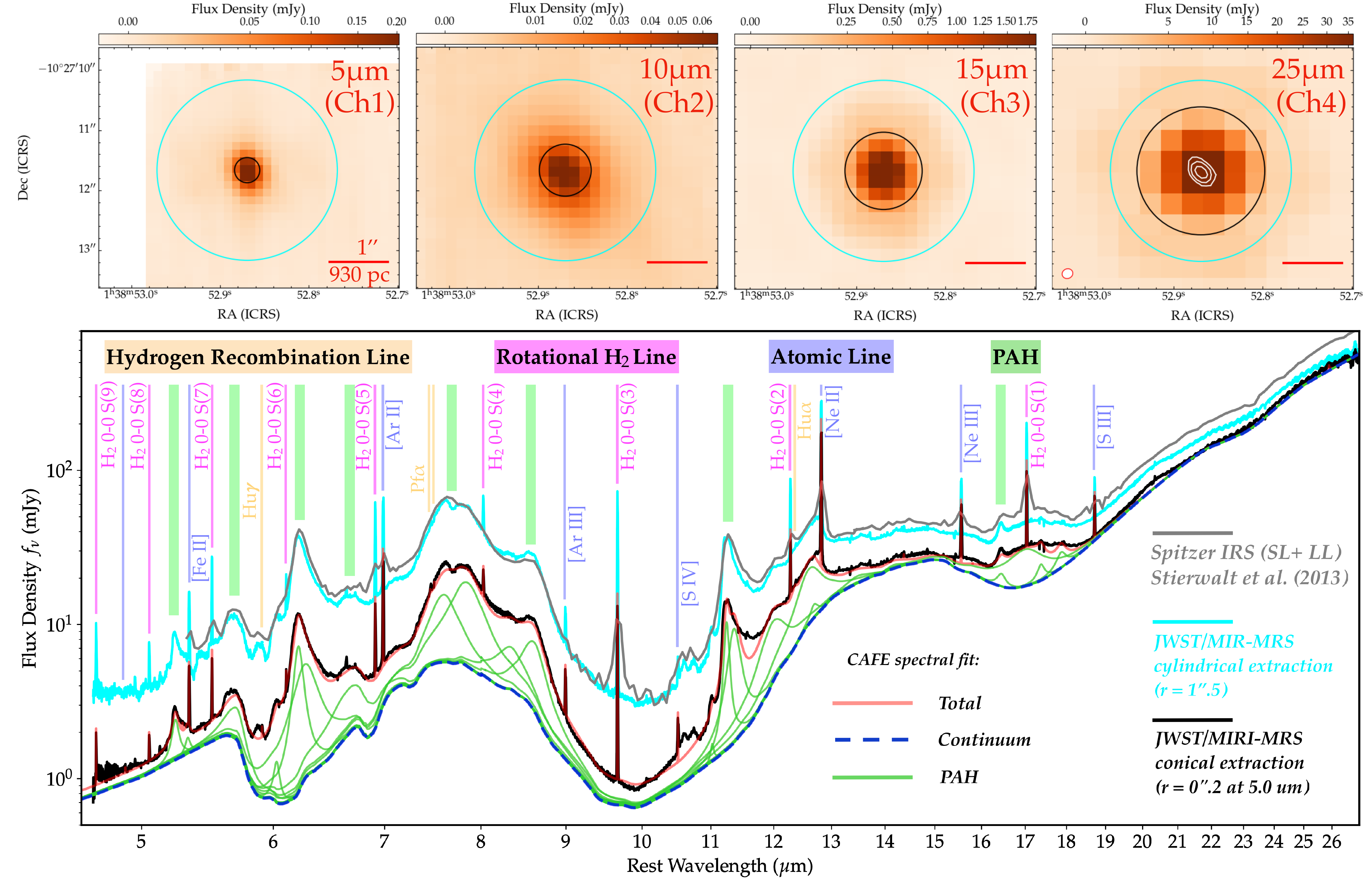}
    \caption{MIR emission of IRASF01364 across the \textit{MRS} wavelength range. \textit{Top:} Median flux density map at rest-frame 5, 10, 15, 25 $\mu$m from ch1, ch2, ch3 and ch4 cubes, respectively. All images are zoomed to the same scale, where the physical size corresponding to 1$''$ is indicated with the red scale bar in the lower right corner of each panel. The cyan circles, with a fixed radius of $r=1\farcs5$, indicate the cylindrical aperture used to extract the integrated MIR spectrum from the central $\sim 3$\,kpc of the galaxy that is covered by all channel cubes. The black circles indicate the expanding conical aperture, with $r=0\farcs2$ at 5\,$\mu$m and $r=1\farcs0$ at 25\,$\mu$m, used to extract the spectrum of the compact ($\sim$\,100\,pc) nuclear starburst, as portrayed by the ALMA 334\,GHz continuum, shown in white contours. The red ellipse indicates the synthesized beam size of the 334\,GHz continuum image (see Section \ref{sec:alma_data}). \textit{Bottom:} \textit{Spitzer/IRS} spectrum from \cite{stierwalt13} (in
    grey), compared to the full  \textit{MRS} spectra extracted with the fixed-radius, cylindrical aperture (cyan) and expanding conical aperture (black). For the latter, the total fitted spectral profile and the underlying
    continuum and PAH profiles from \texttt{CAFE} spectral decomposition are overlaid in red, dark blue
    (dashed) and green, respectively. See Section \ref{sec:miri_analysis} for details. While recovering the emission previously detected with \textit{Spitzer/IRS}, the high spectral resolution of \textit{MRS} allows a much more detailed characterization of the various MIR emission and absorption features present in IRASF01364.  \label{fig:nuc_spec}}
\end{figure*}
\section{Data \& Reduction} \label{sec:data}
\subsection{HST} \label{sec:hst_data}
To visualize the optical morphology of IRASF01364, we download calibrated \textit{HST} images in filters $F336W$, $F435W$ and $F814W$ \citep[ID: 10592, 16914; PI: A.
Evans;][Evans et al. in prep.]{kim13} from the the Mikulski Archive for
Space Telescopes Portal (MAST)\footnote[1]{\url{https://mast.stsci.edu}}. The central wavelengths and widths of the HST filters, in units of $\AA$, are 3361 and 554, 4297 and 1038, 8333 and 2511, respectively, for $F336W$, $F435W$ and $F814W$ \footnote[2]{\url{https://hst-docs.stsci.edu}}.The images are corrected for astrometric offsets using field stars with coordinates provided in Gaia DR3 catalogue \citep{gaiadr3} as initial references, and the corrections were visually fine-tuned using common features seen in the images, resulting in  an average correction of $-0\farcs3$ in RA and $+0\farcs4$ in DEC.  In Figure \ref{fig:fig1}, we show a zoomed-in view of the HST color composite image centered on the system, with $F336W$ (blue), $F435W$ (green) and $F814W$ (red) tracing emission from un-obscured young massive stars, tidal features and dusty materials, respectively. In the same Figure we further show the $F435W-F814W$ map to highlight the heavy and non-uniform dust extinction at and around the galaxy nucleus, underlining the necessity of \textit{JWST/MIRI} observations for uncovering its nature. The fields-of-view (FoVs) of the multi-wavelength IFU datasets (see the next Sections) are also visualized.

\subsection{JWST/MIRI-MRS} \label{sec:miri_data}
Mid-infrared IFU observations of IRASF01364 were conducted with the JWST
Mid-InfraRed Instrument \citep[MIRI;][]{rieke15} in Medium Resolution Spectroscopy (MRS) mode \citep[hereafter as \textit{MRS};][]{wells15} on
August 23, 2023, as part of Cycle-1 GO Program 1717 (PI: V. U). Exposures across
the short (A), medium (B), and long (C) sub-bands of each of the four wavelength
channels resulted in observations that cover the full 4.9 –
28.8\,$\mu$m range. The SLOWR1 readout pattern and standard 4-point dither pattern were
adopted in each sub-band observation of the science target, with an integration
time of 800\,s per exposure. Dedicated background exposures were taken within
each of the three sub-bands using a 2-point dither pattern. \\
\indent The uncalibrated frames were downloaded from MAST and reduced
using the JWST Science Calibration Pipeline version 1.18.0 \citep{bushouse23} and CRDS context of
\texttt{jwst\_1364.pmap}. We utilized the official
MIRI MRS Pipeline Notebook \citep{law25}\footnote[3]{\url{https://github.com/spacetelescope/jwst-pipeline-notebooks}},
which includes steps to apply detector-level corrections, outlier/cosmic-ray shower detection,
wavelength calibration, flux calibration, background subtraction, and
co-addition of dither frames. The final calibrated cubes have FoVs of $5\farcs1\times4\farcs3, 6\farcs3\times5\farcs3, 7\farcs8\times7\farcs0,$ and
$9\farcs5\times9\farcs5$ (see Figure \ref{fig:fig1}), with spaxel sizes of $0\farcs13$, $0\farcs20$, $0\farcs245$, and $0\farcs273$ for channel 1, 2, 3 and 4, respectively. We adopt wavelength-dependent spectral resolutions (R) characterized by \cite{mrs-res2}, which yield R($\lambda$) =  4603 - 128$\lambda$, where $\lambda$ is the observed wavelength in $\mu$m. We perform an astrometric correction by aligning the emission peak of the collapsed channel cubes with that of the 334\,GHz dust continuum image from ALMA (see Section \ref{sec:alma_data}), both of which are dominated by the IR-bright nucleus. The average corrections are $-0\farcs02$ in RA and $-0\farcs08$ in DEC, in line with the expected pointing performance of JWST \citep{mrs-astrometry}. Following \cite{lai23}, we model the point-spread-function (PSF) at each channel using \textit{MRS} observations of the  standard star HD 159222 (PID: 1050; PI: B. Vandenbussche), which yield Full-Width-Half-Maximum (FWHM) of $\sim 0\farcs27-0\farcs31$, $0\farcs38-0\farcs46$, $0\farcs53-0\farcs66$ and $0\farcs80 - 1\farcs05$ for ch1, ch2, ch3 and ch4, respectively.  In Figure \ref{fig:nuc_spec}, we display maps of the median flux values at rest-frame 5, 10, 15, 25\,$\mu$m to illustrate the increasing PSF size towards longer wavelengths. The full spectra extracted from all \textit{MRS} cubes using two different apertures are also shown in the same Figure along with \textit{Spitzer/IRS} spectrum of the entire galaxy \citep[extracted with a $10''\times4''$ slit;][]{stierwalt13} to showcase the high spectral resolution of the former. We describe the spectral extractions in Section \ref{sec:miri_analysis}.
\begin{figure*}[h!]
    \centering
    \includegraphics[width=\linewidth]{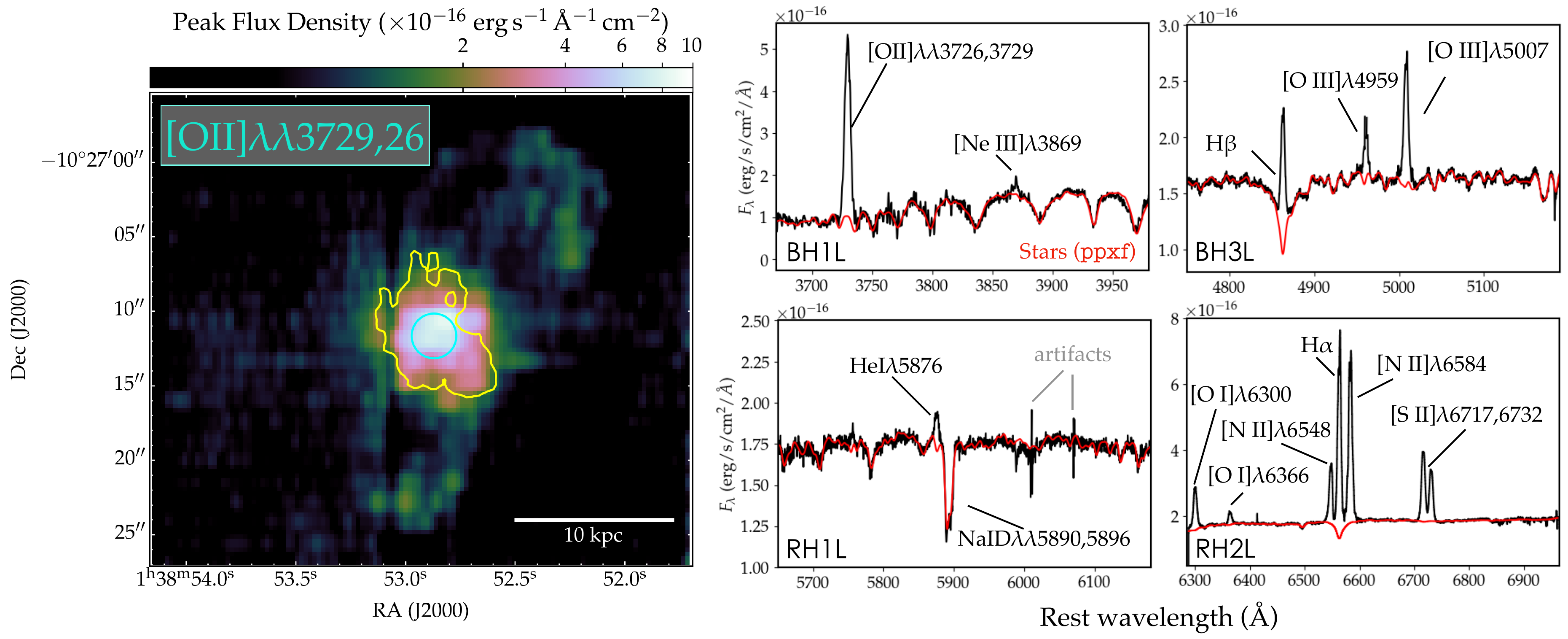}
    \caption{Optical emission of IRASF01364 as observed by Keck/KCWI. \textit{Left:} Map of the peak flux density of the [O\,II]$\lambda\lambda$3729,3726 line, overlaid with the 5$\sigma$ contour of the stellar continuum measured in the adjacent wavelength range, in yellow. For context, the cyan circle marks the aperture used to extract the integrated \textit{MRS} spectrum shown in Figure \ref{fig:nuc_spec}. The same aperture is used to extract the spectra shown in the \textit{right} panels. \textit{Right:} KCWI spectra extracted from the final reduced and aligned BH1L, BH3L, RH1L and RH2L cubes. The fitted stellar spectra derived from \texttt{ppxf} are shown in red. See Section \ref{sec:kcwi_analysis} for details. The [O\,II] emission map reveals extended ionized gas emission around IRASF01364 beyond the bulk of the stellar emission on 10\,kpc-scale. A number of other key emission lines are also detected, allowing further characterization of the ionized gas properties. 
    \label{fig:fig3}}
\end{figure*}
\subsection{ALMA} \label{sec:alma_data}
We utilize archival sub-arcsecond ALMA datasets for IRASF01364 from programs 2017.1.01235.S, 2018.1.00279.S (Band 3 and 7; PI: L. Barcos-Mu\~{n}oz) and 
2019.1.00811.S \citep[Band 6; PI: A. Medling;][]{agostino26} to compare the multi-frequency continuum and bulk molecular gas emission traced by CO (J = 1-0, 2-1, 3-2) with MIR emission probed by \textit{MRS}. To minimize biases related to interferometric spatial filtering, from these ALMA programs we select only observations that recover emission larger than $1''$ (i.e., the Largest Angular Scale, LAS $> 1''$) to more properly compare with \textit{MRS} observations. Details of these datasets are provided in Table \ref{tab:alma_info} in Appendix \ref{sec:ap_alma}. All observations are single-pointing, and we illustrate the Half-Power-Beam-Width (HPBW) of the Primary Beam ($\sim 24''$) at Band 6 in Figure \ref{fig:fig1}. \\
\indent We download the raw datasets from the ALMA Science Archive\footnote[4]{\url{https://almascience.eso.org/aq/}} and recover the calibrated
MeasurementSet (MS) for each dataset by rerunning the \texttt{scriptForPI.py}
script using CASA \citep{casa, casateam} in pipeline mode using the version listed in the associated QA2 report. We then image the continuum emission at each Band using CASA \texttt{tclean} after flagging line emission in each MS, combining MSs from multiple executions and/or array configurations, using the multi-term
multi-frequency deconvolver \citep{rau11} with \texttt{nterms} of 2 to account for the likely non-flat continuum spectral shape. Afterwards, we recover the flagged line emission in each MS, perform continuum subtraction using
\texttt{uvcontsub}, and image the CO line emission cube combining MSs covering the same CO transition, in native velocity channels of 10.5, 10.5 and 7.1\,km\,s$^{-1}$, respectively for the Band 3, 6 and 7 datasets. We adopt the \texttt{MultiScale} deconvolving algorithm \citep{mulitscale} with scales of 5, 10, 30 pixels to model the line emission with components of different sizes, with pixel cell sizes of $0\farcs05, 0\farcs06$ and $0\farcs07$ for the Band 3, 6 and 7 images respectively. For both the continuum and line imaging, we adopt Briggs weighting with robustness of \texttt{2.0} \citep{briggs} to emphasize the extended emission, and manually draw the clean masks iteratively until no significant residual remains down to the 1$\sigma_{\rm rms}$ threshold, where $\sigma_{\rm
rms}$ is the root-mean-square noise measured from an emission-free region in the dirty image, reported in Table \ref{tab:alma_info} along with other properties of the final cleaned images and cubes. The synthesized beams of the final imaging products have FWHM of $\sim 0\farcs3 - 0\farcs5$ across all three Bands. The contours from the Band 7 334\,GHz continuum image, which has the smallest synthesized beam, are also shown in Figure \ref{fig:nuc_spec}.

\subsection{Keck/KCWI} \label{sec:kcwi_data}
To complement the \textit{MRS} observation of the central region of IRASF01364, we further obtained new optical wide-field IFU observations with the Keck Cosmic Wave Imager \citep{kcwi} on the Keck II telescope (2023B\_K454; PI: R. McGurk). The observations were carried out on September 17, 2023 under clear sky conditions with seeing of $0\farcs9$ using a two-point dither pattern with a dedicated off-target sky frame. The BH1, BH3, RH1 and RH2 gratings were adopted to cover wavelength ranges of 3500–4100, 4700–5600, 5900-6485 and 6592-7297~\AA{},
respectively, with a high spectral resolution of $R \sim 3500$ and $\sim
4500$, for RH and BH gratings, respectively. Utilizing the large IFU slicer, we achieve a FoV of 33$''$
$\times$ $20\farcs4$ per pointing with slit width-limited angular resolution
($0\farcs3$ sampling rate along the $1\farcs4$-wide slit). The detector was spatially binned by
a factor of 2$\times$2. Basic reduction and wavelength solutions for the
spectroscopic data were performed using the official KCWI data reduction
pipeline \texttt{KCWI\_DRP} v1.0
\footnote[5]{\url{https://kcwi-drp.readthedocs.io}}, which also performs flux
calibration and sky subtraction using dedicated sky pointings. The individual spectral cubes were cropped and co-added using the
\texttt{CWITools} Python package \citep{osullivan20}. The reduced data cubes have a common FOV of $36''\times 32''$ in pixel
scale of $0\farcs29$, outlined in Figure \ref{fig:fig1} in cyan, and the instrumental
FWHM for the emission lines was measured to be 0.9\,\AA.  We use the WCS-corrected \textit{HST} F336W and F435W images (see Section \ref{sec:hst_data}) as reference for
fine-tuning the astrometry of the KCWI cubes until the UV/optical continuum morphology traced by these datasets visually match. The average corrections are $-0\farcs8$ in RA and $+0\farcs7$ in DEC for the RH gratings and  $-1\farcs1$ in RA and $-2\farcs3$ in DEC for the BH gratings. \\
\indent In Figure \ref{fig:fig3} we showcase the data quality, in particular highlighting the extended ionized gas emission as traced by the [O\,II]$\lambda\lambda$3729,3726 doublet covered by the BH1L cube, as well as spectra extracted from all four cubes, which we analyze in more detail in Section \ref{sec:kcwi_analysis}. As the BH1L spectrum shows the most prominent stellar absorption features, we derive the source redshift from this spectrum by performing spectral fitting with \texttt{ppxf} \citep{ppxf} and the FSPS stellar templates
\citep{conroy09,conroy10}. The resulted redshift is $z = 0.04819\pm0.00012$ accounting for the uncertainties associated with the spectral resolution ($\sim\,34\,$km/s). We experiment with different aperture radii (i.e., 0\farcs5, 1\farcs0, 2\farcs0) for spectral extraction, which yield consistent results. The derived value also agrees well with the redshift (i.e., 0.04823) reported in the NASA/IPAC Extragalactic Database (NED\footnote[6]{\url{https://ned.ipac.caltech.edu/}}), obtained from SDSS DR13 \citep{sdssdr13}. The wavelength ranges spanned by the data cubes cover important emission lines for diagnostics of the excitation, dust extinction and density of the ionized
gas, e.g.,[O\,II]$\lambda\lambda$3729,3726, H$\beta$,
[O\,III]$\lambda\lambda$4959,5007, [O\,I]$\lambda\lambda$6300,6366, H$\alpha$,
[N II]$\lambda\lambda$6548,6584 and [S II]$\lambda\lambda$6717,6732, etc.
Additionally, the RH1L grating covers Na\,ID absorptions at 5890 and
5896~\AA{}. Due to the redshift of IRASF01364, the H$\alpha$-[N\,II] spectral complex coincides with the atmospheric telluric O$_2$\,B-band absorption
feature. We remove this feature via an iterative process using a model
O$_2$\,B-band spectrum assuming $T=296\,$K, downloaded from the GEISA 2015
atmospheric molecular line database\footnote[7]{\url{https://geisa.aeris-data.fr/geisa-2015}}, after applying a
small (1.75~\AA{}) shift and scaling factor to the absorption line widths,
until the continuum level within and outside the H$\alpha$-[N\,II] complex
match.  

\section{Analysis}\label{sec:analysis} 
\subsection{MIRI-MRS}\label{sec:miri_analysis}
\subsubsection{Spectral decomposition \& PAH fitting with \texttt{CAFE}}\label{sec:cafe_fitting}
To conduct an initial assessment of the MIR spectral properties of IRASF01364, we utilize the spectral fitting tool \texttt{CAFE}\footnote[8]{\url{https://github.com/GOALS-survey/CAFE}}\citep{marshall07, pycafe}. The program performs
extraction and stitching of spectra from \textit{MRS} channel cubes, using the longest-wavelength cube as reference, as well as decomposition of the stitched full spectrum
into multi-temperature continuum, emission
lines, and PAH components. Designed to allow flexible modeling of the complex and often blended PAH profiles, the program also computes wavelength-dependent attenuation \citep[e.g.][]{lai24} due to absorptions from silicates, water ice and hydrogenated amorphous carbon for estimating the intrinsic fluxes of individual PAH features.\\
\indent As the MIR emission of IRASF01364 is dominated by the nuclear starburst, we first focus on analyzing the nuclear spectrum across all \textit{MRS} channels extracted with \texttt{CAFE}. To isolate emission from the nuclear starburst, which spans $\sim0\farcs4$ in the 334\,GHz continuum (see Figure \ref{fig:nuc_spec}), we adopt a conical aperture centered on the 334\,GHz emission peak (i.e., RA$=24.72029\deg$, DEC=$-10.45324\deg$) with a radius of $0\farcs2$ at 5.0~$\mu$m and $1\farcs0$ at 25.0\,$\mu$m, to account for the increasing \textit{MRS} PSF with wavelength (see Section \ref{sec:miri_data}). As the chosen aperture size is twice the PSF FWHM at any given wavelength, no aperture correction was performed. The aperture sizes and the extracted spectrum, along with the results of spectral decomposition with \texttt{CAFE} are shown in Figure \ref{fig:nuc_spec}. We later refer to this spectrum as the ``nuclear'' spectrum. The observed and attenuation-corrected fluxes of the PAH features obtained from \texttt{CAFE} are listed in Table \ref{tab:cone_pah}. \\
\indent In Figure \ref{fig:nuc_spec}, we also show, in cyan, another spectrum extracted with a cylindrical aperture with a fixed radius of $r = 1\farcs5$. This is chosen to capture the total MIR emission from the central 3\,kpc, which is the largest area covered by all \textit{MRS} channels. This spectrum closely matches that obtained with \textit{Spitzer/IRS} with a slit size of $\sim 10''\times 4''$ \citep{stierwalt13, stierwalt14}, shown in grey, confirming that the bulk of the mid-IR emission from the system is concentrated in the central few kpc. We later refer to the area defined by this cylindrical aperture as the ``central" region or region C, for which we also extract profiles of the sub-mm CO lines from ALMA datasets (Section \ref{sec:alma_analysis}) and the optical emission lines from Keck/KCWI cubes (Section \ref{sec:kcwi_analysis}) to perform a direct comparison between multi-phase gas tracers observed within the central $\sim3$\,kpc of the galaxy. The PAH fluxes obtained with \texttt{CAFE} from this central region are also reported in Table \ref{tab:cone_pah}.
\begin{figure}[b!]
    \centering
    \includegraphics[width=0.9\linewidth]{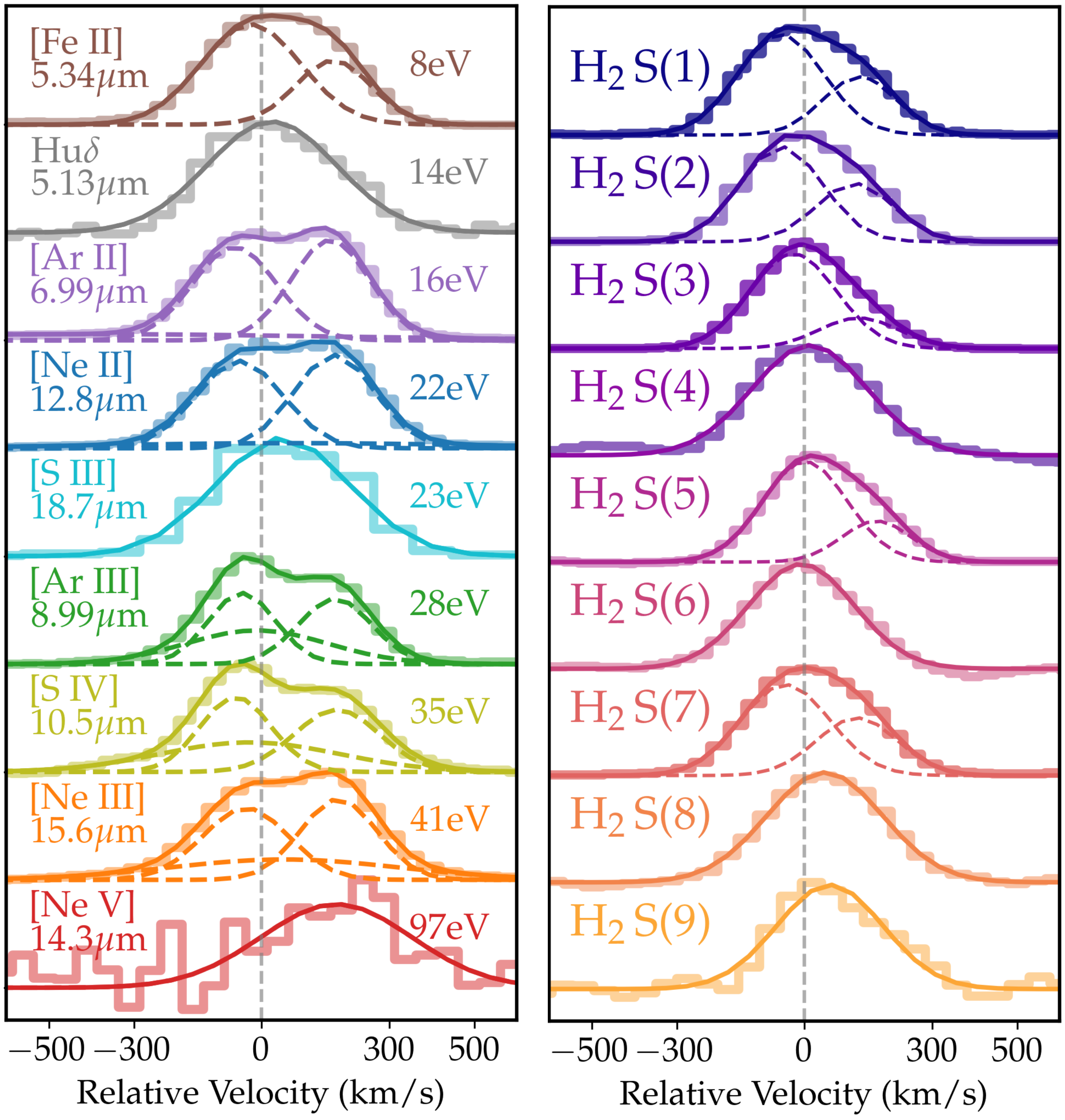}
    \caption{Spectral profiles of atomic (\textit{left}) and rotational \ce{H2}
    lines (\textit{right}) detected in the \textit{MRS} nuclear spectrum, defined
    in Section \ref{sec:cafe_fitting} and shown in Figure \ref{fig:nuc_spec} (in black). All profiles are normalized by the peak line flux density, displayed in velocity space relative to the expected line center based on the redshift of the galaxy (i.e., $z = 0.04819$) and offset vertically for visualization. From top to bottom, the atomic lines are arranged in order of increasing ionizing potential, and the \ce{H2} lines are ordered from \ce{H2}~0-0~S(1) to \ce{H2}~0-0~S(9). The best-fit models are displayed in solid lines, with individual fitted Gaussian components shown in dashed lines. The best-fit parameters are presented in Table \ref{tab:cone_line}. See Section \ref{sec:multi-gauss} for details. Most ionized gas tracers and some \ce{H2} lines show signs of multiple kinematic components, while broad components with FWHM $> 500$\,km\,s$^{-1}$ are only detected in the former.
    \label{fig:fig4}}
\end{figure}
\begin{figure*}[t!]
    \centering
    \includegraphics[width=0.8\linewidth]{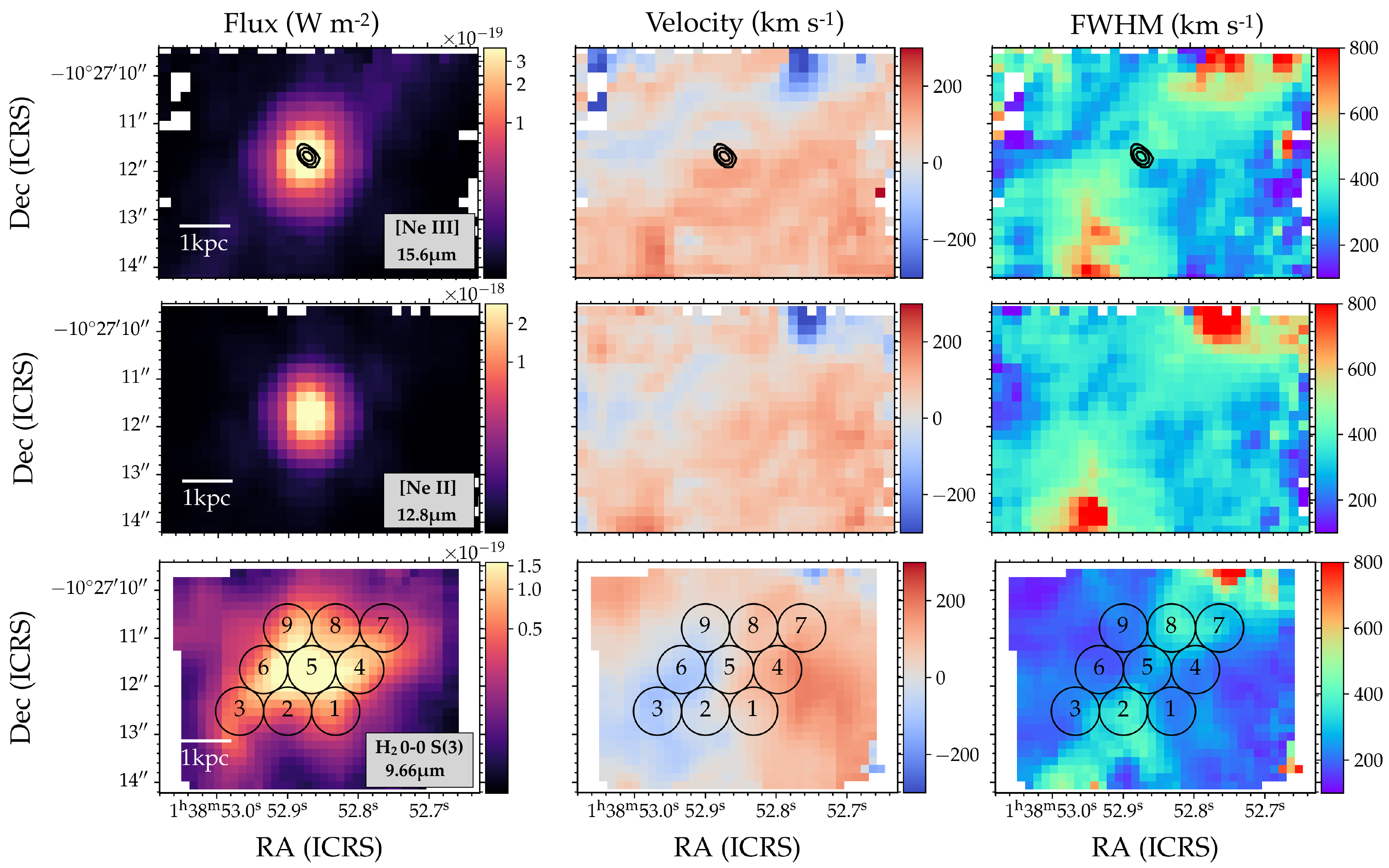}
    \caption{\textit{MRS} maps of the integrated line flux, velocity and FWHM linewidth for [Ne\,III], [Ne\,II] and \ce{H2}~0-0~S(3) lines (see Section \ref{sec:regions}). Spaxels with SNR $< 3$ in each line are masked. ALMA 334\,GHz continuum contours from Figure \ref{fig:nuc_spec} are overlaid in the [Ne\,III] maps, and the spectral extraction regions defined in Section \ref{sec:regions} are overlaid on the \ce{H2}~0-0~S(3) maps. While ionized gas tracers ([Ne\,II], [Ne\,III]) show compact emission at the nucleus, molecular gas traced by the \ce{H2} line shows extended ``X''-shaped morphology on $\sim$\,3\,kpc scales. Ionized and molecular gas  both show enhanced gas dispersion along the SE-NW direction.
    \label{fig:fig5}}
\end{figure*}
\begin{figure}[h!]
    \centering
    \includegraphics[width=0.85\linewidth]
    {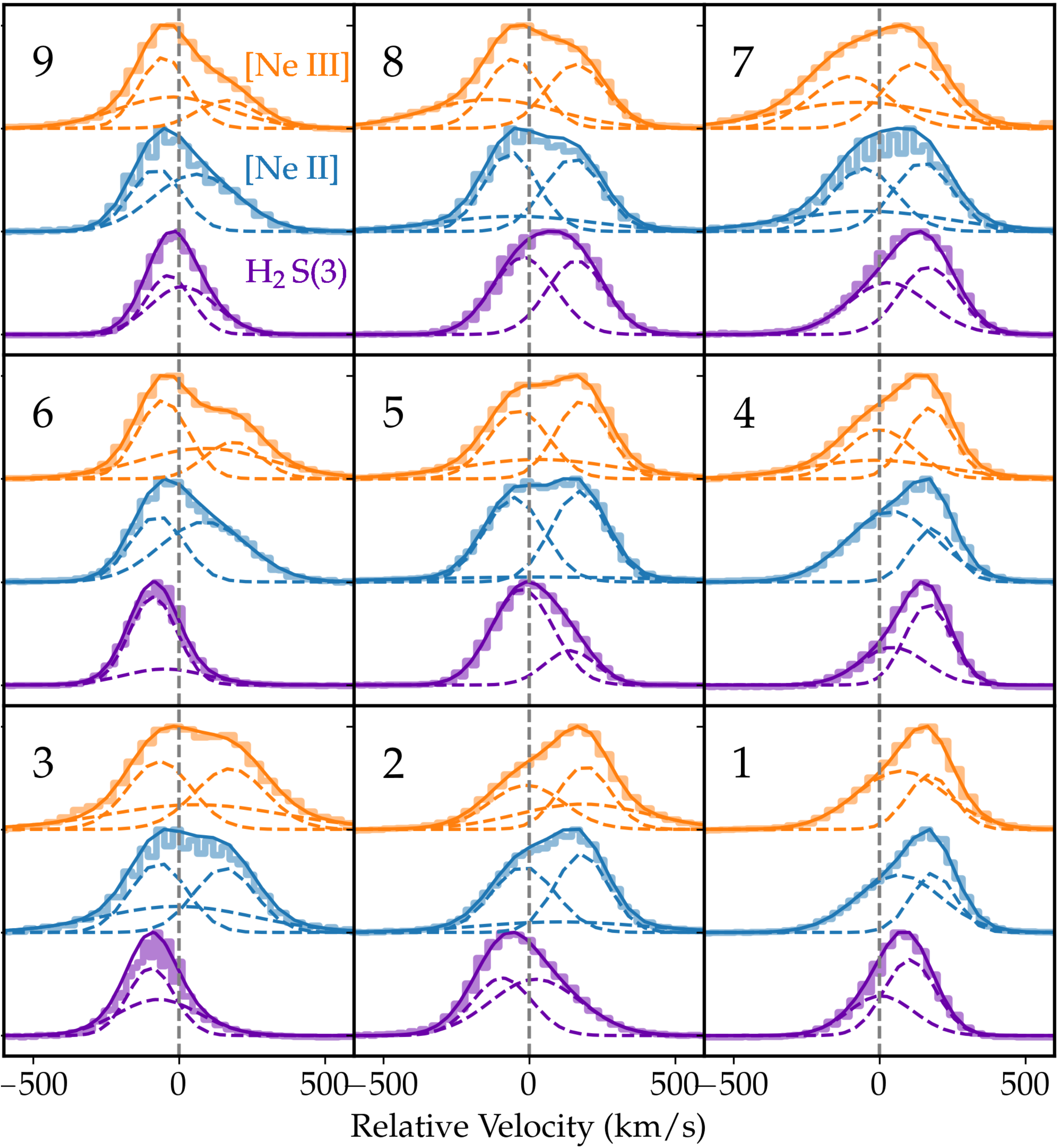}
    \vspace{-0.05cm}
    \caption{Line profiles of [Ne\,III], [Ne\,II] and \ce{H2}~0-0~S(3) extracted from the nine regions defined in Section \ref{sec:regions} and shown in Figure \ref{fig:fig5}. Profiles are normalized by the peak flux density and offset vertically for better visual distinction, with best-fit model and components shown in solid and dashed lines. Broad wings are seen across most regions in the neon lines but not in \ce{H2}~0-0~S(3), which also has overall narrower profiles. 
    \label{fig:fig6}}
\end{figure}
\begin{figure*}[h!]
    \centering
    \includegraphics[width=0.85\linewidth]
    {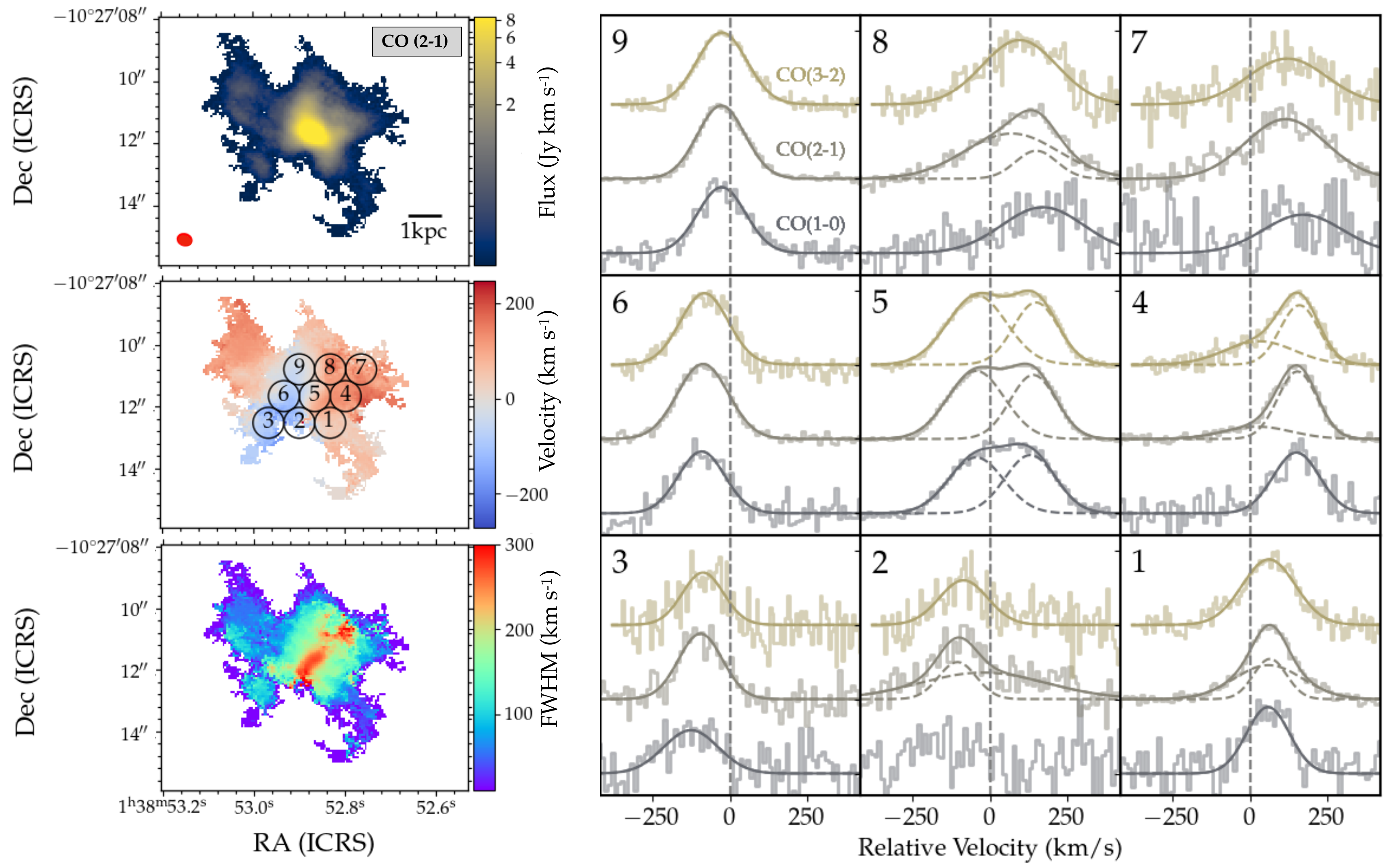}
    \vspace{-0.05cm}
    \caption{Spectral analysis of the ALMA datasets. \textit{Left:} Moment maps of integrated flux, velocity and linewidth FWHM of the CO\,(J=2-1) emission, with the spectral extraction regions defined in Section \ref{sec:regions} (same as in Figure \ref{fig:fig5}) overlaid on the velocity map. The synthesized beam is indicated with a red ellipse in the lower left corner of the flux map. \textit{Right:} Extracted line profiles of multi-J CO lines. Profiles are normalized by the peak flux density and offset vertically for better visual distinction, with best-fit model and components shown in solid and dashed lines. The cold molecular gas traced by CO\,(2-1) shows similar morphology and large-scale velocity gradient along E-W direction as the warm molecular gas (Figure \ref{fig:fig5}). Line profiles are sufficiently characterized by a single component across most regions except for in region 5, where all three transitions show consistent, double-peaked profiles. 
    \label{fig:fig7}}
\end{figure*}
\begin{figure}[h!]
    \centering
    \includegraphics[width=0.9\linewidth]
    {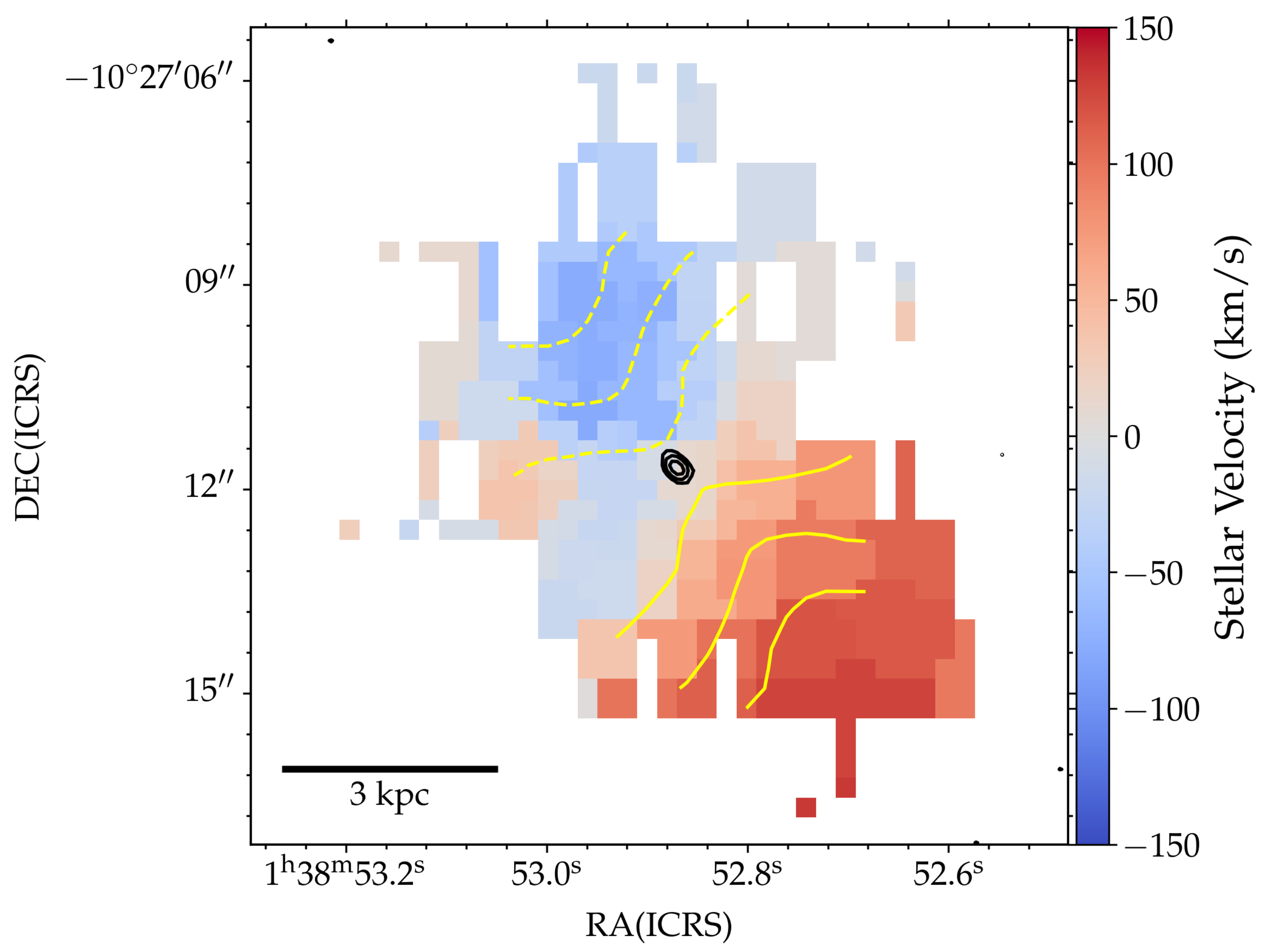}
    \vspace{-0.05cm}
    \caption{Line-of-sight stellar velocity map derived from Keck/KCWI BH1L cube using \texttt{ppxf}. Only spaxels with stellar emission stronger than SNR $> 5$ (i.e., within the yellow contour in Figure \ref{fig:fig3}) are shown. See Section \ref{sec:kcwi_analysis} for details. ALMA 334\,GHz continuum contours from Figure \ref{fig:nuc_spec} are shown in black. The velocity field is reproduced by model of a rotating disk at inclination $i \sim$\,60$^\circ$ and position angle PA $\sim 220^\circ$, shown in yellow contours, in levels of [40, 70, 90]\,km\,s$^{-1}$, with blue-shifted velocities shown in dashed lines (see Section \ref{sec:outflow}). 
    \label{fig:fig8}}
\end{figure}
\subsubsection{Multi-component fitting of emission lines}\label{sec:multi-gauss}
\indent As \texttt{CAFE} is designed to model the full MIR spectrum and PAH features rather than individual emission lines \citep{marshall07}, we further perform more detailed multi-component Gaussian fitting with \texttt{lmfit}\footnote[9]{\url{https://lmfit.github.io/lmfit-py}} \citep{lmfit} of each emission line to more robustly characterize their flux and kinematics. For each line, we first model and subtract the local continuum in adjacent line-free regions between 0.02 to 0.06\,$\mu$m away from the expected central wavelength, using a third-degree Legendre polynomial. Several atomic lines exhibit double-peaked profiles and broad line wings, indicating the presence of multiple kinematic components. Hence, we model the continuum-subtracted line profile three times, with one, two and three Gaussian components, respectively. Each component is restricted to have a central velocity within [-500, 500]\,km\,s$^{-1}$ of the line center, and a linewidth larger than the instrumental linewidth at the observed wavelength (see Section \ref{sec:miri_data}). We adopt the default Levenberg-Marquardt algorithm for the least-square minimization, accounting for the flux uncertainties associated with the spectrum by providing its reciprocal as weights to the residual. We then choose among the three different models the one with the lowest value of the Bayesian Information Criterion (BIC) as the best-fit that sufficiently characterizes the line profile without overfitting. We show the fitting results for Hu$\delta$, \ce{H2}~0-0~S(2) and [Ne\,III]\,15.6~$\mu$m in Appendix \ref{sec:model-selection} as examples to illustrate the model selection process. \\
\indent Besides the lines already identified by \texttt{CAFE} (see Figure \ref{fig:nuc_spec}), we also search for highly-ionized coronal lines from \cite{satyapal21} that may reveal elusive AGN activity in IRASF01364. Notably, the [Ne V]\,14.3\,$\mu$m line with ionizing potentials (IP) of 97\,eV is detected at SNR$> 5$ in both the nuclear and central spectra, while other coronal lines are either not detected or heavily blended with weak PAH or molecular features that prevents reliable identification (e.g., [Mg V]\,13.5\,$\mu$m; Buiten et al. in prep.). We leave more comprehensive coronal line identification and characterization to future work (U et al. in prep.) and focus only on the [Ne\,V]\,14\,$\mu$m line in further analysis (see Section \ref{sec:cl}). In Figure \ref{fig:fig4} we show the line profiles and best-fit models of several emission lines from the nuclear spectrum, i.e., extracted from the conical aperture as defined in Section \ref{sec:cafe_fitting}. The best-fit parameters and the corresponding uncertainties from \texttt{lmfit} for all emission lines identified in the nuclear and central spectra are listed in Table \ref{tab:cone_line}. Most ionized gas tracers show clear double-peaked profiles, with many showing a third broad component with FWHM $> 500$\,km\,s$^{-1}$. While some \ce{H2} lines also show signs of multiple kinematic components, they generally lie close to the expected velocity and have relatively narrow line widths with FWHM $\lesssim 300$\,km\,s$^{-1}$. We note that, unless otherwise specified, we do not apply the MIR attenuation derived from \texttt{CAFE} to the MIR line fluxes reported throughout this work, given that PAH, ionized and molecular gas have different physical origins and spatial distributions as shown in spatially-resolved multi-wavelength studies of nearby galaxies \citep[e.g.,][]{tielens08,egorov25}.
\begin{figure*}[t!]
    \centering
    \includegraphics[width=0.8\linewidth]
    {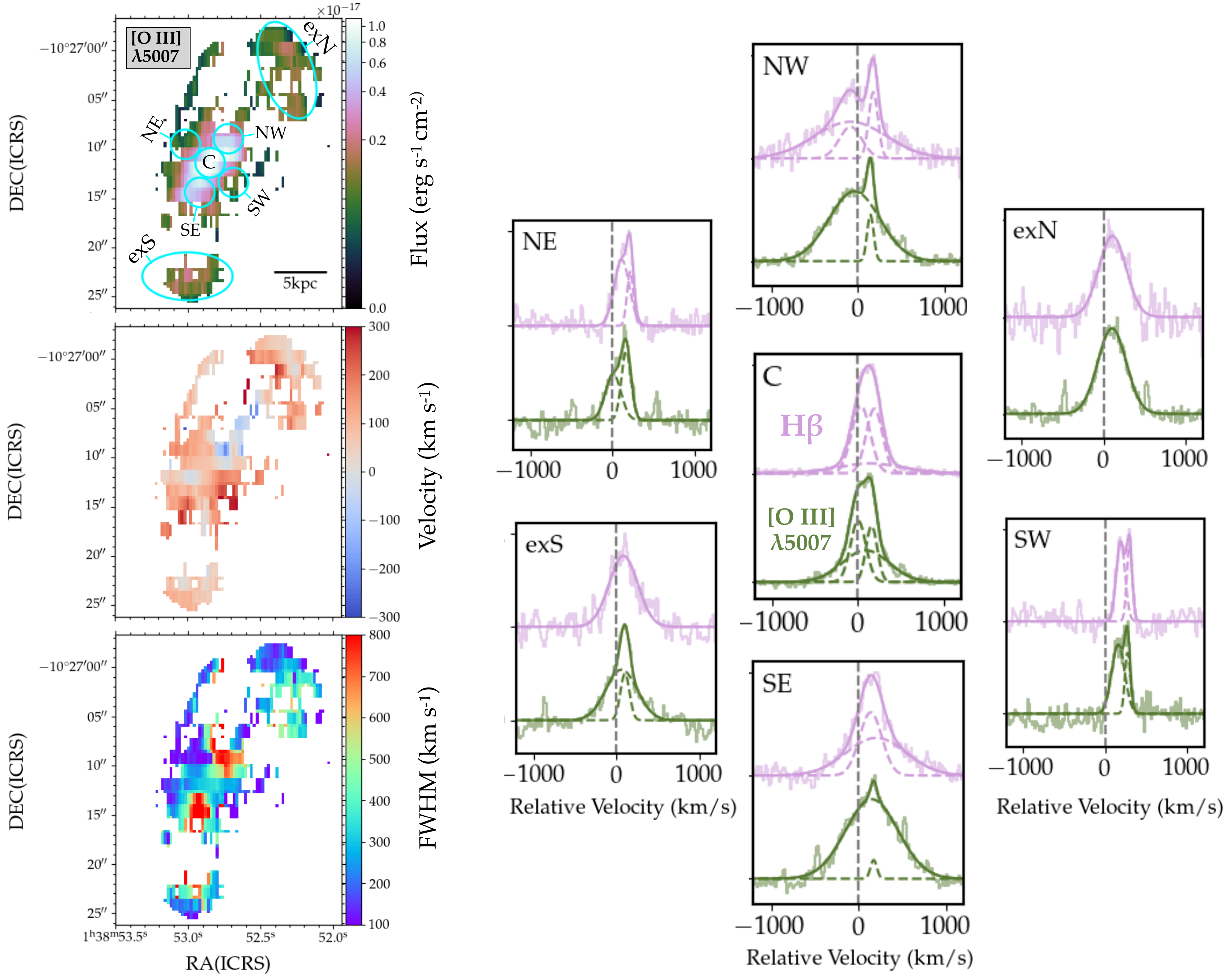}
    \vspace{-0.05cm}
    \caption{Spectral analysis of the Keck/KCWI datasets. \textit{Left:} Maps of integrated flux, velocity and linewidth FWHM of [O\,III]$\lambda$5007 emission, with the spectral extraction regions defined in Section \ref{sec:kcwi_analysis} overlaid on the integrated flux map. \textit{Right:} Extracted line profiles of H$\beta$ and [O\,III]$\lambda$5007, normalized by the peak flux density and offset vertically for better visual distinction, with best-fit model and components shown in solid and dashed lines. The ionized gas as traced by [O\,III]$\lambda$5007 extends beyond $\sim\,10$\,kpc from the nucleus along the SE-NW direction, with no clear sign of rotation in the central region as seen in the stellar kinematics (Figure \ref{fig:fig9}). Line profiles of both H$\beta$ and [O\,III]$\lambda$5007 exhibit broad wings along the SE-NW direction, i.e., kinematic minor axis of the central stellar disk.  
    \label{fig:fig9}}
\end{figure*}
\subsubsection{Emission line maps \& additional spectral fits} \label{sec:regions}
\indent To visualize the spatially-resolved gas distribution and kinematics, we further generate flux, velocity and FWHM linewidth maps of the emission lines via performing, on each spaxel, the same line fitting procedure as described in Section \ref{sec:multi-gauss} with \texttt{lmfit}. Even for the brightest lines, the per-spaxel level signals are insufficient for robust characterization with more than one Gaussian component across the FoV of the \textit{MRS} channel cubes, therefore we show examples of the resulting maps from single-Gaussian fitting, for [Ne\,III], [Ne\,II]  and \ce{H2}~0-0~S(3) in Figure \ref{fig:fig5}. The warm molecular gas as traced by the \ce{H2}~0-0 S(3) line, shows extended ``X''-shaped structures on $\sim 5$\,kpc scales, while the ionized gas traced by the neon lines appear mostly concentrated around the nucleus. The ionized and warm molecular gas also appear to move along perpendicular directions, as suggested by the velocity maps shown in the center column, while all gas tracers show broadened linewidths along the SE-NW direction. We elaborate on these findings in Section \ref{sec:result} and \ref{sec:discussion}. \\
\indent To investigate variations in the MIR spectral properties along the extended molecular emission, we perform additional spectral extractions and decomposition with \texttt{CAFE} in nine non-overlapping cylindrical apertures that cover the spatial extent of the \ce{H2}~0-0 S(3) emission. We set a uniform radius of $0\farcs5$ for these apertures to ensure sufficient SNR for detailed spectral fitting of the emission line profiles while fully encompassing the PSF at the wavelength of the [S\,III]\,18.7\,$\mu$m line (FWHM $\sim 0\farcs86$), which is the longest-wavelength emission line of interest. The regions defined by these apertures are also visualized in Figure \ref{fig:fig5}, and are later referred to as regions 1 to 9, where region 5 is centered on the nucleus. To further compare the resolved kinematics between the ionized and warm molecular gas, in Figure \ref{fig:fig6} we show the line profiles (normalized by the peak flux density) and best-fit models for [Ne\,II], [Ne\,III] and \ce{H2}~0-0~S(3). We present the best-fit parameters of these emission line profiles in Table \ref{tab:reg_line}, and the decomposed fluxes of the brightest PAH features from \texttt{CAFE} in Table \ref{tab:more_pah}. Across regions 1 - 9, the \ce{H2}\,0-0\,S(3) line can be sufficiently modeled with two narrow components while the neon lines consistently show a broad component in region 2, 3, 5, 7, and 8, corresponding to the areas of enhanced gas dispersion from the emission line maps in Figure \ref{fig:fig5}. 
\subsection{ALMA}\label{sec:alma_analysis}
To visualize the distribution and kinematics of the cold molecular gas, we generate moment 0, 1 and 2 maps from the calibrated ALMA CO cubes from Section \ref{sec:alma_data}, using the 3D source-finding tool \texttt{Duchamp} \citep{duchamp}. As an example, in Figure \ref{fig:fig7} we show the maps for the CO (J = 2 - 1) emission at rest-frame frequency ($\nu_{\rm rest}$) of 230.538\,GHz, which exhibits extended structures similar to those of \ce{H2}~0-0~S(3) emission (see Figure \ref{fig:fig5}). To study the spatial variation of the cold molecular gas kinematics, we repeat the analysis described in Section \ref{sec:regions} and extract and model the multi-J CO line profiles across regions 1 to 9. The results are also visualized in Figure \ref{fig:fig7}, and best-fit values are presented in Table \ref{tab:reg_co}, along with those obtained for region C, as defined in Section \ref{sec:cafe_fitting}. The corresponding maps for CO\,(J = 1 - 0) at $\nu_{\rm rest} = 115.271$\,GHz, and CO\,(J = 3 - 2) at $\nu_{\rm rest} = 345.796$\,GHz, which are detected with overall lower SNR, are shown in Appendix \ref{sec:ap_alma}. All three CO lines are sufficiently modeled by a single Gaussian in most regions except for region 5. 
\subsection{KCWI}\label{sec:kcwi_analysis}
To properly recover the flux distribution and kinematics of the emission lines detected in each KCWI cube over the large FoV (e.g., Figure \ref{fig:fig3}), we perform simultaneous modeling of the emission line profile and the underlying stellar continuum at each spaxel with
\texttt{ppxf}. This is done in two stages: First, Voronoi binning is performed on each cube so that each bin reaches SNR of 60 for the brightest line detected within the wavelength range covered by the cube. We then fit the stellar spectrum at each bin after masking the emission lines. Afterwards, we repeat the fitting at each spaxel including both stellar and gas components, where the former is set by rescaling the best-fit stellar template derived for the corresponding Voronoi bin in the first stage, and the latter is modeled by a single Gaussian. At this stage, we set the initial guess for the gas velocities to the stellar velocity derived at each spaxel, and tie the kinematics of emission lines together to further reduce degeneracy, given that some lines are blended (e.g., [O\,II], [S\,II] doublets). In Figure \ref{fig:fig8} we show the stellar velocity field derived from the BH1L cube where the stellar features are most prominent. In Figure \ref{fig:fig9}, we show the emission line maps for [O\,III]$\lambda$5007, which best showcase the large-scale kinematics of the extended ionized gas emission detected out to $\sim\,10$\,kpc from the nucleus. \\
\indent To investigate the ionized gas kinematics in more detail, we extract spectra from seven regions covering different parts of the ionized gas emission, which are respectively referred to by their locations as the ``exS'', ``SE'', ``SW'',``C'',``NE'', ``NW'', and ``exN'' regions, and illustrated in Figure \ref{fig:fig9}. The ``C'' region was initially defined in Section \ref{sec:cafe_fitting}. Region SE, SW, C, NE and NW are circular with radii of $r=1\farcs5$ for matched-scale comparison across area of bright [O\,III]$\lambda$5007 emission, while region exS and exN are elliptical and larger, with semi-axes of $\sim 2\farcs5 \times 5''$, to encompass the fainter extended emission. We then conduct multi-component Gaussian fitting of the brightest emission lines detected in each spectrum following the same procedures described in Section \ref{sec:multi-gauss}, after fitting and subtracting the underlying stellar component with \texttt{ppxf} (see Figure \ref{fig:fig3} for examples). We set the line flux ratios of the [O\,III]$\lambda$5007/$\lambda$4959 and [N\,II]$\lambda$6584/$\lambda$6548 to the theoretical value of 3, and restrict those of the [O\,II]$\lambda$3729/$\lambda$3726 and [S\,II]$\lambda$6716/$\lambda$6731 to vary only between the theoretical limits of 0.38 - 1.45 and 0.44 - 1.45 \citep{sanders16}, respectively. The line profiles of [O\,III]$\lambda$5007 and H$\beta$ extracted from all seven regions are shown in the right panels of Figure \ref{fig:fig9}. We report the best-fit parameters of individual components for the [O\,III]$\lambda$5007 and H$\beta$ lines in Table \ref{tab:reg_kcwi}. Both lines show prominent broad components with FWHM $\gtrsim 600$\,km/s in region C, SE and NW. Detailed modelling of the kinematics of other key diagnostic lines is challenging given either the much fainter emission (e.g., [O\,I]$\lambda$6300, [S\,II]$\lambda\lambda$6716,31) or heavily blended profiles (H$\alpha$ and [N\,II]$\lambda\lambda$6548,84, [O\,II]$\lambda\lambda$3726,29), therefore we present only the total flux of these other lines in Table \ref{tab:reg_kcwi_other}.\\
\begin{figure*}[t!]
    \centering
    \includegraphics[width=1.0\linewidth]
    {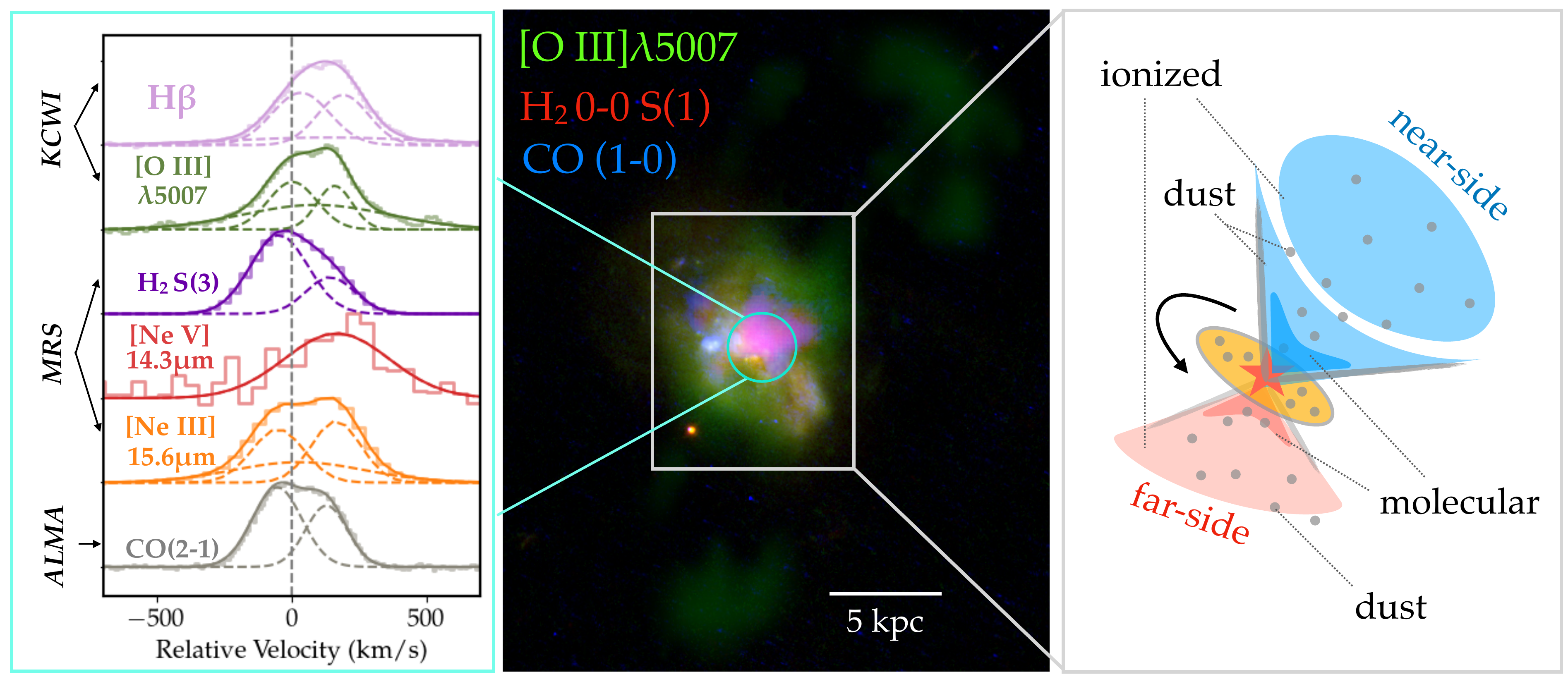}
    \vspace{-0.05cm}
    \caption{Multi-phase gas morphology and kinematics derived from Keck/KCWI, \textit{JWST/MIRI-MRS}, and ALMA data. \textit{Left:} Comparison of line profiles of emission lines detected in the multi-wavelength datasets in the central 3\,kpc in region C, defined in Section \ref{sec:cafe_fitting} (see also Figure \ref{fig:nuc_spec}). Profiles are normalized by the peak flux density and offset vertically for better visual distinction, with best-fit model and components shown in solid and dashed lines. \textit{Middle:} Color-composite overlay of
    ALMA CO\,(1-0), \textit{MRS} \ce{H2}\,0-0\,S(1) and Keck/KCWI [O\,III]$\lambda$5007 emission 
    on top of the F336W/F435W/F814W tri-color \textit{HST} image from Figure \ref{fig:fig1}, showcasing the
    multi-phase biconical structure emerging from the dust-obscured nucleus of IRASF01364. The cyan ellipses highlight the regions defined in Section \ref{sec:kcwi_analysis} for analyzing the large-scale ionized gas emission. 
    \textit{Right:} Simplified representation of the molecular and ionized outflow bicones with respect to the stellar disk (in yellow ellipse) and the nucleus (red star). The black curved arrow indicates the rotation of the stellar disk inferred from Figure \ref{fig:fig8}. The cold and warm molecular gas are co-spatial and trace the inner edges of the ionized bicone, and appear brighter and more extended on the near-side NW cone.
    \label{fig:fig10}}
\end{figure*}
\section{Results} \label{sec:result}   
\subsection{A Multi-phase Biconical Outflow}\label{sec:outflow}
Despite the different gas phases and spatial scales probed by our multi-wavelength datasets, the emission line maps separately presented in Figure \ref{fig:fig5}, \ref{fig:fig7} and \ref{fig:fig9} all exhibit consistent, prominent linewidth enhancement along the SE-NW direction, which is the kinematic minor axis of the stellar rotation shown in Figure \ref{fig:fig8}. We model the observed stellar velocity field using $^{3\mathrm{D}}$Barolo
\citep{3dbarolo}, and overlay the best-fit results in Figure \ref{fig:fig8}, which yield disk inclination angle of $i \sim 60^\circ$ and position angle PA$\sim 220^\circ$. This indicates a highly-inclined stellar disk with a minor axis along SE-NW direction, which may also produce broadened linewidths in the gas tracers due to the large velocity gradient along the line-of-sight. However, the spectral analyses presented in Figure \ref{fig:fig6} and \ref{fig:fig9} reveal that the linewidth enhancement arises from the superposition of at least two distinct kinematic components at and beyond the nucleus. In particular, Figure \ref{fig:fig9} shows that, in regions SE and NW, both the [O\,III]$\lambda$5007 and H$\beta$ lines exhibit a prominent, broad component with FWHM of $\gtrsim$\,700\,km\,s$^{-1}$ (see Table \ref{tab:reg_kcwi}), indicating highly turbulent ionized gas motion at $\sim 5$\,kpc away from the nucleus and beyond the bulk of the stellar content. This, combined with the biconical shape of the linewidth-enhanced regions from the multi-wavelength emission line maps, signifies the presence of a multi-phase biconical outflow along the SE-NW direction. \\
\indent In the left panel of Figure \ref{fig:fig10}, we show the line profiles of several gas tracers extracted from region C, which is covered by all our datasets. Similar to what is seen in the nuclear \textit{MRS} spectrum (Figure \ref{fig:fig4}), broad line wings indicative of outflowing gas are present in most ionized gas tracers but not the molecular ones, and the high-ionization [Ne\,V]\,14$\mu$m line shows a distinct, red-shifted profile. These differences imply that they do not necessarily trace the same kinematic structure(s), despite being measured over the same spatial scale. We leave a more in-depth discussion of the origins of the observed line profiles in the central region to Section \ref{sec:discussion}, where we further infer the energetics of the multi-phase gas involved in the biconical outflow. \\
\indent In the central panel of Figure \ref{fig:fig10} we visualize the morphology of the multi-phase gas derived from the multi-wavelength datasets. The extended [O\,III]$\lambda$5007 emission unambiguously traces the ionized bicone along the SE-NW direction beyond the bulk of the stellar emission probed by \textit{HST}. The warm and cold molecular gas traced by \ce{H2} and CO both exhibit ``X''-shaped morphology and outline the inner edges of the ionized bicone. Given that ALMA observations have a much larger FoV than those from \textit{MRS}, the close spatial correspondence between \ce{H2} and CO indicates that molecular gas is mostly concentrated at the base of the bicone, similar to what was found in the outflows of nearby starbursts \citep[e.g.,][]{leroy15, beirao15, bolatto21}. The brighter and more extended molecular emission in the NW indicates that it occupies the near-side. This is supported by the line profiles presented in Figure \ref{fig:fig6} and \ref{fig:fig9}, where the ionized gas tracers consistently present blue-shifted broad wings in region 8 and NW, and red-shifted ones in region 2 and SE. We provide a sketch of the inferred outflow configuration in the right panel of Figure \ref{fig:fig10}, where we also highlight the fact that dust is present in the bicone based on the prominent obscuration along the NW cone observed in the \textit{HST} data shown in Figure \ref{fig:fig1}. In the following Sections we present key results drawn from our analyses in Section \ref{sec:analysis} to characterize the spatially-resolved and integrated multi-phase ISM properties across IRASF01364 in the context of this newly discovered galactic outflow.
\subsection{Dust Extinction} \label{sec:dust_ext}
The non-uniform and extended distribution of dust and heavy nuclear obscuration are among the most defining features of the optical morphology of IRASF01364, as shown in Figure \ref{fig:fig1}. We quantify the spatially-resolved variation in visual extinction using the H$\alpha$ and H$\beta$ line flux measured with the KCWI dataset (Table \ref{tab:reg_kcwi_other}) across the regions defined in Section \ref{sec:kcwi_analysis} and shown in Figure \ref{fig:fig9}. Following \cite{fluetsch21}, we adopt the extinction law for actively star-forming galaxies from \cite{calzetti00} (i.e., $R_V = 4.05$), and assume the intrinsic H$\alpha$/H$\beta$ ratio of 2.86 for Case B recombination with temperature of $T = 10^4$\,K \citep{hummer87}. \\
\indent The resulting values are highest in region SW, covering a dusty tidal tail of the merger, with $A_V \sim 3.0$. Regions SE and C have consistent $A_V \sim 2.1 - 2.4$, while regions NW and exN have significantly lower values, with $A_V \sim 0.75 - 0.8$. Region exS has a high derived $A_V$ of $2.67\pm0.40$, but the faint and highly blended H$\alpha$ and [N\,II] doublet emission here may have biased the measured line ratio. These values further lend support to the picture presented Figure \ref{fig:fig10} where the NW cone lies on the near-side, and the SE cone on the far-side. The latter is much fainter due to heavier obscuration by the disk and/or other foreground materials associated with the tidal tail \citep{fluetsch21}. The high extinction in the SW region likely comes from a combination of pre-existing dusty gas in the tidal tail and partial obscuration by dust along the edges of the bicone. \\
\indent To further quantify the level of dust obscuration at the nucleus, we
measure the apparent depth of the silicate feature at 9.7\,$\mu$m from
the \textit{MRS} spectra, defined as $s_{9.7} = \ln(f_{9.7}/C_{\rm 9.7})$,
where $f_{9.7}$ is the flux density measured at 9.7\,$\mu$m, and $C_{\rm
9.7}$ is the flux density of the underlying continuum at 9.7\,$\mu$m, extrapolated with a power-law from 5 to
14\,$\mu$m, following a method commonly adopted in the literature \citep[e.g.][]{spoon07, stierwalt13, rich23}.
The resulting $s_{9.7}$ values are $-2.35$, $-2.12$ and $-1.66$, in the expanding cone
aperture, regions 5 and C, respectively. Compared to \textit{Spitzer/IRS} measurements made with a $\sim 10''\times 4''$ slit, which yield $s_{9.7} = -1.27$ for IRASF01364, and $s_{9.7} > -1.0$ for more than 60\% of local LIRGs \citep[e.g.][]{stierwalt13}, these values indicate highly concentrated dusty materials at the nucleus of IRASF01364. Following \cite{chiar07} and \cite{guver09}, these values indicate $A_V \gtrsim 60$ and column density of $N{\rm_H} \gtrsim 1.3\times10^{23}$\,cm$^{-2}$. This result is consistent with the nuclear dust extinction derived by \cite{koala-goals} at 2.2\,$\mu$m using Br$\gamma$/Br$\delta$ line ratios, which yield $A_V \sim 12 A_{2.2} \simeq 46\pm26.4$ assuming extinction law from \cite{calzetti00}. The high uncertainties were due to faint Br$\delta$ emission. It is possible that the intrinsic nuclear extinction may be even higher, at up to $A_V \sim $200 and $N_H \sim 5\times10^{23}$\,cm$^{-2}$, as inferred from the high silicate optical depth ($\tau_{\rm sil} \sim 12$) obtained from \texttt{CAFE} (see Appendix \ref{ap:more_pah}) assuming a mixed geometry \citep{marshall07}. However, potential degeneracy among the underlying multi-temperature dust model components could also impact the derived optical depths. Hence, we adopt the values derived from $s_{9.7}$ as the main reference for the nuclear extinction, as this quantity is measured directly from the observed spectrum, independent of assumptions about how multiple dust components and their associated opacities are distributed within the galaxy. Lastly, we highlight that the $A_V$ derived from the KCWI spectrum of region C is more than 30 times lower, demonstrating the limitation of optical lines in probing the heavy nuclear obscuration in local LIRGs. 
\subsection{Detection of high IP lines}\label{sec:cl}
As shown in Figure \ref{fig:fig4} and \ref{fig:fig10}, we detect the [Ne\,V]\,14$\mu$m line with IP $ = 97\,$eV in both the nuclear and central \textit{MRS} spectra, as well as [O\,IV]\,26$\mu$m with IP$ = 55\,$eV (see Table \ref{tab:cone_line}). This marks the first detections of high IP ($> 50$\,eV) lines in IRASF01364. Previous non-detections with \textit{Spitzer/IRS} \citep{petric11, inami13} can likely be explained by a combination of spatial and spectral dilution, as well as limited sensitivity. As star formation does not provide sufficient energy to quadruply ionize Ne, the detection of [Ne\,V] on kpc-scales is generally considered as a strong evidence for AGN activity \citep[e.g.][]{satyapal07, petric11,
satyapal21, bierschenk24}. However, the [Ne\,V]/[Ne\,II] and [O\,IV]/[Ne\,II] ratios measured from the \textit{MRS} spectra are low, at $\sim 0.005$ and $\sim 0.2$, respectively, which fall in the range previously measured among starburst-dominated local LIRGs with \textit{Spitzer/IRS} \citep{petric11, inami13, ps10}. This indicates that the AGN does not contribute significantly to the ionized gas excitation at the nucleus, which is consistent with the low AGN fraction ($\sim 7\%$) derived from multi-wavelength nuclear SED fitting by \cite{gao25}. We further discuss the properties of the putative AGN in Section \ref{sec:agn} based on the [Ne\,V]\,14$\mu$m detection. \\
\indent We do not detect the [Ne\,V]\,24$\mu$m line, and estimate a flux upper-limit of $5\times 10^{-19}$ W\,m$^{-2}$ from three times the root-mean-square of the nuclear spectrum measured around the expected redshifted wavelength (25.37\,$\mu$m). This yields a lower-limit of $\sim 0.6$ for the [Ne\,V]14$\mu$m/24$\mu$m ratio, which is widely adopted in AGN studies as a diagnostic of the electron density \cite[e.g.,][]{dudik07, ps10, ramos25}. However, our derived lower-limit is too low to yield any meaningful constraint on the electron density, which may range from $\lesssim\,$100 to up to 10$^7$\,cm$^{-3}$, as shown by modelling results \citep[e.g.,][]{ramos25}. We also do not detect any high IP optical lines in the KCWI spectra, likely due to dust extinction (see Section \ref{sec:dust_ext}). A more detailed \textit{JWST} investigation of the high IP lines in a sample of local LIRGs including IRASF01364 will be presented in a future work (U et al. in prep.). 
\subsection{PAH Equivalent Width and Band Ratios}
\begin{figure}
    \centering
    \includegraphics[width=1.0\linewidth]{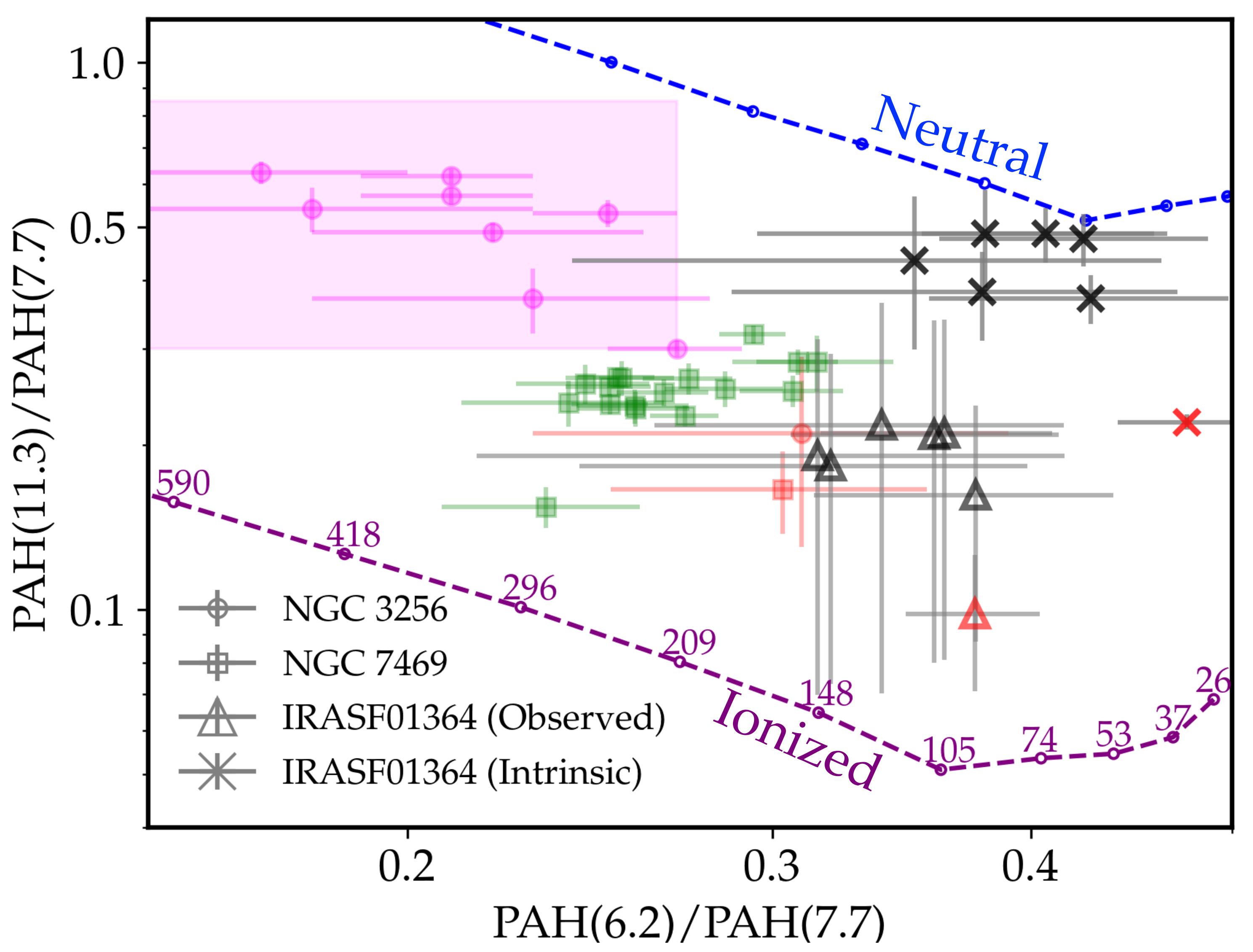}
    \caption{PAH band ratios measured between the 11.3
    and 7.7$\mu$m features with respect to the ones between 6.2 and
    7.7\,$\mu$m features, from region 1 - 9. Region 3 and 7 are not shown due to poor coverage of the 6.2 and 7.7\,$\mu$m features. Observed and attenuation-corrected values are shown in triangles and crosses, respectively. Measurements for NGC\,7469 from \cite{lai22} are shown in squares, and those for NGC\,3256 from \cite{bohn24} are shown in circles and the pink shaded region. Red-colored symbols mark the AGN. Dashed lines mark the theoretical tracks from \cite{draine21} for neutral
    (blue) and ionized grains (purple) with varying
    numbers of carbon atoms (indicated in purple texts). Similar to the other two LIRGs, in IRASF01364, the dust grains are more ionized at the AGN than in the extranuclear regions. No clear differences in grain sizes are observed between the AGN and non-AGN regions.
    \label{fig:fig11}}
\end{figure}

Broad emission features between 3 - 18\,$\mu$m from PAH molecules mark the most common trait of MIR spectra of star-forming galaxies \citep{smith07}. These molecules trace small dust grains in the photodissociation regions \citep[PDR;][]{bakes94} and are sensitive probes of the radiation field and grain properties \citep[e.g.][]{draine21}. In particular, the equivalent width (EW) of PAH features are observed to be lower in AGN environments \citep[e.g., EW(PAH\,6.2) $< 0.27$\,$\mu$m;][]{stierwalt13, inami18}, due to dilution by the strong AGN continuum, and/or grain destruction by the harsh radiation field \citep{smith07,diamond10}. As the 3.3\,$\mu$m feature is not covered in our \textit{MRS} dataset, we follow \cite{stierwalt13} and measure the apparent EW of the 6.2\,$\mu$m PAH feature from the nuclear \textit{MRS} spectrum, and from regions 5 and region C, using a spline fit of the local continuum. This yields a consistent value $\sim 0.29-0.33$\,$\mu$m across these different apertures, which is slightly lower than previous measurement with \textit{Spitzer/IRS}
(i.e., 0.39\,$\mu$m) and still
puts the nucleus among the class of composite sources (i.e., $0.27 <$ EW(PAH\,6.2) $< 0.54$\,$\mu$m), consistent with the scenario of a non-dominant AGN (see also Section \ref{sec:cl}). \\
\indent PAH features are also detected across region 1 - 9, suggesting the potential presence of small dust grains outside the nucleus and in the outflow, which have been observed in several nearby galactic outflows \citep{sutter25, lopez26} and predicted by simulation of multiphase outflow \citep{richie26}. In Figure \ref{fig:fig11} we show the PAH band flux ratios between 6.2 and 7.7\,$\mu$m, and those between 11.3 and 7.7\,$\mu$m, derived from \texttt{CAFE} fitting of the \textit{MRS} spectra extracted from region 1 - 9 (see Appendix \ref{ap:more_pah}). Region 3 and 7 are excluded as the 6.2 and 7.7\,$\mu$m features are poorly constrained. Following \cite{lai22}, we overlay theoretical tracks from \cite{draine21}, which illustrate the utility of the 11.3/7.7 and 6.2/7.7 ratios as tracers of grain ionization state and size, respectively. While the 6.2/7.7 ratios for the extranuclear regions (in black) are roughly consistent within uncertainties between the observed (in triangles) and intrinsic (i.e., attenuation-corrected; in crosses) measurements, the former show significantly reduced 11.3/7.7 ratios indicating a higher fraction of ionized grains compared to the latter. This difference is even more prominent at the nucleus (i.e., region 5, in red). This is attributed to the large optical depth near the 11.3\,$\mu$m band due to the deep 9.8\,$\mu$m silicate absorption feature (Section \ref{sec:dust_ext}). Hence, properly accounting for the effect of MIR attenuation is crucial for understanding dust grain properties in heavily-obscured centers of local LIRGs \citep[][]{lai24,donnelly25}. \\
\indent For comparison, we also overlay spatially-resolved measurements from two other local LIRGs, NGC\,3256 \citep[in circles for outflows and pink shaded area for star-forming regions;][]{bohn24} and NGC\,7469 \citep[in squares;][]{lai22}, which were also obtained with \texttt{CAFE}. For all three systems, the AGN (in red symbols) show consistently lower 11.3/7.7 ratios than the non-AGN regions, indicating more ionized dust grains towards the AGN. The 6.2/7.7 ratios are generally higher in IRASF01364 than in the other two systems, potentially pointing to smaller grain sizes in the former. Considering the uncertainties, no clear differences in the 6.2/7.7 ratios are observed between the AGN and non-AGN regions in these systems. 

\subsection{Spatially-resolved ionized gas excitation}
\subsubsection{MIR}\label{sec:mir_ratio}
\begin{figure}
    \centering
    \includegraphics[width=1.0\linewidth]{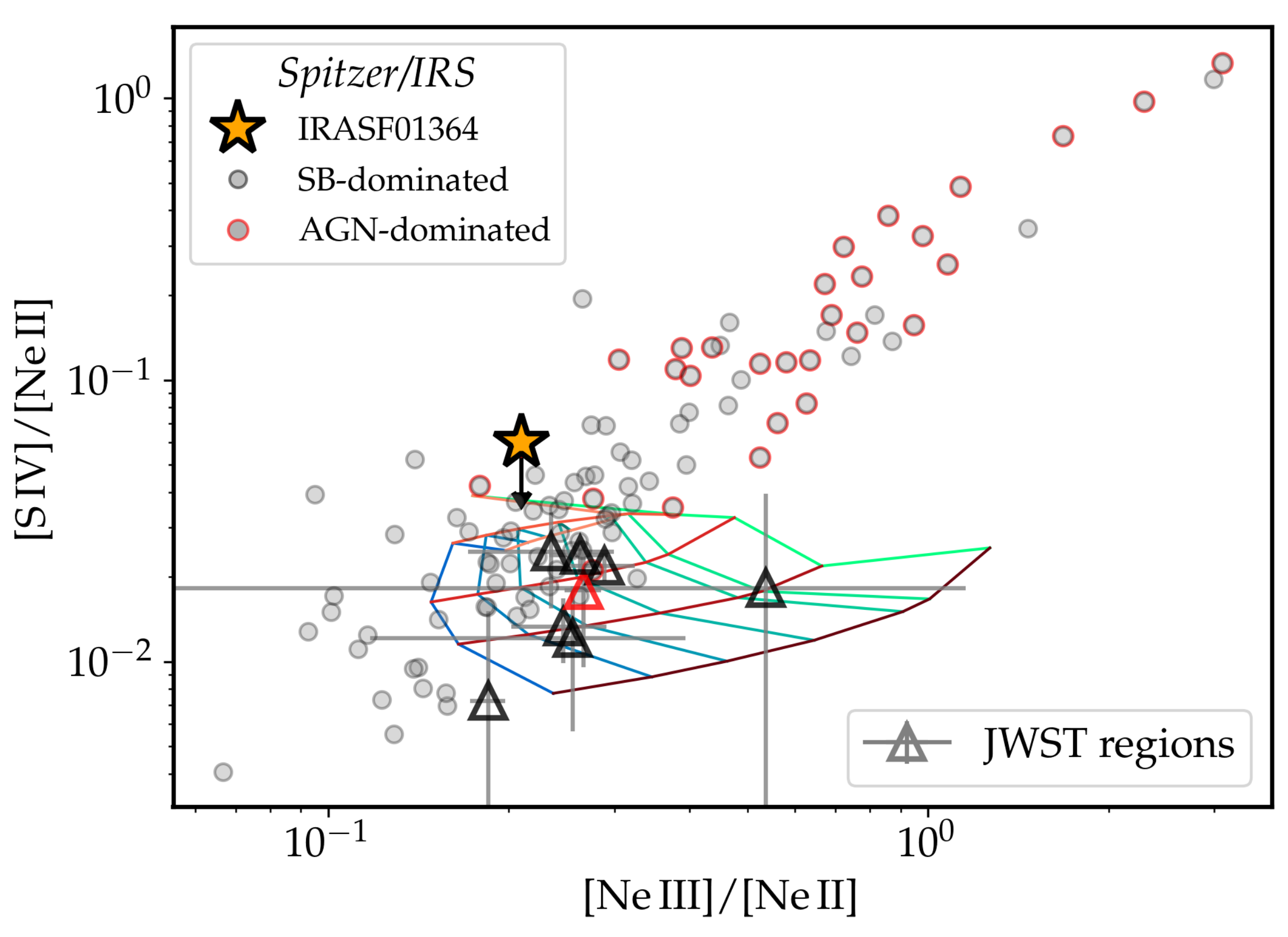}
    \caption{[S\,IV]/[Ne\,II] vs. [Ne\,III]/[Ne\,II] ratios across region 1 - 9, shown in triangles. Region 5 (the nucleus) is colored in red. Region 7 is not shown due to poor detection of the [S\,IV]\,10$\mu$m line. Measurements with \textit{Spitzer/IRS} for the GOALS sample from \cite{inami13} are shown in grey circles, with AGN-dominated systems outlined in red. The star marks the \textit{Spitzer/IRS} measurement for IRASF01364, where only an upper limit for [S\,IV]/[Ne\,II] was obtained. Model grids for fast radiative shock with solar metallicity and pre-shock electron density of 100\,cm$^{-3}$ \citep{allen08, mappingsv} are overlaid in colored lines, with orange-brown lines representing shock velocities from 100 - 200\,km\,s$^{-1}$ and blue-green lines representing magnetic field strengths from 1 - 100\,${\rm \mu}$G. All regions show line ratios similar to starburst-dominated systems, which are also consistent with the range predicted by models of fast radiative shock.
    \label{fig:fig12}}
\end{figure}

Following \cite{inami13}, we utilize the [S\,IV]/[Ne\,II] vs. [Ne\,III]/[Ne\,II] ratios to investigate the ionized gas excitation across regions 1 - 9, shown in Figure \ref{fig:fig12}. As [S\,IV]\,10\,$\mu$m, [Ne\,III]\,15.6$\mu$m and [Ne\,II]\,12.8$\mu$m have distinct IPs of 35, 41 and 22\,eV, respectively, elevated [S\,IV]/[Ne\,II] and/or [Ne\,III]/[Ne\,II] ratios could be used to identify regions of high excitation. In the Figure we also show \textit{Spitzer/IRS} measurements from \cite{inami13} for the GOALS sample in grey circles, with AGN-dominated systems outlined in red. Our spatially-resolved \textit{MRS} measurements for IRASF01364 (in triangles) are consistent with previous unresolved measurement with \textit{Spitzer/IRS} (marked in star) which did not detect the [S\,IV]\,10\,$\mu$m line. All regions show similarly low line ratios ([S\,IV]/[Ne\,II]$\sim$0.01, [Ne\,III]/[Ne\,II]$\sim$0.2) and overlap with starburst-dominated GOALS sources, with no clear differences between the nucleus and extra-nuclear regions. While in this work we do not apply MIR attenuation corrections to the atomic line fluxes, we note that such a correction would increase [S\,IV]/[Ne\,II] values by up to a factor of 2, which would still lie within the starburst-dominated regime.\\
\indent As discussed in \cite{inami13}, these low ratios could result from either photo-ionization by young starburst ($\lesssim 3.5$\,Myr) or shock excitation. For context, we emulate \cite{inami13} and overlay model grids of fast radiative shock from \cite{allen08}, with shock velocity of 100 - 200\,km\,s$^{-1}$, magnetic field strength of 1 - 100\,${\rm \mu}$G, solar metallicity and pre-shock electron density of 100\,cm$^{-3}$. The model grids were generated with the MAPPINGS V shock modelling code \citep{mappingsv}, which updates the calculation from \cite{allen08} with more realistic conditions, and were obtained from the Mexican Million Models Shock Database\footnote[10]{\url{http://3mdb.astro.unam.mx:3686/}} \citep{alarie19}. Line ratios of all regions are within the range predicted by the shock model, suggesting that a dominant contribution from shocks across the central region of IRASF01364 is plausible. 
\subsubsection{Optical}
\begin{figure*}
    \centering
    \includegraphics[width=0.8\linewidth]{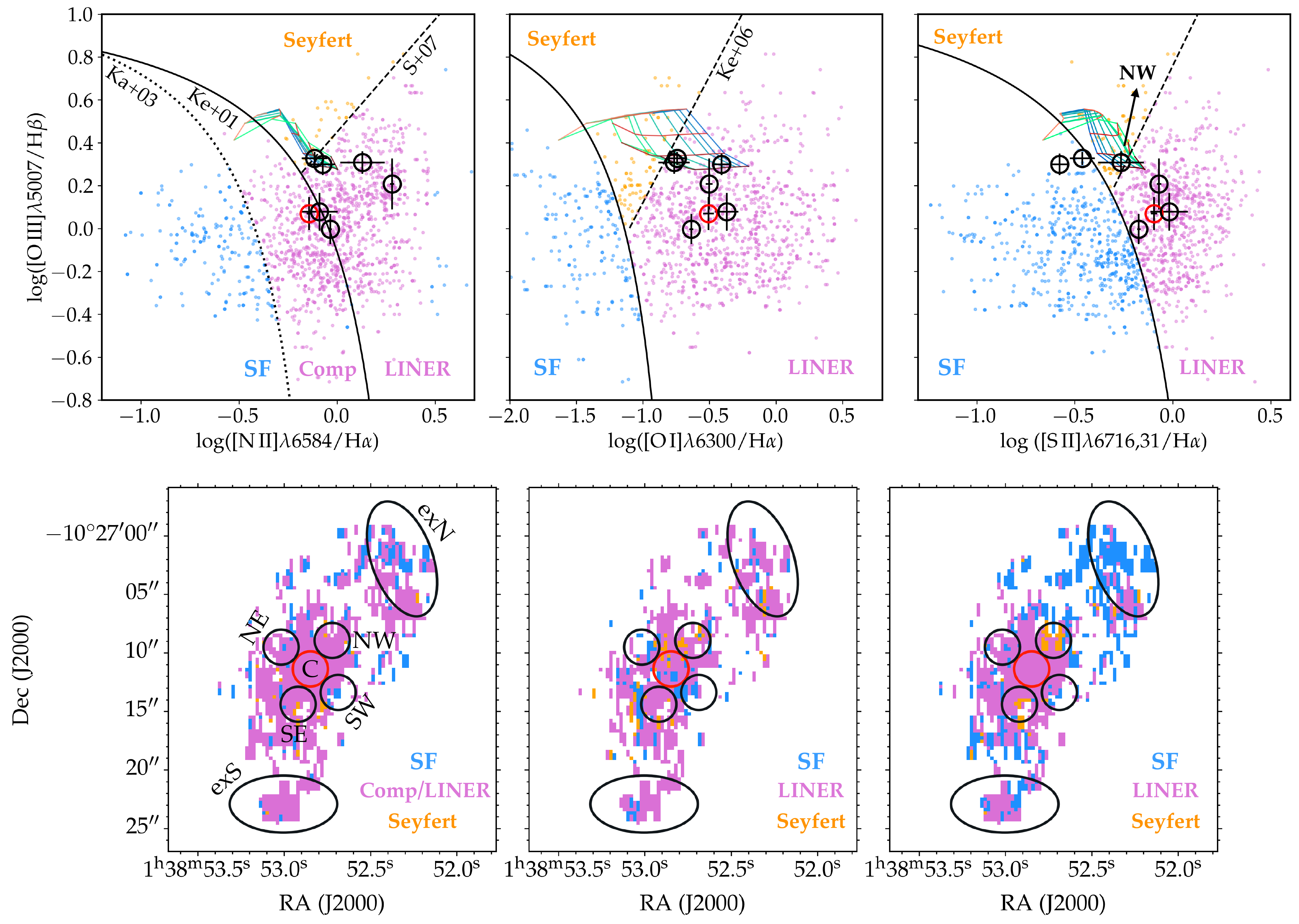}
    \caption{Classification of the excitation mechanisms of the ionized gas in
    IRASF01364 based on the optical BPT diagrams \citep{bpt1981,vo87} using $\log$
    ([O\,III]$\lambda$5007/H$\beta$) vs. $\log$
    ([N\,II]$\lambda$6584/H$\alpha$), $\log$([O\,III]$\lambda$5007/H$\beta$) vs.
    $\log$ ([O\,I]$\lambda$6300/H$\alpha$),
    $\log$([O\,III]$\lambda$5007/H$\beta$) vs. $\log$
    ([S\,II]$\lambda$6716,31/H$\alpha$), from left to right, respectively.
    \textit{Top}: measured line ratios at each spaxel are shown in blue, purple
    and orange, for classifications of ``star formation'', ``composite/LINER'',
    ``Seyfert'' excitation, respectively. The demarcation lines are defined by \cite{kauffmann03} (Ka+03), \cite{kewley01}(Ke+01), \cite{schawinski07} (S+07), and \cite{kewley06}(Ke+06). Model grids for fast radiative shock with the same parameters as in Figure \ref{fig:fig12} are overlaid. \textit{Bottom}: spatial distribution
    of spaxels with different classifications. In all panels, circles
    mark measurements/locations of the regions defined in Section \ref{sec:kcwi_analysis}, with region C highlighted in red. Most optical line emission in the IRASF01364 can be attributed to composite/LINER excitation. Region NW shows Seyfert-like line ratios in the [S\,II]/H$\alpha$ diagram, which may be due to AGN photo-ionization or shocks.
    \label{fig:fig13}}
\end{figure*}

To further investigate the excitation of large-scale ionized gas emission detected with KCWI, we construct the optical Baldwin–Phillips–Terlevich (BPT) diagnostic diagrams \citep{bpt1981,vo87} using the observed [O\,III]/H$\beta$, [N\,II]/H$\alpha$, [O\,I]/H$\alpha$ and [S\,II]/H$\alpha$ line flux ratios for all spaxels with SNR $> 5$ detection in the [N\,II], [O\,I] and [S\,II] line, respectively, shown in Figure \ref{fig:fig13}. Extinction correction is not applied here as its effect on the line ratios is expected to be trivial given the proximity in wavelengths between the emission line pairs, and also because not all regions have reliable $A_{V}$ estimates (see Section \ref{sec:dust_ext}). We classify and color-code each spaxel into ``SF (star formation)'' (blue), ``composite'' or ``LINER'' (low-ionization nuclear excitation region) (purple) and ``Seyfert'' (orange), following criteria presented by \cite{kewley06}. The [O\,III]/H$\beta$ vs. [N\,II]/H$\alpha$ diagram also includes the empirical Seyfert/LINER demarcation by \cite{schawinski07}. Measurements from individual regions defined in Section \ref{sec:kcwi_analysis} are also shown (in circles), with region C colored in red. \\
\indent A majority of the spaxels fall into the composite/LINER category in all three diagrams, which may be explained by a combination of several mechanisms including a low-luminosity AGN \citep{ferland83}, evolved stars \citep{binette94} and shock ionization \citep{dopita95}. The only region that falls into the Seyfert category in the [S\,II]/H$\alpha$ diagram is region NW on the near-side outflow cone (see Figure \ref{fig:fig10}), suggesting AGN excitation of the outflowing ionized gas. However, the observed Seyfert-like line ratios are also consistent with those predicted by the MAPPINGS V fast radiative shock model grids (see Section \ref{sec:mir_ratio} and Figure \ref{fig:fig12}) for shock velocities of $\sim 150 - 200$\,km\,s$^{-1}$. Overall, Figure \ref{fig:fig12} and \ref{fig:fig13} indicate that shock excitation is sufficient to explain the observed MIR and optical line ratios.
\subsection{Multi-phase gas mass in the central region}
While the multi-wavelength IFU datasets do not share the same FoVs, their common coverage in region C allow a matched-scale evaluation of the multi-phase gas content within the central $\sim 3\,$kpc of the system. In the following sections we derive the total multi-phase gas masses contained within region C based on the line fluxes measured in Section \ref{sec:analysis}, without identifying their physical origins (disk rotation, outflow, etc.). This latter topic is discussed in detail in Section \ref{sec:kinematics}.
\subsubsection{Ionized gas}\label{sec:result_ionized}
After correcting for the visual dust extinction estimated in Section \ref{sec:dust_ext}, the [O\,III]$\lambda$5007 and H$\beta$ line luminosities are  $(4.9\pm2.6)\times10^{40}$\,erg\,s$^{-1}$ and $(5.3\pm3.0)\times10^{40}$\,erg\,s$^{-1}$, respectively. Given the measured [S\,II]$\lambda$6716/$\lambda$6731 line ratio ($\sim\,1.38$; Table \ref{tab:reg_kcwi_other}), the electron density at region C is $n_e \sim\, 50\,$\,cm$^{-3}$. Assuming that the ionized gas fills the entire volume covered by region C, and a typical electron temperature of 10$^4\,$K, the above values yield an ionized gas mass of (3.9$\pm2.0)\times10^5$\,M$_\odot$ from [O\,III]$\lambda$5007 and $(9.0\pm5.1)\times10^6$\,M$_\odot$ from H$\beta$, following Eq.(5) and (7) in \cite{carniari15}. It is possible that the electron density is underestimated when using the [S\,II]$\lambda$6716/$\lambda$6731 line ratio \citep[e.g.,][]{baron19}, in which case the values reported here would serve as upper limits. We note that the average flux ratios derived from all
the KCWI apertures yield $n_e$ values of 100 - 300 cm$^{-3}$, based on the prescription of \cite{sanders16} which is consistent with
previous estimates for the GOALS galaxy sample using MIR [S\,III] 33.5/18.7 ratios observed with \textit{Spitzer/IRS}
\citep{inami13} on similar physical scales. Adopting this range of $n_e$ would decrease the estimated ionized gas mass by a factor of 2 - 6.
\subsubsection{Warm molecular gas} \label{sec:result_warmh2} 
\begin{figure}
    \centering
    \includegraphics[width=0.9\linewidth]{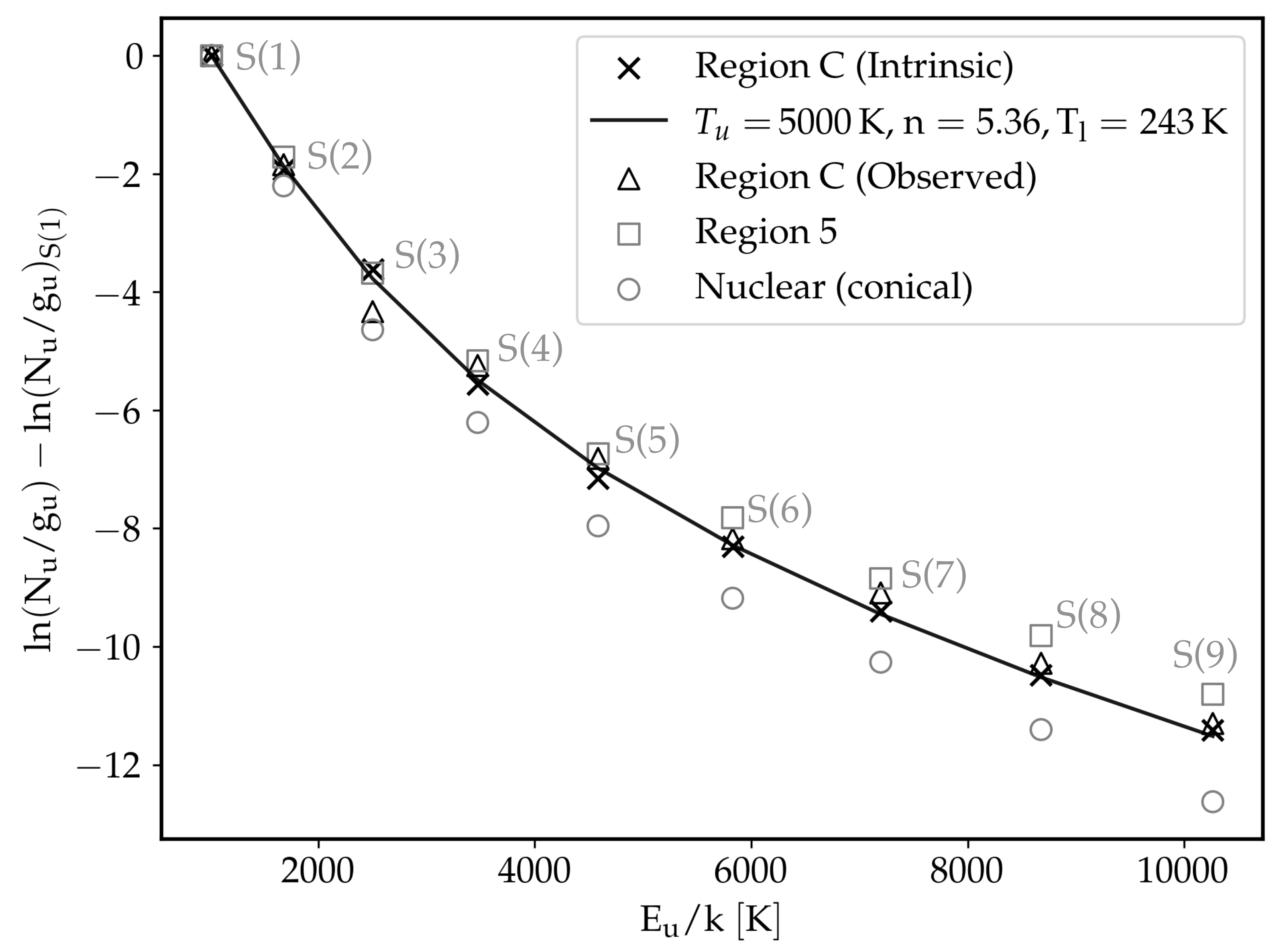}\\
    \caption{\ce{H2} excitation diagram and best-fit result assuming a continuous power-law temperature model as prescribed in \cite{togi16}, with an upper temperature limit set to 5000\,K. Measurements from region C (Figure \ref{fig:fig9}) are shown in triangles and crosses, for observed and extinction-corrected values, respectively. Extinction-corrected measurements from the nuclear region (Figure \ref{fig:nuc_spec}) and region 5 (Figure \ref{fig:nuc_spec}) are shown in empty gray circles and squares, respectively. Extinction correction mostly affects the measurement of the \ce{H2}\,0-0\,S(3) line. Compared to region C, region 5 shows somewhat higher contribution from S(6) - S(9) lines, while the nuclear (conical) region shows much lower contribution from S(3) - S(9) lines, indicating that the warmer molecular gas components are concentrated at the nucleus. \label{fig:h2diagram}}
\end{figure}
\indent We use the nine rotational transitions of \ce{H2} detected in the \textit{MRS} spectrum to constrain the temperature and mass of the warm ($\gtrsim$\,100\,K) molecular gas within region C, by adopting the continuous power-law temperature model described in
\cite{togi16}. The model is characterized by the lower
and upper temperature limits ($T_l, T_u$) and the index ($n$) of the power-law distribution, and allows a smooth extrapolation between the transitions to recover the total molecular gas mass down to $\lesssim 100$\,K. In Figure \ref{fig:h2diagram}, we show the best-fit model with $T_u$ set to 5000\,K following \cite{bohn24}, for the extinction-corrected measurements (in crosses) that yields $T_l \simeq 243\pm3\,K$ and $n = 5.36\pm0.02$. The spectral index is in the range of values derived for nearby star-forming galaxies by \cite{togi16}, but somewhat steeper than previously derived in local LIRGs on global scales \citep{zakamska10, ps14}, suggesting a relatively higher contribution from colder gas components in IRASF01364. However, these values are similar to those derived by \cite{bohn24} in the resolved nuclear outflows of NGC\,3256. We note that using observed values without correcting for MIR attenuation (shown in triangles) does not affect the derived values, as the correction only affects the \ce{H2}\,0-0\,S(3) line, as shown in Figure \ref{fig:h2diagram}. \\
\indent For comparison, we also show measurements from the nuclear conical aperture (circles) and region 5 (squares), which respectively show excess and deficit in emission from warmer components, particularly those traced by the S(6) to S(9) lines, compared to region C. This indicates that the warmer molecular gas components are mostly concentrated at the innermost area encompassed by region 5 and the short-wavelength portion of the nuclear conical aperture, while the colder components are more diffuse and fill the central few kpcs covered by region C.  Accounting for the heavy element correction factor of 1.36, the best-fit model shown in Figure \ref{fig:h2diagram} yields a total warm ($T \gtrsim 240\,$K) molecular gas of $M_{\ce{H2,warm}} \simeq 3.4\times10^{8}$\,M$_\odot$ and column density of $N_{\ce{H2,warm}} \simeq 3.5\times10^{21}$\,cm$^{-2}$ in region C.   
\\

\subsubsection{Cold molecular gas} \label{sec:results_coldh2}
The main source of uncertainty in estimating the cold ($T \sim 10 - 100\,$K) molecular gas mass from CO observations lies in the CO-to-\ce{H2} conversion factor, $\alpha_{\rm CO}$, which has been shown to be environment-dependent \citep[e.g., see review by][]{bolatto13}. The line ratios between different CO transitions offer clues on the physical condition of the molecular gas in different environments. \\
\indent The CO line luminosity at each transition is given by \cite{solomon97}:
\begin{equation}
    \frac{L'_{\rm CO}}{\rm K\,km s^{-1}\,pc^{2}} = 3.25\times 10^{7} \frac{D_{\rm L}^2}{\nu_{\rm obs}^2(1+z)^3} S \Delta v
\end{equation}
where the luminosity distance $D_{\rm L}$ is in Mpc, the observed frequency
$\nu_{\rm obs}$ is in GHz, and the integrated line flux $S\Delta v$ is in
Jy\,km\,s$^{-1}$. The resulting line luminosities for region C in CO(1-0), CO(2-1) and CO(3-2) are (3.6$\pm$1.5)$\times 10^{9}$, (2.9$\pm$0.1)$\times 10^{9}$, (3.0$\pm$0.1)$\times 10^{9}\, \rm K\,km s^{-1}\,pc^{2}$, respectively, which yield line ratios of $r_{\rm 21}\sim 0.8$ (i.e.,
$L'_{\rm CO(2-1)}/L'_{\rm CO(1-0)}$), $r_{\rm 32}\sim 1.0$ and $r_{\rm 31}\sim0.8$. These values are higher than those found in nearby disk galaxies
\citep[e.g.][]{leroy22} but similar to 
global measurements for local U/LIRGs \citep{montoya23}, potentially due to the outflow reducing the CO optical depth. Hence we adopt the constant CO-to-\ce{H2} conversion factor of $\alpha_{\rm CO}
\sim 1.7\pm0.5$\,$M_\odot$/($\rm K\,km\,s^{-1}\,pc^{2}$) for CO\,(1-0) from
\cite{montoya23} based on calibration using neutral carbon [C\,I] measurements, which yield cold molecular gas mass of $M_{\rm H_2, cold} \sim (6\pm3) \times 10^{9}$\,M$_\odot$. This value is also consistent with the total mass estimated from single-dish CO(1-0) measurements by \cite{herrero19} using a similar $\alpha_{\rm CO}$, indicating that most of the molecular gas in IRAS\,F01364 is concentrated in the central few kpcs. Roughly half of this mass is contained within region 5, which is three times smaller than region C in radius. The inferred gas column densities within regions C and 5 are $(6\pm3)\times10^{22}$\,cm$^{-2}$ and $\sim 3\times10^{23}$\,cm$^{-2}$,
respectively. These values are about 10 times higher than those inferred for the
warm molecular gas from rotational \ce{H2} lines in the last Section.
\section{Discussion} \label{sec:discussion} 
\subsection{Identifying the kinematic components of the outflow}\label{sec:kinematics}
As mentioned in Section \ref{sec:outflow}, the diversity and complexity of line profiles of multi-phase gas tracers observed in region C (Figure \ref{fig:fig10}) indicates that they may originate from different kinematic features. This is also hinted at in Figure \ref{fig:fig5} where we observe different morphologies and velocity gradients between the ionized gas traced by the MIR neon lines and warm molecular gas traced by the \ce{H2} emission. In this Section we attempt to provide a coherent picture to interpret the various line profiles observed across the different gas phases in the central 3\,kpc of IRASF01364, in the context of the newly uncovered multi-phase outflow (see Section \ref{sec:outflow}) and similar biconical outflows described in the literature. \\
\begin{figure}[t!]
    \centering
    \includegraphics[width=0.9\linewidth]{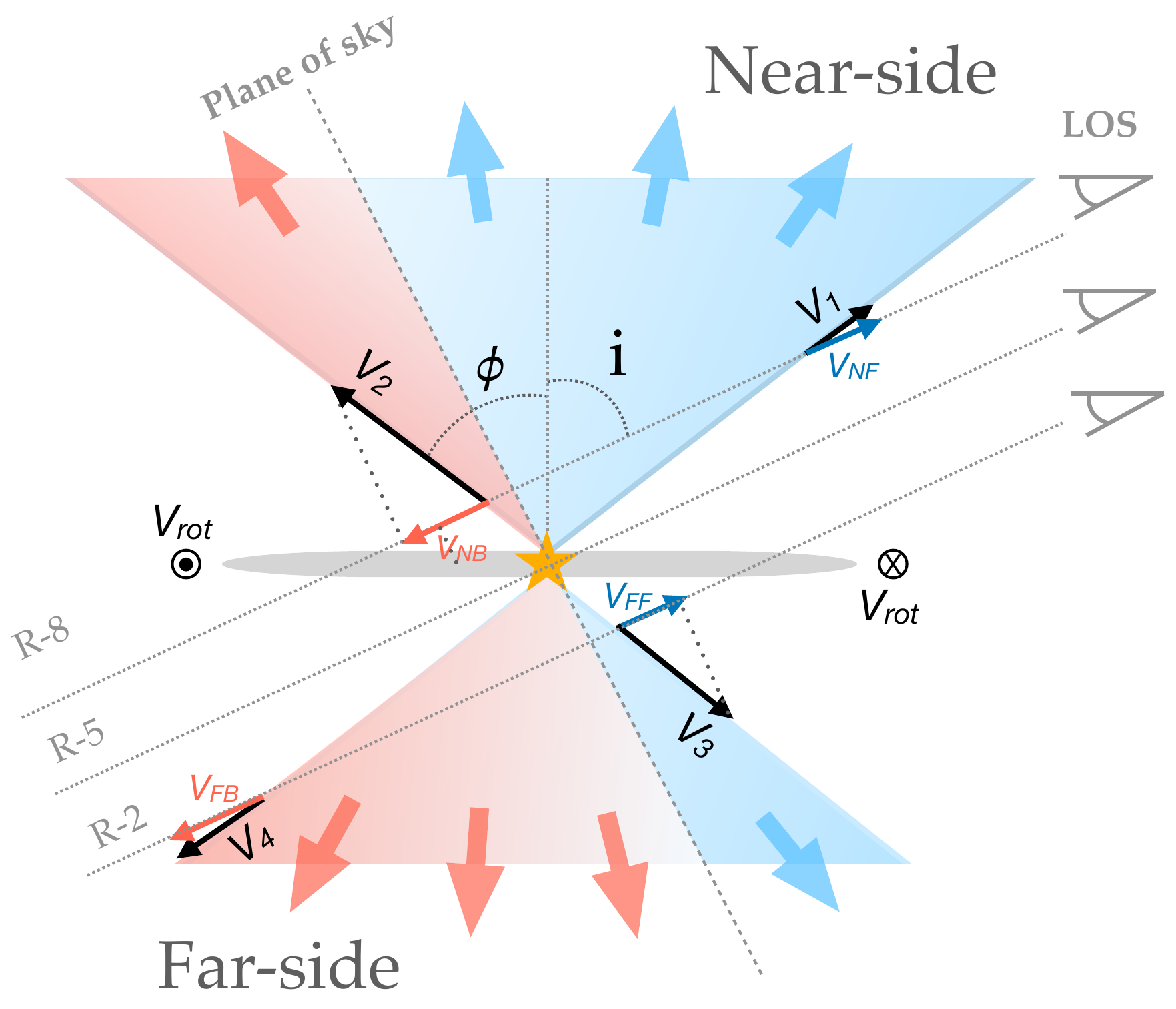}\\
    \caption{Schematic of the kinematic model for the biconical
    outflow observed in IRASF01364, modified from \cite{heckman90} and
    \cite{veilleux01}. The three gray dotted lines represent the potential
    lines-of-sight (LOS) towards regions 8, 5, and 2 (R-8, R-5, R-2) along
    the bicone, and the plane-of-sky is indicated by the dashed gray line. The yellow star and elongated ellipse respectively represent the nuclear starburst and the surrounding gaseous disk rotating at $v_{\rm rot}$, viewed at an inclination angle $i$. The half opening angle the bicone is given by $\phi$, and blue-shifted and
    red-shifted projected velocities are colored in blue and red, respectively. Each component is labeled based on the structure of origin (e.g. ``NF" --``near-side
    cone, front wall", ``FB'' -- ``far-side cone, back-wall''). Deprojected velocities are colored in black. \label{fig:fig15}}
\end{figure}
\indent As highlighted in Figure
\ref{fig:fig10}, nearly all multi-wavelength gas tracers in the spectrum of region C feature a similar double-peaked line profile. These profiles have also been commonly observed in biconical outflows of nearby starburst galaxies \citep[e.g., M\,82, NGC\,253, NGC\,2992;][]{heckman90, shopbell98, veilleux01, leroy15}, and are attributed to the superposition of the nuclear disk and/or walls of the bicone along a tilted line-of-sight (LOS). To interpret the potential origins of the observed multiple kinematic components, in Figure \ref{fig:fig15} we provide a modified version of the schematic presented by \cite{heckman90}, based on the outflow configuration inferred previously in Section \ref{sec:outflow} and focusing on the
LOS towards regions 8, 5 and 2 (hereafter as R-8, R-5 and R-2) along the bicone. 
\subsubsection{Outflow velocities}\label{sec:narrow}
\indent As discussed in \cite{heckman90} and illustrated in Figure \ref{fig:fig15}, gas flowing along the front and back walls of a biconical outflow from a inclined disk would naturally produce two distinct narrow kinematic components along the LOS, at red-shifted and blue-shifted projected velocities, respectively. For example, along R-8 through the near-side NW cone, we would observe two velocity components coming from the front and back wall, respectively, at $v_{NF}  = v_1 \cos (i - \phi)$ (NF - ``near-side, front wall'') and $v_{NB} = v_2 \cos (i + \phi)$ (NB - ``near-side, back wall''), where $v_1$ and $v_2$ are the intrinsic velocities at the observed location, $i$ is the inclination of the nuclear disk, and $\phi$ is the half opening angle of the near-side cone. Similarly, we may observe two components at $v_{FF}  = v_3 \cos (i + \phi)$ (FF - ``far-front'') and $v_{FB} = v_4 \cos (i - \phi)$ (FB - ``far-back'') for the far-side SE cone at R-2, in the simplistic case that the bicone is perfectly symmetrical. \\
\indent Now focusing on the observed MIR neon line profiles (see Figure \ref{fig:fig6}), if we assume that the two narrow components observed in R-8 and R-2 originate from outflowing gas along the walls, from Table \ref{tab:reg_line} we then have $|v_{NF}| \simeq 60$\,km\,s$^{-1}$, $|v_{NB}| \simeq 150$\,km\,s$^{-1}$, $|v_{FF}| \simeq 10$\,km\,s$^{-1}$ and $|v_{FB}| \simeq 180 $\,km\,s$^{-1}$. We estimate $\phi\sim 40 - 50^{\circ}$ from the FWHM linewidth map of [O\,III]$\lambda5007$ (Figure \ref{fig:fig9}) based on the opening angle of the area with FWHM $\gtrsim 500$\,km\,s$^{-1}$, and adopt $i\sim 60^{\circ}$ derived from the stellar velocity field in Figure \ref{fig:fig8} (see Section \ref{sec:outflow}). The resulting de-projected velocities would be $v_1 \sim 60$\,km\,s$^{-1}$, $v_2 \sim 580$\,km\,s$^{-1}$, $v_3 \sim 40$\,km\,s$^{-1}$ and $v_4 \sim 190$\,km\,s$^{-1}$. These values would suggest that the outflow is decelerating away from the disk in the near-side cone, but somehow accelerating in the far-side cone, which we consider unlikely as both sides should exhibit similar behavior. \\
\indent Given the clear rotation pattern seen in the stellar emission within the central 3\,kpc, it is possible that gas rotating with the stellar disk at a velocity $v_{\rm rot}$ also contributes to the observed line profiles. As shown in Figure \ref{fig:fig8}, such rotation produces blue-shifted narrow component towards the disk far-side (DF), and red-shifted narrow component towards the disk near-side (DN), with $v_{DF} = v_{DN} = v_{\rm rot}\sin\,i$ assuming the gaseous disk is symmetric. In this case, the blue-shifted narrow component in R-8 indicates $|v_{DF}| \simeq 60$\,km\,s$^{-1}$, and the red-shifted narrow component in R-2 indicates $|v_{DN}| \simeq 180$\,km\,s$^{-1}$,  which are also consistent with the narrow components observed in R-5. The much higher value of $|v_{DN}|$ could result from the LOS intercepting the disk near the edge in R-2 compared to R-8, yielding $v_{\rm rot} \sim 200$\,km\,s$^{-1}$. \\
\indent While the above scenarios have attributed the narrow components distinctly to either gas flowing along the walls of the outflow cones or that is rotating around the stellar disk, we note that realistically, they likely represent contributions from both structures, which would require high spatial and spectral resolution to disentangle. This is illustrated in Appendix \ref{ap:bbarolo}, where we perform kinematic modeling of the molecular gas disk on the CO(2-1) dataset, which has the highest spatial and spectral resolution among the multi-wavelength IFU datasets utilized in this work. This was done with $^{3\mathrm{D}}$Barolo, using disk inclination and position angle previously derived by \cite{agostino26} from $\sim\,$10-pc resolution ALMA CO(2-1) observations of the nuclear molecular disk. We describe in more detail the procedures and results in Appendix \ref{ap:bbarolo}. The results shown in Figure \ref{fig:bbarolo} highlights the challenge in robustly separating the outflow and disk components, which span the same velocity range. This is also seen in the emission line channels maps shown in Appendix \ref{sec:chanmaps}. For an example, the [Ne\,III] line, shown in Figure \ref{fig:miri_chan}, shows clear signature of disk rotation as the bright, compact nuclear emission moves from NE to SW from between $\sim -300$ to 300 \,km\,s$^{-1}$. Meanwhile, extended fainter emission corresponding to the NW and SE cone also appear in these channels, and hence the narrow components associated with the walls of the outflow cones would be blended with the disk emission both spatially and spectrally, especially given the \textit{MRS} spectral resolution of $\sim 120$\,km\,s$^{-1}$ (in FWHM) at the observed wavelengths of [Ne\,III].\\
\indent Nevertheless, the broad component with FWHM $\sim 600 - 1000$\,km\,s$^{-1}$ consistently seen in R-2, R-5, R-8 in the ionized gas tracers (Table \ref{tab:reg_line}) allows for a more definitive characterization of the outflow velocity in the ionized gas phase, given by $v_{\rm out, ion} = |v_{\rm broad}| + {\rm FWHM_{\rm broad}}/2 \sim 300 - 600$\,km\,s$^{-1}$, following definitions commonly adopted in the literature \citep[e.g.,][]{rupke05-agn, fluetsch21}. For the CO lines, a broad component is detected along the bicone marginally in R-2 and more strongly in R-8 (see Figure \ref{fig:fig7} and Table \ref{tab:reg_co}), which yield outflow velocity in the cold molecular gas phase of $v_{\rm out, cold\,\ce{H2}} \sim 200 -300$\,km\,s$^{-1}$. These values are in agreement with the fact that emission is seen along the bicone up to $\sim \pm 500$\,km\,s$^{-1}$ in the [Ne\,III] channel maps in Figure \ref{fig:miri_chan}, but only up to $\sim \pm 200$\,km\,s$^{-1}$ in the CO(2-1) channel maps shown in Figure \ref{fig:co21_chan}. For the \ce{H2}\,0-0\,S(3) line (Table \ref{tab:reg_line}), no broad components are identified in any regions potentially due to insufficient spectral resolution ($\sim 90$\,km/s in FWHM). As the channel maps in Figure \ref{fig:miri_chan} show emission up to $\sim \pm 350 $\,km\,s$^{-1}$ along the bicone, from which we adopt a warm molecular gas outflow velocity of $v_{\rm out, warm\,\ce{H2}} \sim 350$\,km\,s$^{-1}$. 
\subsubsection{Molecular vs. ionized gas }\label{sec:broad} 
\indent As already highlighted in previous Sections, broad wings characterized by a FWHM linewidth of $> 500$\,km\,s$^{-1}$ are seen consistently in both MIR neon lines in regions 2, 3, 5, 7, 8 (see also Table \ref{tab:reg_line}) along the outflow direction, with the broadest component measured at the nucleus, reaching FWHM $\sim 1000 $\,km\,s$^{-1}$ in both region 5 and the nuclear aperture (see Table \ref{tab:cone_line}). On larger scales, shown in Figure \ref{fig:fig10}, the ionized gas emission traced by [O\,III]$\lambda5007$ and H$\beta$ in regions NW and SE along the bicone are completely dominated by a broad component with FWHM $\sim 300 - 500$\,km\,s$^{-1}$. In all these regions, the broad component exhibits consistent blue-shifted velocities towards the NW and red-shifted velocities towards SE, which we interpret as the presence of outflowing ionized gas in between the walls of the bicone.\\
\indent As illustrated in Figure \ref{fig:fig15}, outflowing materials traveling at a range of velocity inside the bicone would produce a broad component that is dominantly blue-shifted in the near-side cone seen at R-8 and region NW and red-shifted in the far-side cone seen at R-2 and region SE. Along a LOS closer to the nucleus (e.g., R-5 or region C), the superposition of materials flowing in opposite directions would produce a broad component near systemic velocity. This picture is consistent with what we observe in the ionized gas tracers, except for the [Ne\,V]\,14$\mu$m line, which we discuss later in Section \ref{sec:nev_line}. \\
\indent On the other hand, this broad component is largely absent in the molecular gas tracers, where the broadest component with FWHM of $\sim 300 - 500\,$km\,s$^{-1}$ is detected only in CO (2-1) at R-8 and marginally at R-2. As discussed in Section \ref{sec:narrow}, this likely indicates a slower outflow in the molecular gas phase, whose line profiles are not easily distinguishable from that of the disk. In addition, as we illustrate in Figure \ref{fig:fig15}, slow-moving molecular gas accumulating near the disk will produce dominantly red-shifted emission in the near-side cone and blue-shifted emission in the far-side cone, in contrast to what is seen in the ionized gas tracers. These predictions are in agreement with what is seen in the channel maps in Appendix \ref{sec:chanmaps}, and can be attributed to the tilted outflow cones resulting in LOS intercepting the NB and FF walls much closer to the disk than the NF and FB walls. As such, emission coming from the NB and FF walls will consistently dominate over those coming from molecular gas in the center of the outflow and along the NF and FB walls, due to the rapidly decreasing molecular gas density with elevation. This may also explain the prominent ``X''-shaped morphology observed in the CO and \ce{H2} lines (i.e., Figure \ref{fig:fig10}, Figure \ref{fig:miri_chan}).
\subsubsection{Kinematics of the [Ne\,V]\,14$\mu$m line} 
\label{sec:nev_line}
\indent As shown in Figure \ref{fig:fig4} and \ref{fig:fig10}, unlike other ionized gas tracers, the [Ne\,V]\,14$\mu$m line tracing the most highly-ionized gas only exhibits a single red-shifted narrow component at $\sim 180\,$\,km\,s$^{-1}$ with FWHM $\sim 300\,$km\,s$^{-1}$. Similar profiles have been observed in optical studies of high IP lines in nearby AGN hosting radio jet \citep[e.g.,][]{ra06,ra17}, where the highly-ionized gas traces a dense energetic outflow and is driven mechanically by the jet. Extended coronal line emission associated with such jet-driven outflows have also been observed in [Ne\,V] with MRS in local Seyferts and quasars \citep{davies24, kader26, bianchin26}. While VLA observations at 15 and 33\,GHz do not show evidence of a radio jet down to $\sim 100\,$pc scales \citep{bm17, song22}, we cannot rule out the possibility of a small-scale one-sided jet \citep[e.g.,][]{mezcua14} driving the observed red-shifted emission of [Ne\,V]\,14$\mu$m. Alternatively, this component may trace highly-excited gas emanating from the inner NB wall of the biconical outflow, given that it shares similar velocity as the red-shifted narrow components seen in the other ionized gas tracers with lower IP. The $\sim$1.6 times higher [Ne\,V]\,14$\mu$m line flux measured from region C compared to that from the nuclear spectrum (Table \ref{tab:cone_line}) potentially indicates that the [Ne\,V] emission is extended. However, assuming the emission is unresolved and considering \textit{MRS} aperture correction factor (1.13 from \cite{mrs-res2}) given the aperture sizes specified in Section \ref{sec:miri_analysis}, the evidence for extended [Ne\,V] emission is marginal. A more detailed search and characterization of other tracers with similarly high IPs would be required to investigate the kinematics of the highly-ionized gas, which we defer to future works. 
\subsubsection{Potential contributions from the merger} 
\label{sec:kin_sum}
\indent Given the merger nature of IRASF01364, one may also question whether tidal effects play a role in driving the observed gas kinematics. For an example, the distinctly different velocity fields observed between the ionized and molecular may also result from disk counter-rotation/misalignment triggered by the merger \citep[e.g.,][]{lu21}. However, we highlight that the disk inclination and position angle derived from the stellar velocity field shown in Figure \ref{fig:fig10} are in agreement with those derived from previous 10\,pc-scale observations of the nuclear ionized and molecular gas disks of IRASF01364 in Pa$\alpha$, rovibrational \ce{H2}, and CO(2-1) lines by \cite{medling15} and \cite{agostino26}. Hence, there is no evidence of strong disk misalignment among stars, ionized and molecular gas.\\
\indent Nevertheless, the effect of merger is clearly present, which manifests in the extended molecular loop and tail observed towards the NE and SW direction respectively in the CO (2-1) maps shown in Figure \ref{fig:fig7}. The narrow, double-peaked profiles of [O\,III]$\lambda5007$ and H$\beta$ in the NE and SW regions along the major axis of the stellar disk shown in Figure \ref{fig:fig9} likely result from overlapping emission between the disk and tidal features. However, given the consistent disk rotation observed in this work and previous high-resolution studies of the nuclear disk \citep[i.e.,][]{medling15,agostino26}, as well as broadened line profiles along the outflow direction observed across the multi-phase tracers, we argue that the multi-phase kinematics of the central 3\,kpc of the system focused in this work are likely dominated by disk rotation and the outflow rather than tidal features.

\subsection{Outflow Energetics}\label{dis:energy}
Following the scenario proposed above, in this Section, we aim to quantify the outflow properties, as measured mainly within the central $\sim 3\,$kpc of
IRASF01364, i.e., in region C, where gas emission is the brightest across gas phases. This region also covers the bulk of the CO emission (e.g., see Figure \ref{fig:more_alma}) tracing the cold molecular gas outflow, which dominates the total outflow energetics, as we show later in this Section. We follow
\cite{lutz20} and \cite{fluetsch21} and define outflow properties (outflow
dynamical time $\tau_{\rm dyn}$, mass outflow rate $\dot{M}_{\rm out}$, outflow
kinetic energy $E_{\rm kin}$ and kinetic power $\dot{E}_{\rm kin}$) as follows:
\begin{gather*} 
\tau_{\rm dyn} = R_{\rm out}/v_{\rm out}\\
\dot{M}_{\rm out} = M_{\rm out}/\tau_{\rm dyn}\\
{E}_{\rm kin} = 0.5M_{\rm out}v^{2}_{\rm out}\\
\dot{E}_{\rm kin} = 0.5\dot{M}_{\rm out}v^{2}_{\rm out}
\end{gather*}
where $v_{\rm out}$ and $R_{\rm out}$ are the outflow velocity and
radius. Unless otherwise noted, we adopt $R_{\rm out} $ to be the deprojected physical radius of
region C, which is 1.6\,kpc for a disk inclination of 60$^{\circ}$. We discuss below the outflow properties inferred from different gas tracers. \\
\indent \textbf{Cold Molecular Gas:} Via modeling the disk emission using $^{3\mathrm{D}}$Barolo, as shown in Appendix \ref{ap:bbarolo}, we estimate that the outflow may contributes $\sim 40\%$ of the CO(2-1) emission captured in region C. This would yield 
an outflow mass in cold molecular gas of $M_{\rm out, cold\,\ce{H2}} \sim 1.2 - 3.6\,\times10^9$\,M$_\odot$, based on the total cold molecular gas mass derived in Section \ref{sec:results_coldh2}. While we have adopted $\alpha_{\rm CO} = 1.7$\,M$_\odot$\,(K\,km\,s$^{-1}$\,pc$^2$)$^{-1}$ in Section \ref{sec:results_coldh2}, for the outflowing molecular gas this value may be much lower due to turbulence reducing the CO optical depth \citep[e.g.,][]{bolatto13}. Previous works on outflows in local LIRGs have commonly adopted $\alpha_{\rm CO} = 0.8$\,M$_\odot$\,(K\,km\,s$^{-1}$\,pc$^2$)$^{-1}$ \citep[e.g.,][]{cicone14, fluetsch19}, and an even lower value of $\sim 0.4 - 0.6$\,M$_\odot$\,(K\,km\,s$^{-1}$\,pc$^2$)$^{-1}$ has been estimated in the outflow of NGC\,3256 from CO absorption features observed with \textit{JWST/NIRSpec} by \cite{ps24}. Adopting this latter value, and an outflow velocity of $v_{\rm out, cold\,H_2} \sim 200 - 300$\,km\,s$^{-1}$ estimated in Section \ref{sec:narrow}, we then have $M_{\rm out, 
cold\,H_2} \sim 3 - 12 \times 10^{8}$\,M$_\odot$\,yr$^{-1}$, $\tau_{\rm dyn,  cold\,H_2} \sim 5 - 8$\,Myr, $\dot{M}_{\rm out,
cold\,H_2} \sim 38 - 240$\,M$_\odot$\,yr$^{-1}$, $E_{\rm kin, cold\,H_2} \sim 0.1 - 1.1
\times 10^{57}$\,erg and $\dot{E}_{\rm kin, cold\,H_2} \sim 0.5 - 6.8 \times
10^{42}$\,erg\,s$^{-1}$. \\
\indent \textbf{Warm Molecular Gas:} Following discussions in Section \ref{sec:narrow}, we infer from the channel maps (i.e., Figure \ref{fig:miri_chan}) an outflow velocity in the warm molecular gas phase up to $v_{\rm out, cold\,H_2} \sim 350$\,km\,s$^{-1}$. As detailed modelling of the disk emission is challenging given the relatively coarse spectral resolution, here we simply assume that the warm molecular gas also contributes $\sim 40\%$ of the total \ce{H2} emission captured in region C, following the distribution of the cold molecular gas traced by CO(2-1). This is based on the similar spatial distributions and velocity profiles observed between the warm and cold molecular gas tracers, shown in Figure \ref{fig:fig10}. Following this assumption, we have $M_{\rm out, warm\,H_2} \sim 1.4\times10^{8}$\,M$_\odot$ based on the total warm molecular gas mass derived in Section \ref{sec:result_warmh2}. The inferred
outflow properties are $\tau_{\rm dyn, warm\,H_2} \sim 4.5$\,Myr, $\dot{M}_{\rm
out, warm\,H_2} \sim 31$\,M$_\odot$\,yr$^{-1}$, $E_{\rm kin, warm\,H_2} \sim 1.7\times10^{56}$\,erg and $\dot{E}_{\rm kin, warm\,H_2} \sim
1.2\times10^{42}$\,erg\,s$^{-1}$. \\
\indent \textbf{Ionized Gas:} Following discussions in Section \ref{sec:narrow}, we assume that the outflow is mainly traced by the broad components identified in the ionized gas tracers, which yields
$v_{\rm out,ion} \sim 500 - 600$\,km\,s$^{-1}$ from 
[Ne\,II], [Ne\,III], [O\,III] and H$\beta$ (Table \ref{tab:cone_line} and \ref{tab:reg_kcwi}). These broad components account for 20 - 50\% of the total line fluxes of these tracers within region C, which correspond to an outflow mass in ionized gas of 
$M_{\rm out, ion} \sim 0.8 - 7.0\,\times10^6$\,M$_\odot$ based on the total mass calculated in Section
\ref{sec:result_ionized} from H$\beta$. The inferred ionized outflow properties are $\tau_{\rm
dyn,ion} \sim 3$\,Myr, $\dot{M}_{\rm
out, ion} \sim 0.3 - 2.3$\,M$_\odot$\,yr$^{-1}$,
$E_{\rm kin, ion} \sim 0.2 - 2.5 \times 10^{55}$\,erg and $\dot{E}_{\rm kin, ion}
\sim 0.2 - 2.6 \times 10^{41}$\,erg\,s$^{-1}$. As discussed in Section \ref{sec:narrow}, it is possible that the outflow walls also contributes, to some degree, to the narrow components observed in region C, in which case the estimated $\dot{M}_{\rm out, ion}, E_{\rm kin, ion}$ and $\dot{E}_{\rm kin, ion}$ may further increase by up to a factor of two. We note that while the ionized gas emission extends out to a projected distance of $\sim 10\,$kpc beyond the nucleus, it is unclear whether they belong to the same outflow structure, and hence not considered here. \\
\indent The values calculated above together indicate that molecular gas, especially in the cold phase
($T \lesssim 100$\,K), heavily dominates the mass and energetics of the galactic
outflow in IRASF01364, consistent with conclusions drawn from previous
multi-phase studies of outflows in other local U/LIRGs \citep{fluetsch21}. The
values of the mass outflow rates and kinetic energies in the molecular phase are
also comparable to values derived from previous studies \citep[e.g.][]{cicone14,
ps18,lutz20, fluetsch21} of local U/LIRGs, with $\dot{M}_{\rm out}$ nearly 10 times higher than what is seen in nearby starbursts \citep[e.g.][]{leroy15,krieger19,bolatto21}. However, the local U/LIRGs hosting $\dot{M}_{\rm out} > 100\,$M$_\odot$\,yr$^{-1}$ in the
sample of \cite{lutz20} all have a much higher AGN contribution to the
bolometric luminosity ($> 25\%$;) than IRASF01364 \citep[$ \sim 5 - 7\% ;$][]{ds17, gao25}. We note that the assumed CO-to-\ce{H2} conversion factor has a key impact on the estimated energetics. Sensitive, multi-tracer observations will provide critical constraints on the derived outflow properties. We discuss the potential origin of this powerful outflow in Section \ref{dis:origin}.
\subsection{A low-luminosity, dust-obscured AGN}\label{sec:agn} 
\indent The detection of [Ne\,V]\,14.3$\mu$m line from \textit{MRS} provides the first direct
evidence of AGN activity in the nucleus of IRASF01364. The line flux measured from region C is $(4.5\pm0.8)\times 10^{-16}$\,erg\,s$^{-1}$\,cm$^{-2}$, corresponding to an AGN bolometric luminosity of $L_{\rm bol} \simeq\, 1.2 - 1.8\times10^{43}$\,erg\,s$^{-1}$, adopting calibrations by \cite{satyapal07} and \cite{spinoglio22}. Using the relationship between [Ne\,V]\,14.3$\mu$m and 2$-$10\,keV luminosity obtained by \citet{bierschenk24}, we expect intrinsic 2$-$10\,keV luminosity of $L^{\rm int}_{\rm 2-10\,keV}\sim 2.1\times10^{42}$\,erg\,s$^{-1}$. Given the observed 2$-$10\,keV luminosity from \textit{Chandra} is $L^{\rm obs}_{\rm 2-10\,keV}\sim 1.5\times10^{41}$\,erg\,s$^{-1}$ \citep{iwasawa11}, we estimate the obscuring column density $N_{\rm H}$ using {\textsc RXTorusD} \citep{ricci23sim}, a physical X-ray absorption and reflection model. We adopt a photon index of $\Gamma \sim 1.8$, typical for nearby AGN \citep{ricci17}, which yields $N_{\rm H} \sim 1.6\times10^{23}\rm\,cm^{-2}$, indicating that the AGN is dust-obscured. Similar results are obtained by Torres-Alb\`{a} et al. (in prep.) via direct modelling of the \textit{Chandra} spectrum \citep{iwasawa11}. The inferred obscuring column density is consistent with nuclear obscuration derived from the 9.7\,$\mu$m absorption feature (Section \ref{sec:dust_ext}) and the gas column density from CO (Section \ref{sec:results_coldh2}), pointing to a potential shared origin, although the latter is highly dependent on the uncertain $\alpha_{\rm CO}$. \\
\indent While dust-obscured AGN are frequently found in local gas-rich mergers \citep{ricci17b, koss18, ricci21}, the low inferred $L^{\rm int}_{\rm 2-10\,keV}$ makes IRASF01364 one of the few local IR-luminous mergers found so far
to host a low-luminosity AGN with $L_{\rm 2-10\,keV} < 10^{43}$\,erg\,s$^{-1}$
\citep[e.g.][]{ricci21}. The AGN also seems to play a minimal role in photo-ionizing the surrounding gas, based on the MIR and optical line ratios shown in
Figure \ref{fig:fig12} and \ref{fig:fig13}, consistent with the low bolometric AGN fraction ($\sim 7\%$) inferred from X-ray-to-mm SED fitting by \cite{gao25}. Following \cite{ricci23}, we infer an extreme obscuring column density of $\sim 10^{26}$\,cm$^{-2}$ from comparing the observed 2-10\,keV emission to the ALMA 100\,GHz flux ($\sim 2$\,mJy) measured on $\sim\,300$\,pc-scale. While this may suggest a potentially more powerful AGN that is more obscured than inferred from the mid-IR, the starburst likely dominates both the X-ray and 100\,GHz emission of IRASF01364, hence complicating the interpretation. Observations at much higher resolutions would be necessary to isolate emission from the low-luminosity AGN. Upcoming deep observations with \textit{XMM-Newton} will
also further shed light on the nature of the AGN (ID: 096261; PI: N. Torres-Alb\`{a}). \\
\indent Given a central blackhole mass of $\sim 10^{8} - 10^{9}$\,M$_\odot$ estimated from kinematic modelling \citep{medling15,agostino26}, the inferred Eddington ratio
(i.e.,$\lambda_{\rm edd}= L_{\rm bol}/L_{\rm edd}$) would be extremely low $\lesssim 10^{-3}$, indicating highly-inefficient accretion. This agrees with the recent finding from \cite{rieke25} that most late-stage IR-luminous mergers appear to have low accretion rates, challenging the picture of a merger-quasar evolutionary sequence where the final phase of the gas-rich merger is characterized by the emergence of a highly accreting AGN \citep[e.g.][]{hopkins06}. Continuous search with \textit{JWST} in galaxy mergers will potentially uncover many more cases of low-luminosity AGN that do not dominate the excitation nor total energy budget of the system, as already suggested from previous surveys with \textit{Spitzer} \citep[e.g.,][]{petric11,stierwalt13,inami13} and \textit{NuSTAR} \citep{ricci21}.
\subsection{Star formation rate of the nuclear starburst}\label{dis:sf} 
\indent Due to the relative faintness of MIR hydrogen recombination lines, many studies have explored and advocated for the usage of other bright MIR features
as extinction-free tracers of star formation rates (SFR), such as
the [Ne\,II]\,12.8$\mu$m and [Ne\,III]\,15.6$\mu$m lines
\citep[e.g.][]{ho07,inami13,zhuang19,whitcomb20} and PAH features
\citep[e.g.][]{shipley16,xie19,hernan20}.\\  
\indent From the observed line flux measurements from region C reported in Table \ref{tab:cone_line}, we derive a total SFR of $ \sim 8 - 10$\,M$_\odot$\,yr$^{-1}$ from [Ne\,II] and [Ne\,III], accounting for
contribution from the AGN using the [Ne\,V]\,14.3$\mu$m line, following calibrations provided in \citep{zhuang19}. The values derived from observed fluxes of 6.2, 7.7 and 11.3\,$\mu$m PAH
features (Table \ref{tab:cone_pah}) based on prescriptions from \cite{shipley16} are lower, at 
$\sim 2 - 6$\,M$_\odot$\,yr$^{-1}$. Assuming Case B recombination \citep{hummer87}, we
derive SFR $\sim 4$\,M$_\odot$\,yr$^{-1}$ from the observed Pf$\alpha$ and Hu$\alpha$ line flux, with H$\alpha$-based calibrations
from \cite{murphy11}.  We note that the
inferred intrinsic H$\alpha$ luminosity is about 5 - 10 times higher than the
extinction-corrected H$\alpha$ luminosity calculated within region C from the
KCWI datasets, further confirming the limitation of optical lines for probing the heavy
nuclear obscuration in IRASF01364. \\
\indent The general agreement between values derived from PAH and hydrogen
recombination lines is expected as both prescriptions were calibrated on H$\alpha$ measurements \citep{murphy11,shipley16}. The AGN and shock from the outflow may have played a role in
reducing the PAH emission via destruction of small grains \citep{zhang22,gb24}. Meanwhile, since both hydrogen recombination lines and neon lines trace the ionizing photon rates \citep{ho07, murphy11}, the relatively lower values derived from the former are likely due to their faint emission blending partially with the underlying strong MIR continuum and PAH emission, hence we consider the estimates from the much brighter neon lines to be more reliable.\\
\indent Previous studies of IRASF01364 based on high-resolution ($<0\farcs1 = 90$\,pc) NIR Br$\gamma$ \citep{koala-goals} and radio continuum observations \citep{bm17,song22} yield nuclear SFR of 50 - 60\,M$_\odot$\,yr$^{-1}$, which is about five times higher than the estimates based on MIR neon lines. There are several potential causes for the discrepancy: First, the MIR lines may still suffer significant attenuation due to the heavy nuclear obscuration in IRAS\,F01364. If we were to assume that the  wavelength-dependent attenuation derived from \texttt{CAFE} applies equally to PAH and ionized gas, the inferred intrinsic SFR from the neon lines would be $\sim 30$\,M$_\odot$\,yr$^{-1}$, which approaches the values reported in the literature. However, as we note in Section \ref{sec:miri_analysis}, it is unlikely that the PAH and ionized gas are subject to the same attenuation.\\
\indent Alternatively, as shown in a recent work by \cite{robinson26} using results from radiative transfer modeling \citep{efs22}, SFR calibrations based on MIR neon lines or PAH features measured in normal star-forming galaxies consistently underestimate the SFR for local ULIRGs by up to 1 dex, which is attributed to a distinct mode of star formation in these more extreme systems. This is also observed by \cite{inami13}, who found that the neon line luminosities strongly correlate with the total IR luminosities at low IR luminosities but show clear deviation towards high IR luminosities. Applying the scaling relations derived by \cite{robinson26} (their Eqn. 34), we obtain from the [Ne\,II] line flux, a SFR of $\sim 40 - 50$\,M$_\odot$\,yr$^{-1}$ for the nuclear starburst, which is similar to values based on previous high-resolution NIR and radio continuum observations. Given that the total SFR of the system is $\sim 130$\,M$_\odot$ \citep{howell10}, and the nucleus dominates the MIR emission as shown in Figure \ref{fig:nuc_spec}, we hence adopt a nuclear SFR $\sim 40 - 60$\,$_\odot$\,yr$^{-1}$. A systematic, matched-resolution comparison among multi-wavelength SFR tracers across a range of nuclear environments in local LIRGs would further help elucidate the origin of the discrepancy identified in this work and the literature.
\subsection{Connecting the outflow, AGN and nuclear starburst} \label{dis:origin} 
As noted in Section \ref{dis:energy}, the total mass outflow rate and kinetic energy/power are dominated by those of the molecular gas phase, which are highly dependent on the unconstrained $\alpha_{\rm CO}$ factor. This hinders a direct comparison among the energetics of the outflow, starburst, and AGN to infer the origin and driving mechanisms of the outflow. Nevertheless, we note that the outflow orientation along the galaxy
minor axis and the derived outflow velocities (Section \ref{sec:kinematics}) match what is commonly observed in nearby
starbursts \citep[e.g.][]{leroy15, bolatto21, lamperti22}.  Adopting the SFR estimated from previous high-resolution NIR and radio continuum observations of $40 - 60$\,M$_\odot$\,yr$^{-1}$, the kinetic power delivered by
supernovae (SNe) from the nuclear starburst would be $\dot{E}_{\rm
kin, SN} = 7 \times 10^{41} ({\rm SFR/M_\odot\,yr^{-1}})\sim 4 \times
10^{43}$\,erg\,s$^{-1}$ \citep{veilleux05}, which is about two times higher than the estimated bolometric luminosity of the obscured AGN from Section \ref{sec:agn}.  In addition to the mechanical energy ejected by SNe, the dusty compact starburst also provides ideal conditions for supplying strong
IR radiation pressure that is efficient in driving the massive molecular outflow
\citep[e.g.][]{murray05, huang20}. Indeed, given a starburst radius of $\sim 140\,$pc
estimated from the radio continuum \citep{song22}, the inferred
star formation rate surface density ($\Sigma_{\rm SFR}$) would be $\gtrsim
10^3$\,M$_\odot$\,yr$^{-1}$/kpc$^2$, reaching the limit of radiation pressure supported
maximal starburst \citep{bm17, thompson05}.  Thus, the combined power from SNe and radiation pressure from the optically thick starburst may be sufficient in driving the observed outflow with a total kinetic energy of $\dot{E}_{\rm kin} = \dot{E}_{\rm kin, ion} +\dot{E}_{\rm kin, warm\,\ce{H2}}+\dot{E}_{\rm kin, cold\,\ce{H2}} \sim 2 - 8 \times10^{42}$\,erg\,s$^{-1}$, without evoking AGN contribution. The mass-loading factor, given by the ratio between the total mass outflow rate $M_{\rm out}$ ($\sim M_{\rm out, warm\,\ce{H2}} + M_{\rm out, cold \,\ce{H2}} \simeq 70 - 270$\,M$_\odot$\,yr$^{-1}$) and SFR, is about 1 - 5, which is also similar to those previously estimated in other starbursts \citep[e.g.,][]{leroy15,bolatto13,fluetsch19}.\\
\indent On the other hand, as discussed by \cite{cicone14} and \cite{lutz20}, AGN may still play a significant role in driving massive molecular outflows in starburst-dominated systems due to short-term AGN variability. Observational studies have shown that AGN power can vary rapidly, i.e., ``flicker'', over a timescale of $\sim\,$10$^{5}$\,yr \citep[e.g.,][]{schawinski07, keel12}, which is much shorter than the timescale of outflows and starbursts \citep[$> 10^6$\,Myr; e.g.,][]{zubovas20}. A more active AGN in the past would also help explain the high SMBH mass measured from previous 10-pc scale studies \citep{medling15,agostino26}, given the low accretion level estimated in Section \ref{sec:agn}. In this scenario, AGN winds driven by powerful AGN activity in the recent past may have supplied significant energy to efficiently drive materials out of the disk. In addition, as predicted by simulations by \cite{richings18a,richings18b} for AGN with $10^{44-47}$\,erg\,s$^{-1}$, some molecular gas may have also formed in-situ in the shocked layer of the AGN wind with moderate outflow velocity of few 100\,km\,s$^{-1}$, as the swept-up gas cools rapidly within $\sim 1$\,Myr. The extended ionized gas emission detected in the KCWI datasets out to $> 10\,$kpc beyond the nucleus (i.e., region exN and exS; Figure \ref{fig:fig9} and \ref{fig:oiii_chan}) may be the relics of such AGN-driven wind. Alternatively, they may be tidal features photo-ionized directly by the powerful AGN, as have been observed in other interacting and merging galaxies \citep{keel24, finlez25}. However, the emission in exN and exS do not show clear signature of AGN photo-ionization from the BPT diagram in Figure \ref{fig:fig13}, as commonly seen in  ``Extended Emission Line Regions'' (EELRs) found in
nearby quasars and active galaxies \citep[e.g.][Bianchin et al. in
prep.]{lintott09, keel12, sartori16, finlez22, venturi23, finlez25}. Hence, in the case that a more active AGN was present in the past, it likely did not reach sufficient power to dominate the overall gas excitation. \\
\begin{figure}[t!]
    \centering
    \includegraphics[width=0.9\linewidth]{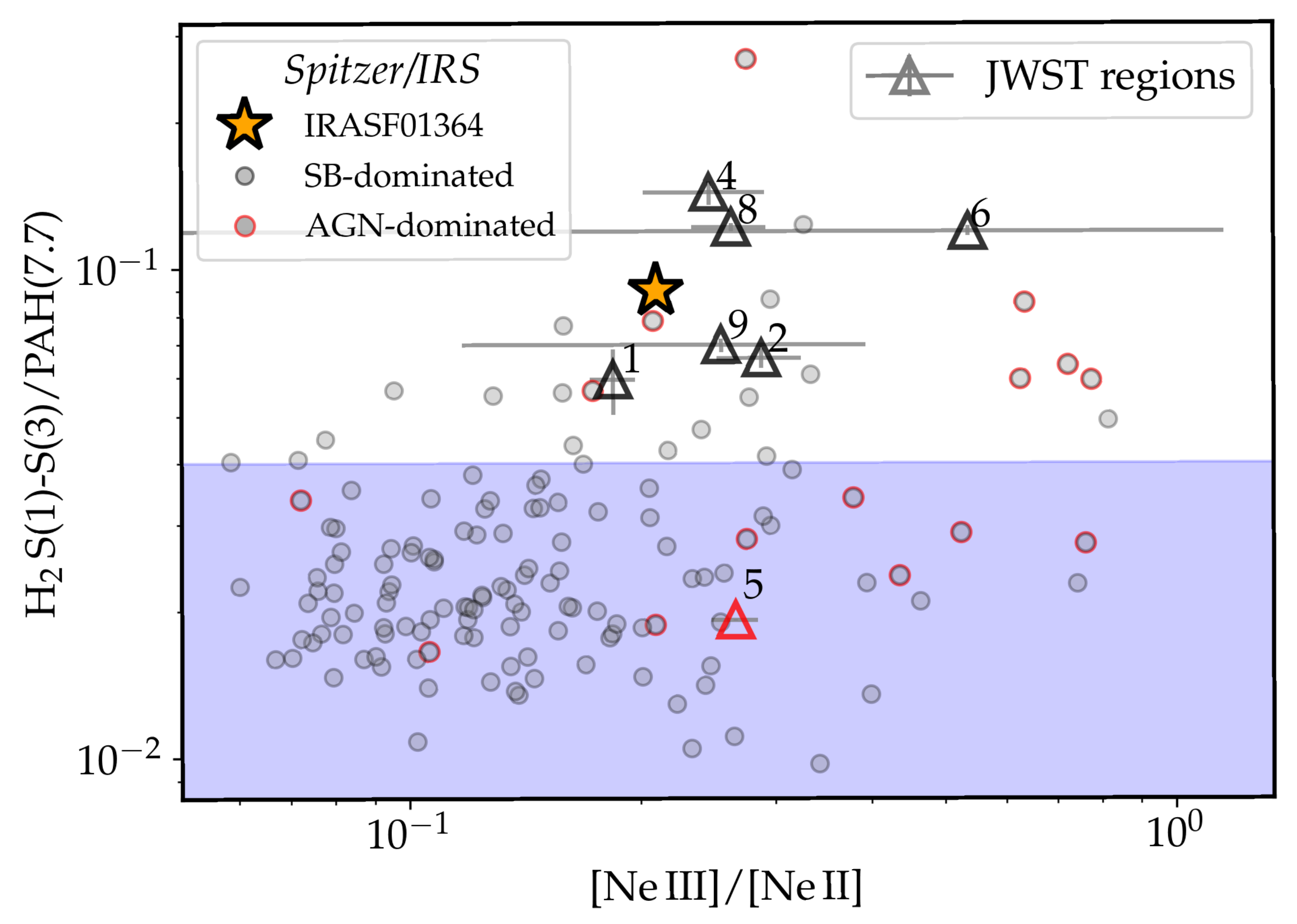}\\
    \caption{Ratio between the summed fluxes of \ce{H2},0-0\,S(1) - S(3) and the 7.7\,$\mu$m PAH feature, with respect to the [Ne III]/[Ne II] line flux ratio. Measurements from regions  1 - 9 are shown in triangles, and region 5 (the nucleus) is colored in red. Region 3 and 7 are excluded due to poor detection of the 7.7\,$\mu$m PAH feature. Measurements with \textit{Spitzer/IRS} for the GOALS sample from \cite{inami13} and \cite{stierwalt14} are shown in grey circles, with AGN-dominated systems outlined in red. Star marks the \textit{Spitzer/IRS} measurement for IRASF01364. The purple shaded area indicate threshold from PDR modelling \citep{guilard12}. All extranuclear regions show excess \ce{H2} emission consistent with previous \textit{Spitzer/IRS} measurement but the nucleus show low \ce{H2}/PAH ratio consistent with prediction for PDR.  \label{fig:fig16}}
\end{figure}
\indent Regardless of the driving mechanisms, the MIR and optical diagnostics presented in Figure \ref{fig:fig12} and \ref{fig:fig13} suggests that shock associated with the outflow may be responsible for gas excitation throughout the system, which likely gives rise to the excess \ce{H2} emission seen with \textit{Spitzer/IRS} \citep{stierwalt14}. In Figure \ref{fig:fig16} we show the resolved measurements of the \ce{H2}/PAH ratio and [Ne\,III]/[Ne\,II] ratio from region 1 - 9, where we also overlay previous \textit{Spitzer/IRS} measurement from \cite{stierwalt14} and \cite{inami13} for the GOALS sample (circles), including IRASF01364 (star). While measurements for the nuclear region (region 5) are consistent with the model-predicted limit for a photo-dissociation region \citep{guilard12}, marked in the shaded area, all extranuclear regions lie above this limit, and approach the values measured on the global scale. The \ce{H2}/PAH ratios also do not show clear dependence on the [Ne\,III]/[Ne\,II] ratio, indicating that the enhancement in \ce{H2} is not related to the hardness of the radiation field due to the starburst/AGN. This provides a spatially-resolved confirmation that excess \ce{H2} emission comes from shocked gas in the outflow rather than direct excitation from the starburst or AGN. In this regard, follow-up \textit{MRS} observations of bright \ce{H2} emitters identified by \cite{stierwalt13} will likely uncover many more cases of powerful galactic outflows such as the one in IRASF01364.

\section{Summary}\label{sec:summary} 
We combine IFU observations from \textit{JWST-MIRI/MRS}, Keck/KCWI and
ALMA to study the multiphase ISM properties in the late-stage merger
and local LIRG IRAS F01364-1042. This system was initially highlighted by \cite{stierwalt13} due to its bright \ce{H2} emission detected with \textit{Spitzer/IRS}. The new, spatially resolved multi-wavelength datasets allow us to uncover and characterize, for the first time, a multi-phase galactic biconical outflow and a dust-obscured AGN in this starburst-dominated system. We summarize the main results as follows: 
\begin{enumerate}
     \item The key tracers of the ionized, warm, and cold molecular gas all show enhanced line width along the minor axis (SE-NW direction) of an inclined rotating stellar disk. The region of enhanced linewidths exhibits a clear biconical shape, signifying the presence of a multi-phase biconical outflow extending at least 5\,kpc beyond the nucleus. Dust is present in the outflow, as visible from archival \textit{HST} images, and from the derived visual extinction based on H$\alpha$/H$\beta$ ratios measured from the KCWI datasets, which also supports a picture where the NW cone lies on the near-side, as inferred from the relatively stronger gas emission towards the NW direction.  
     \item The multi-phase gas exhibits different line profiles, with the ionized gas commonly showing a broad component with FWHM $> 500$\,km\,s$^{-1}$. We adopt a simplified kinematic model based on similar biconical outflows studied in the literature to explain the observed line profiles, which we attribute to the superposition of a gaseous nuclear disk and outflowing gas along the bicone. The molecular gas is likely more concentrated near the disk compared to the ionized gas, producing the apparent perpendicular motion observed between two. Based on this model, the observed line profiles and channel maps, we infer an outflow velocity of $\sim\,$500 - 600\,km\,s$^{-1}$, $\sim 350$\,km\,s$^{-1}$, and $\sim\,$200 - 300\,km\,s$^{-1}$ in the ionized, warm, and cold molecular gas phase, respectively. 
     \item Focusing on measurements from the central $\sim$\,3\,kpc, we infer $\sim\,$40\,\% flux contribution from the outflow, and estimate the multi-phase outflow energetics, which is heavily dominated by the cold molecular gas phase, similar to outflows found in other local LIRGs. We estimate a high cold molecular gas mass outflow rate of $\sim\,$38 - 240\,M$_\odot$\,yr$^{-1}$, and kinetic power of $\sim 0.5 - 6.8 \times 10^{42}$\,erg\,s$^{-1}$, which are similar to those measured in AGN-dominated systems from the literature. However, these value are subject to high uncertainties related to the CO-to-\ce{H2} conversion factor.
     \item Detection of faint [Ne\,V]\,14$\mu$m emission with \textit{JWST/MRS} at the nucleus confirms, for the first time, the presence of a low-luminosity AGN in IRASF01364, which was suggested by previous multi-wavelength studies. From the observed [Ne\,V]\,14$\mu$m flux we estimate an AGN bolometric
     luminosity of $L_{\rm bol} = 1.2 - 1.8\times10^{43}$\,erg\,s$^{-1}$. Comparison between the observed and the inferred intrinsic X-ray luminosity suggests the AGN is dust-obscured with an obscuring column density of $N_H \sim 10^{23}$\,cm$^{-2}$. The AGN is also accreting inefficiently based on the high SMBH mass estimated from previous 10-pc scale observations, indicating that the bulk of the SMBH growth took place in the past. 
     \item We estimate the SFR using MIR tracers measured on kpc-scales from the \textit{JWST/MRS} dataset, and found lower values compared to previous NIR/radio continuum observations at higher resolution, potentially due to MIR attenuation or distinct physical conditions present in the compact nuclear starburst. Adopting the SFR from the latter, we estimate kinetic power from supernovae of $\sim 4 \times 10^{43}$\,erg\,s$^{-1}$. Given the additional, high expected radiation pressure from the dusty starburst, star formation alone is sufficient for powering the observed powerful outflow. However, a more active AGN in the recent past may also have played a role by enhancing the molecular gas content in the outflow and producing the bright excess \ce{H2} emission in the MIR, as predicted by simulations. We confirm that the excess \ce{H2} seen on global scale with \textit{Spitzer/IRS} arises from outside the nucleus, likely a result from widespread shock associated with the outflow. 
\end{enumerate}

\noindent The results from this work highlight the importance of spatially-resolved multi-wavelength observations in investigating the nature of dust-obscured starbursts. In particular, we show that the ionized and molecular gas can have distinct behaviors, with ionized gas exhibiting stronger signatures of turbulence associated with the outflow, which may contribute to the differences observed between ionized and molecular gas tracers in dynamical studies of high$-z$ star-forming galaxies \citep[e.g.,][]{rizzo24}. Additionally, we emphasize that in the case of IRASF01364, the molecular gas distribution and kinematics observed on kpc-scale does not solely trace disk rotation, and caution should be taken when interpreting the cold gas dynamics when information from other gas phases is not available. Last but not least, this work showcases the capability of \textit{JWST/MRS} to unveil low-luminosity AGN hidden in dusty, starburst-dominated systems, which may represent a common phase in gas-rich galaxy mergers. 
\begin{acknowledgements}
    We would like to thank the anonymous referee for their thorough and constructive feedback that significantly improved the quality of this work. YS would like to thank F. Kemper, A. Prieto, A. Rodr\'{i}guez Ardila and M. Romano for their helpful inputs on interpreting the PAH features, coronal line kinematics and AGN/outflow properties. VU and MB acknowledges funding support from the Space Telescope Science Institute program \#JWST-GO-01717.001-A. VU acknowledges support from the National Science Foundation (NSF) Astronomy and Astrophysics Research Grant (AAG) \#AST-2536603, NASA Astrophysics Data Analysis Program (ADAP) grant \#80NSSC23K0750, and NASA Astrophysics Decadal Survey Precursor Science (ADSPS) grant \#80NSSC25K0169. MB thanks the financial support of the IAU-Gruber Foundation Fellowship, and acknowledges support from the Juan de La Cierva scholarship with reference JDC2023- 052684 -I, funded by MICIU/AEI/10.13039/501100011033 y por el FSE+ and from the Agencia Estatal de Investigaci\'on of the Ministerio de Ciencia, Innovaci\'on y Universidades (MCIU/AEI) under the grant ``Tracking active galactic nuclei feedback from parsec to kiloparsec scales'', with reference PID2022$-$141105NB$-$I00 and the European Regional Development Fund (ERDF). ET acknowledges support from ANID through CATA-BASAL FB210003, and FONDECYT Regular 1241005 and  1250821. MSG and CML acknowledges that this research project was supported by the Hellenic Foundation for Research and Innovation (HFRI) under the "2nd Call for HFRI Research Projects to support Faculty Members \& Researchers" (Project Number: 03382). TG acknowledges support from the Australian Research Council through Discovery Project DP210101945, funded by the Australian Government. DBS gratefully acknowledges support from NSF Grant 2407752. This work is based in part on observations made with the NASA/ESA/CSA James Webb Space Telescope. The data were obtained from the Mikulski Archive for Space Telescopes at the Space Telescope Science Institute, which is operated by the Association of Universities for Research in Astronomy, Inc., under NASA contract NAS 5-03127 for JWST. These observations are associated with program \#GO-01717. This paper also makes use of the following ALMA data: ADS/JAO.ALMA\#2017.1.01235.S, ADS/JAO.ALMA\#2019.1.00811.S, 2018.1.00279.S. ALMA is a partnership of ESO (representing its member states), NSF (USA) and NINS (Japan), together with NRC (Canada), NSTC and ASIAA (Taiwan), and KASI (Republic of Korea), in cooperation with the Republic of Chile. The Joint ALMA Observatory is operated by ESO, AUI/NRAO and NAOJ. The National Radio Astronomy Observatory is a facility of the National Science Foundation operated under cooperative agreement by Associated Universities, Inc. Some of the data presented herein were obtained at Keck Observatory, which is a private 501(c)3 non-profit organization operated as a scientific partnership among the California Institute of Technology, the University of California, and the National Aeronautics and Space Administration. The Observatory was made possible by the generous financial support of the W. M. Keck Foundation. The authors wish to recognize and acknowledge the very significant cultural role and reverence that the summit of Maunakea has always had within the Native Hawaiian community. We are most fortunate to have the opportunity to conduct observations from this mountain.
\end{acknowledgements}

%
\bibliographystyle{aa} 
\bibliography{refs.bib} 
\begin{appendix}



\onecolumn
\section{Information on the ALMA datasets}\label{sec:ap_alma} 
We provide details on the ALMA observations used in this work in Table \ref{tab:alma_info}, along with the characteristics of the reduced imaging products. The reduction and imaging procedures of these datasets are described in Section \ref{sec:data}. Emission line maps for the CO (J = 2-1) line, which achieves the highest SNR, are shown in Figure \ref{fig:fig7}. Maps for the CO (J = 1-0) and (J=3-2) lines are shown below in Figure \ref{fig:more_alma} for reference. 

\begin{table*}[h!]
\small
\caption{\label{a1} Information on ALMA observations and imaging products}
\centering
\begin{tabular}{ccccc}
\hline \hline
\multicolumn{5}{c}{Observations}\\
\hline
Project & PI & Band & MOUS & LAS \\
\hline
\multirow{2}{*}{2017.1.01235.S} & \multirow{2}{*}{L. Barcos-Mu\~{n}oz} & \multirow{2}{*}{3} & uid://A001/X1292/X3a & $1\farcs26$\\
&  &  & uid://A001/X1292/X3c &  $8\farcs65$\\
\hline
\multirow{2}{*}{2019.1.00811.S} & \multirow{2}{*}{A. Medling} & \multirow{2}{*}{6} & uid://A001/X1465/X228b & $3\farcs88$\\
  &  &  & uid://A001/X1465/X2285 &  $3\farcs92$\\
\hline
\multirow{2}{*}{2018.1.00279.S} & \multirow{2}{*}{L. Barcos-Mu\~{n}oz} & \multirow{2}{*}{7} & uid://A001/X13b3/X10c & $1\farcs15$\\
 &  &  & uid://A001/X1358/X39 &  $4\farcs24$\\
\hline \hline
\multicolumn{5}{c}{Imaging Products}\\
\hline
Band & Target Emission & $\theta_M \times \theta_m$ & PA   & $\sigma_{\rm rms}$ \\
 & & & ($\deg$)  & (mJy/beam) \\
\hline
 \multirow{2}{*}{3}& 105\,GHz Continuum &  $0\farcs37\times0\farcs28$ & $-$74 & 0.03\\
 & CO (J=1-0) & $0\farcs35\times0\farcs27$ & $-$73  & 0.54\\
 \hline
  \multirow{2}{*}{6}& 228\,GHz Continuum &  $0\farcs32\times0\farcs27$ & 40 & 0.05\\
 & CO (J=2-1) & $0\farcs42\times0\farcs33$ & 73  & 0.52\\
 \hline
  \multirow{2}{*}{7}& 334\,GHz Continuum & $0\farcs18\times0\farcs16$   & $-$60 & 0.20 \\
 & CO (J=3-2) & $0\farcs44\times0\farcs37$ & 90  & 2.20\\
 \hline
\end{tabular}
\tablefoot{$\theta_M$ and $\theta_m$ are the major and minor axis length of
the synthesized beam. The reported $\sigma_{\rm rms}$, in unit of mJy/beam, are
given per velocity channel width of 10.5, 10.5 and 7.1 km\,s$^{-1}$ for CO J=1-0, J=2-1
and J=3-2 cubes. In the last two columns we also list the Member ObsUnitSet
(MOUS) of the associated ALMA observations and the Largest Angular Scale (LAS),
or the maximum recoverable scale, of each dataset.\label{tab:alma_info} }
\end{table*}
\begin{figure}[h!]
    \centering
    \vspace{-8mm}
    \includegraphics[width=0.8\linewidth]{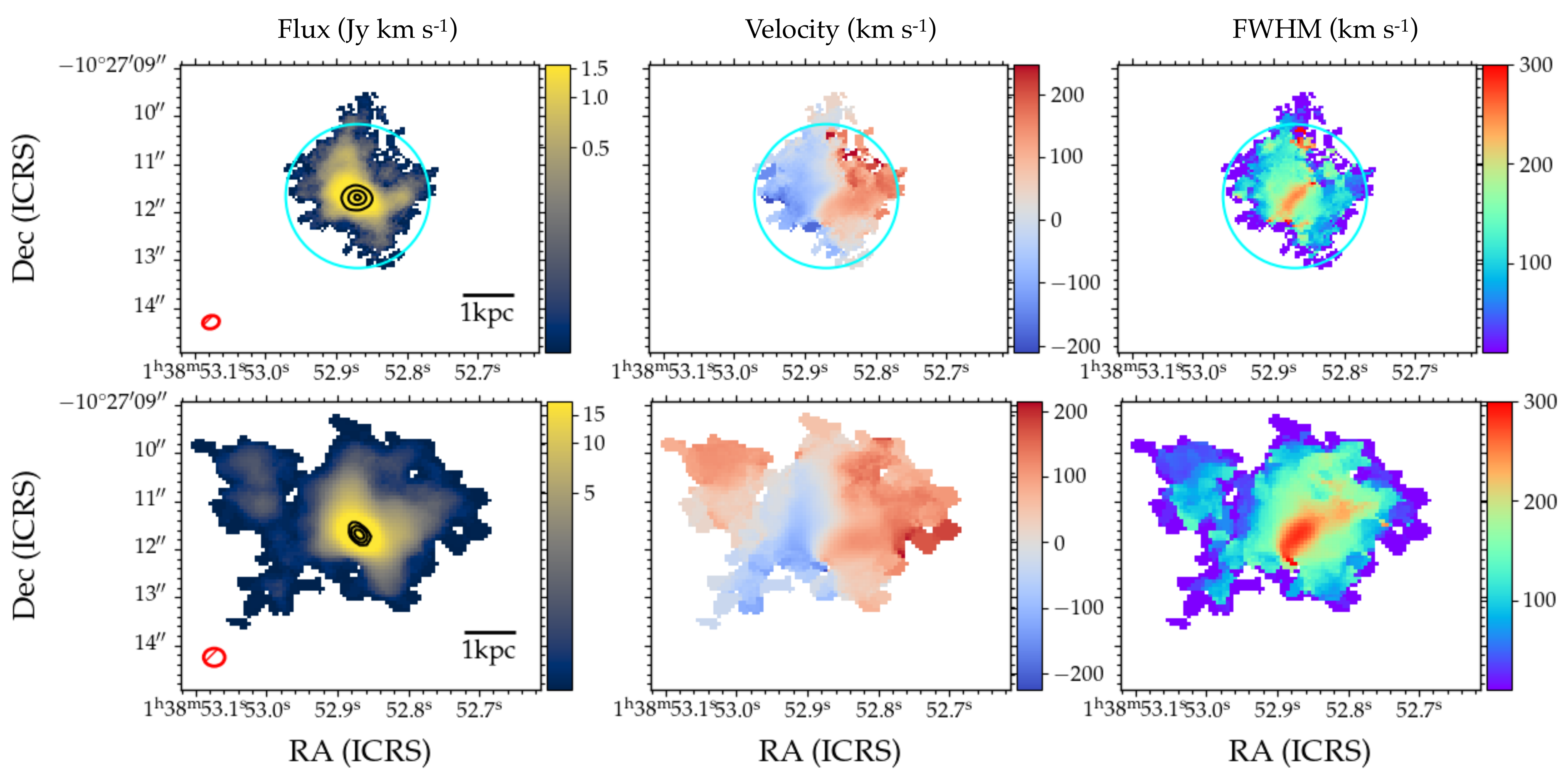}
    \caption{Flux, velocity and linewidth FWHM of the CO (J = 1 - 0) and CO (J = 3 - 2) emission, shown in the upper and lower panels, respectively. The 105\,GHz and 334\,GHz continuum emission in levels of [10, 30, 50]$\sigma_{\rm rms}$, and [5, 10, 20]$\sigma_{\rm rms}$, are overlaid in black contours over the CO (J = 1 - 0) and CO (J = 3 - 2) flux maps, respectively, where $\sigma_{\rm rms}$ are reported in Table \ref{tab:alma_info}. Cyan circle mark region C, which encompasses the bulk of the cold molecular gas emission traced by CO (J = 1 - 0).}
    \label{fig:more_alma}
\end{figure}
\vspace{-8mm}
\section{Fluxes of PAH features obtained from \textit{MRS} spectra with \texttt{CAFE}}\label{ap:more_pah}
We provide the observed and attenuation corrected fluxes of PAH features obtained from fitting the \textit{MRS} spectra using \texttt{CAFE} \citep{pycafe}, as described in Section \ref{sec:cafe_fitting}. In Table \ref{tab:cone_pah}, we report values for all features identified with \texttt{CAFE} from the nuclear and central spectra, described in Section \ref{sec:cafe_fitting} and shown in Figure \ref{fig:nuc_spec}. The fitted optical depths of absorption features of silicate at 9.7$\mu$m ($\tau_{\rm Sil}$),  water ice at 6$\mu$m ($\tau_{\rm H_2O}$), and hydrogenated amorphous carbon ($\tau_{\rm HAC}$) at $6.8\,\mu$m derived from \texttt{CAFE} are listed in the lower section of the Table. In Table \ref{tab:more_pah}, we list values for only the brightest PAH features at 6.2, 7.7 and 11.3\,$\mu$m from spectra extracted with region 1 - 9. These regions are defined in Section \ref{sec:miri_analysis} and shown in Figure \ref{fig:fig5}. The 6.2 and 7.7\,$\mu$m PAH features are poorly constrained in region 3 and 7, hence not reported here. For the other regions, flux ratios measured between these PAH features are visualized in Figure \ref{fig:fig11}. The fitted $\tau_{\rm Sil}$, $\tau_{\rm H_2O}$, and $\tau_{\rm HAC}$ derived from \texttt{CAFE} are listed in right side of the Table.
\begin{table}[t!]
\small
\caption{\label{tab:cone_pah} PAH features in the \textit{MRS} nuclear and central spectra} 
\centering
\begin{tabular}{l|cc|cc}
\hline\hline
 & \multicolumn{2}{c|}{Nuclear} & \multicolumn{2}{c}{Central (3\,kpc)}\\\hline
Feature & $f_{obs}$ & $f_{int}$ & $f_{obs}$ & $f_{int}$ \\
(1) & (2) & (3)  &  (4) & (5) \\
\hline
PAH6.2     & 0.21$\pm$0.01  & 0.90$\pm$0.04   & 0.66$\pm$0.06  & 2.54$\pm$0.21\\
PAH7.7(C)  & 0.70$\pm$0.02  & 2.17$\pm$0.06  & 1.75$\pm$0.11  & 5.31$\pm$0.34\\
PAH8.3     & 0.09$\pm$0.01  & 5.31$\pm$0.34  & 0.22$\pm$0.02  & 0.81$\pm$0.09\\
PAH8.6     & 0.11$\pm$0.01  & 0.45$\pm$0.02  & 0.28$\pm$0.03  & 1.19$\pm$0.13\\
PAH11.3(C) & 0.09$\pm$0.01  & 0.61$\pm$0.02  & 0.22$\pm$0.02  & 1.46$\pm$0.11\\
PAH12.6(C) & 0.21$\pm$0.01  & 0.79$\pm$0.04  & 0.33$\pm$0.04  & 1.20$\pm$0.13\\
PAH17.0(C) & 0.28$\pm$0.01  & 1.30$\pm$0.06  & 0.35$\pm$0.03  & 1.59$\pm$0.12\\
\hline
$\tau_{\rm Sil}$ &   \multicolumn{2}{c|}{12.0}  &  \multicolumn{2}{c}{11.6} \\ 
$\tau_{\rm H_2O}$ &  \multicolumn{2}{c|}{1.47} &  \multicolumn{2}{c}{1.07} \\
$\tau_{\rm HAC}$ & \multicolumn{2}{c|}{0.46} &  \multicolumn{2}{c}{0.22} \\
\hline
\end{tabular}
\tablefoot{(1) Name of the PAH feature or complex (``C''), where the number represents the central wavelength of the emission in $\mu$m. (2) \& (3): observed and attenuation-corrected PAH flux and uncertainties derived by \texttt{CAFE} from the nuclear spectrum, in units of 10$^{-15}$\,W\,m$^{-2}$. (4) \& (5): same as (2) and (3), but from spectra extracted within the central region (i.e., region C). Last three rows list the optical depths of the absorption features due to silicate, water ice and hydrogenated amorphous carbon derived with \texttt{CAFE}.}
\end{table}

\begin{table}[t!]
\small
\caption{\label{tab:more_pah} PAH features in \textit{MRS} spectra extracted from region 1 - 9} 
\centering
\begin{tabular}{c|cc|cc|cc|ccc}
\hline\hline
 & \multicolumn{2}{c|}{PAH6.2} & \multicolumn{2}{c|}{PAH7.7(C)} & \multicolumn{2}{c|}{PAH11.3(C)} & \multicolumn{1}{c}{\multirow{2}{*}{$\tau_{\rm Sil}$}} & \multicolumn{1}{c}{\multirow{2}{*}{$\tau_{\rm H_2O}$}}& \multicolumn{1}{c}{\multirow{2}{*}{$\tau_{\rm HAC}$}}\\\cline{1-7}
Region & $f_{obs}$ & $f_{int}$ & $f_{obs}$ & $f_{int}$ & $f_{obs}$ & $f_{int}$ \\ \hline
1 & 0.44$\pm$0.04 & 1.97$\pm$0.19 & 1.21$\pm$0.10 & 4.85$\pm$0.42 & 0.25$\pm$0.02 & 2.36$\pm$0.21 & 7.60 & 0.46 & 0.00\\ 
2 & 0.38 $\pm$ 0.05 & 1.68 $\pm$ 0.21 & 1.05 $\pm$ 0.09 & 3.96 $\pm$ 0.33 & 0.22 $\pm$ 0.22 & 1.88 $\pm$ 0.17 & 10.3 & 0.69 & 0.05\\
3 & - & - & - & - & 0.05 $\pm$ 0.05 & 0.28 $\pm$ 0.23 & - & - & -\\
4 & 0.44 $\pm$ 0.1 & 1.62 $\pm$ 0.37 & 1.4 $\pm$ 0.3 & 4.61 $\pm$ 1.0 & 0.27 $\pm$ 0.27 & 2.0 $\pm$ 0.48  & 7.79 & 0.37 & 0.00 \\
5 & 4.03 $\pm$ 0.21 & 18.1 $\pm$ 0.1 & 10.7 $\pm$ 0.6 & 38.2 $\pm$ 2.1 & 1.05 $\pm$ 1.05 & 8.4 $\pm$ 0.52 & 14.5 & 1.16 & 0.30 \\
6 & 0.45 $\pm$ 0.09 & 1.29 $\pm$ 0.26 & 1.41 $\pm$ 0.19 & 3.4 $\pm$ 0.47 &  0.26 $\pm$ 0.26 & 1.29 $\pm$ 0.22 & 5.90 & 0.56 & 0.04\\
7 & - & - &- & - & 0.07 $\pm$ 0.07 & 0.08 $\pm$ 0.03 & - & - &- \\
8 & 0.32 $\pm$ 0.06 & 1.14 $\pm$ 0.2 & 0.94 $\pm$ 0.12 & 3.0 $\pm$ 0.4 & 0.2 $\pm$ 0.2 & 1.46 $\pm$ 0.22 & 5.25 & 0.32 & 0.00 \\
9 & 0.67 $\pm$ 0.09 & 2.89 $\pm$ 0.37 & 1.77 $\pm$ 0.18 & 6.77 $\pm$ 0.7 & 0.29 $\pm$ 0.29 & 2.51 $\pm$ 0.27 & 12.1 & 0.61 & 0.04 \\
\hline
\end{tabular}
\tablefoot{$f_{obs}$ and  $f_{int}$ are observed and attenuation-corrected PAH flux and uncertainties derived by \texttt{CAFE}, in units of 10$^{-16}$\,W\,m$^{-2}$. Last three columns list the derived optical depths of the absorption features due to silicate, water ice and hydrogenated amorphous carbon.}
\end{table}
\vspace{-8mm}
\section{Selection of best-fit models of emission line profiles}\label{sec:model-selection}
\begin{figure}[h!]
    \centering
    \includegraphics[width=0.7\linewidth]{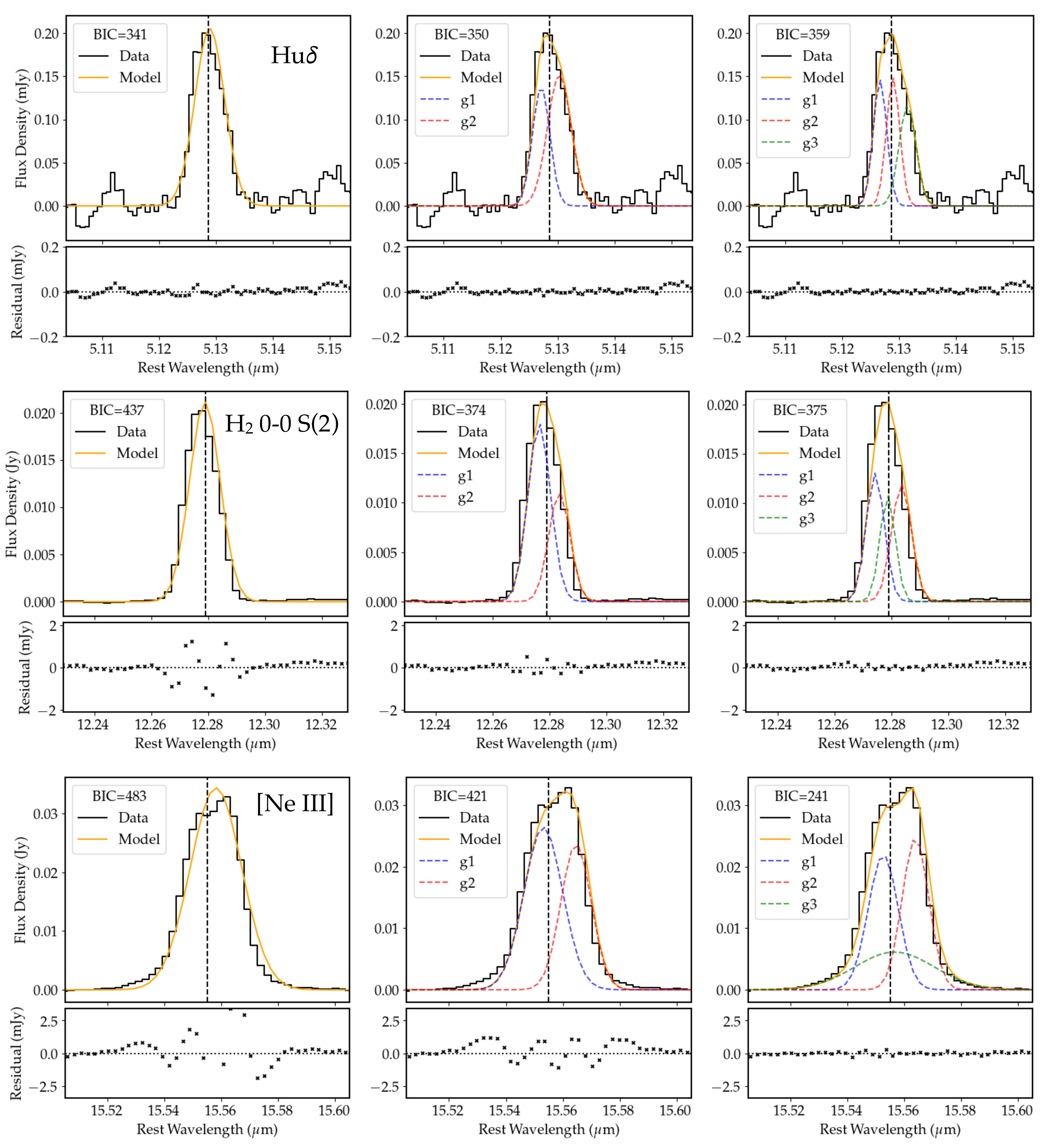}
    \caption{Continuum-subtracted line profiles of the Hu$\delta$, \ce{H2}\,0-0\,S(2), and [Ne\,III] lines detected in the nuclear \textit{MRS} spectrum (Figure \ref{fig:nuc_spec}), shown in black curves, respectively in the top, middle and bottom panels. The model fitting results (in orange curves) and residual using one, two and three Gaussian components are shown from left to right. The individual Gaussian components (i.e., ``g1'', ``g2'', ``g3'') for multi-Gaussian models are shown in dashed blue, red and green curves.}
    \label{fig:modelbic}
\end{figure}

We illustrate in Figure \ref{fig:modelbic} below the model selection process described in Section \ref{sec:miri_analysis} for fitting the emission line profiles. We consider the model with the lowest BIC value as the best-fit model, which correspond to a single Gaussian model, two-Gaussian model and three-Gaussian model for Hu$\delta$, \ce{H2}\,0-0\,S(2), and [Ne\,III] lines shown in Figure \ref{fig:modelbic}. In the case where the BIC values are the same or have only slight difference of $< 50$, we choose the model with less components as the best-fit to avoid over-fitting. We also exclude models that yield a best-fit component with SNR $< 5$. 

\section{Best-fit parameters for emission lines detected in the multi-wavelength datasets}\label{ap:best-fits}
We provide the best-fit parameters of continuum-subtracted line profiles of individual emission lines from the multi-wavelength IFU datasets presented in this work, using the methodology described in Section \ref{sec:miri_analysis} and Appendix \ref{ap:best-fits} above. In Table \ref{tab:cone_line} we list the values for all mid-IR emission lines identified from the nuclear and central \textit{MRS} spectra, defined in Section \ref{sec:cafe_fitting}. Table \ref{tab:reg_line} lists the values for [Ne\,II], [Ne\,III] and \ce{H2}\,0-0\,S(3) lines extracted from region 1 - 9, defined in Section \ref{sec:miri_analysis} and shown in Figure \ref{fig:fig5}. The continuum-subtracted line profiles of these three lines are also visualized in Figure \ref{fig:fig6}. Table \ref{tab:reg_co} lists the values for CO J=1-0, 2-1 and 3-2 lines extracted from region 1 - 9 and also the central region, from the multi-frequency ALMA datasets described in Appendix \ref{sec:ap_alma}. These line profiles are visualized in Figure \ref{fig:fig7}. Table \ref{tab:reg_kcwi} and \ref{tab:reg_kcwi_other} report values for the brightest emission lines detected in the KCWI spectra extracted in the seven regions (i.e., exS, SE, SW, C, NE, SW, exN) defined in Section \ref{sec:kcwi_analysis}. The line profiles of H$\beta$ and [O]\,III are visualized in Figure \ref{fig:fig9}. For the other lines, robust decomposition is challenging due to relatively fainter emission and/or line blending, hence only the integrated line fluxes are reported (i.e., in Table \ref{tab:reg_kcwi_other}). The flux ratios of these lines are visualized in Figure \ref{fig:fig13}.
\begin{table*}[h!]
\small
\caption{\label{tab:cone_line} Emission lines detected in the \textit{MRS} nuclear and central spectra} 
\centering
\begin{tabular}{cll|ccc|ccc}
\hline\hline
                           &                        &                       & \multicolumn{3}{c}{Nuclear Region}          & \multicolumn{3}{c}{Central Region}          \\
Line                       & $\lambda_{\rm rest}$                    & IP                    & $f_{\rm obs}$           & $v_{\rm cen}$        & FWHM         & $f_{\rm obs}$           & $v_{\rm cen}$        & FWHM        \\
(1)  & (2) & (3) & (4)  & (5)  & (6) & (7) &  (8) & (9)  \\ \hline
\ce{H2}\,0-0\,S(9)         & 4.69                   &                       & 0.62$\pm$0.06  & 60$\pm$8    & 285$\pm$19   & 5.16$\pm$0.17 & 33$\pm$4    & 360$\pm$9    \\ \hline
\ce{H2}\,0-0\,S(8)         & 5.05                   &                       & 0.43$\pm$0.02  & 48$\pm$2    & 306$\pm$5    & 2.67$\pm$0.10 & 39$\pm$3    & 321$\pm$9    \\ \hline
Hu$\delta$                          & 5.13                   & 13.6                  & 0.16$\pm$0.01  & 27$\pm$8    & 360$\pm$20   & 0.74$\pm$0.11 & 142$\pm$30  & 646$\pm$72 \\ \hline
\multirow{2}{*}{[Fe\,II]}           & \multirow{2}{*}{5.34}  & \multirow{2}{*}{7.9}  & 1.66$\pm$0.08  & -25$\pm$8   & 262$\pm$9    & 5.19$\pm$0.37 & -35$\pm$11  & 249$\pm$14   \\
                                    &                        &                       & 0.85$\pm$0.08  & 169$\pm$7   & 198$\pm$8    & 3.26$\pm$0.38 & 163$\pm$10  & 200$\pm$12   \\ \hline
\multirow{2}{*}{\ce{H2}\,0-0\,S(7)} & \multirow{2}{*}{5.51}  & \multirow{2}{*}{}     & 1.54$\pm$0.12  & -40$\pm$11  & 233$\pm$10   & 8.99$\pm$0.54 & -46$\pm$12  & 246$\pm$13   \\
                                    &                        &                       & 0.73$\pm$0.12  & 131$\pm$13  & 189$\pm$12   & 4.25$\pm$0.57 & 135$\pm$13  & 194$\pm$14   \\ \hline
Hu$\gamma$                          & 5.91                   & 13.6                  & 0.17$\pm$0.01  & 121$\pm$9   & 423$\pm$22   & 0.43$\pm$0.06 & 87$\pm$25   & 548$\pm$50   \\ \hline
\ce{H2}\,0-0\,S(6)                  & 6.11                   &                       & 0.80$\pm$0.02  & -6$\pm$2    & 284$\pm$5    & 5.18$\pm$0.14 & -4$\pm$3    & 306$\pm$6    \\ \hline
\multirow{2}{*}{\ce{H2}\,0-0\,S(5)} & \multirow{2}{*}{6.91}  & \multirow{2}{*}{}     & 2.99$\pm$0.13  & 0$\pm$6     & 218$\pm$7    & 14.9$\pm$0.5  & -30$\pm$4   & 213$\pm$4    \\
                                    &                        &                       & 1.00$\pm$0.14  & 167$\pm$11  & 178$\pm$10   & 9.15$\pm$0.48 & 151$\pm$5   & 196$\pm$6    \\ \hline
\multirow{3}{*}{[Ar\,II]}           & \multirow{3}{*}{6.99}  & \multirow{3}{*}{15.8} & 1.19$\pm$0.06  & -63$\pm$1   & 218$\pm$2    & 2.81$\pm$0.17 & -60$\pm$1   & 216$\pm$2    \\
                                    &                        &                       & 1.10$\pm$0.06  & 162$\pm$1   & 198$\pm$2    & 2.52$\pm$0.15 & 164$\pm$1   & 190$\pm$1    \\
                                    &                        &                       & 1.18$\pm$0.12  & -48$\pm$22  & 1211$\pm$64  & 2.86$\pm$0.32 & -73$\pm$28  & 1220$\pm$76  \\ \hline
Pf$\alpha$                          & 7.46                   & 13.6                  & 1.34$\pm$0.11  & 153$\pm$12  & 445$\pm$28   & 2.82$\pm$0.19 & 146$\pm$9   & 407$\pm$21   \\ \hline
\ce{H2}\,0-0\,S(4)                  & 8.03                   &                       & 2.77$\pm$0.11  & 12$\pm$4    & 303$\pm$9    & 13.1$\pm$0.2  & 2$\pm$2     & 318$\pm$4    \\ \hline
\multirow{3}{*}{[Ar\,III]}          & \multirow{3}{*}{8.99}  & \multirow{3}{*}{27.6} & 0.17$\pm$0.49  & -45$\pm$3   & 164$\pm$10   & 0.30$\pm$0.02 & -55$\pm$2   & 162$\pm$7    \\
                                    &                        &                       & 0.19$\pm$0.55  & 173$\pm$6   & 182$\pm$12   & 0.35$\pm$0.02 & 153$\pm$3   & 197$\pm$7    \\
                                    &                        &                       & 0.39$\pm$0.16  & -10$\pm$35  & 384$\pm$22   & 0.82$\pm$0.26 & 2$\pm$18    & 514$\pm$48   \\ \hline
\multirow{2}{*}{\ce{H2}\,0-0\,S(3)} & \multirow{2}{*}{9.66}  & \multirow{2}{*}{}     & 2.85$\pm$11.59 & -26$\pm$10  & 221$\pm$7    & 18.1$\pm$0.6  & -35$\pm$5   & 237$\pm$6    \\
                                    &                        &                       & 0.85$\pm$0.31  & 128$\pm$25  & 202$\pm$18   & 7.10$\pm$0.66 & 155$\pm$8   & 193$\pm$8    \\ \hline
\multirow{3}{*}{[S\,IV]}            & \multirow{3}{*}{10.51} & \multirow{3}{*}{34.8} & 0.05$\pm$0.02  & -61$\pm$4   & 173$\pm$12   & 0.13$\pm$0.14 & -60$\pm$6   & 188$\pm$13   \\
                                    &                        &                       & 0.06$\pm$0.01  & 177$\pm$7   & 217$\pm$12   & 0.18$\pm$0.14 & 170$\pm$7   & 197$\pm$16   \\
                                    &                        &                       & 0.13$\pm$0.05  & -28$\pm$40  & 489$\pm$50   & 0.29$\pm$0.15 & -43$\pm$53  & 532$\pm$69   \\ \hline
\multirow{2}{*}{\ce{H2}\,0-0\,S(2)} & \multirow{2}{*}{12.28} & \multirow{2}{*}{}     & 4.00$\pm$0.24  & -50$\pm$9   & 209$\pm$9    & 11.6$\pm$0.54 & -53$\pm$5   & 210$\pm$6    \\
                                    &                        &                       & 1.67$\pm$0.25  & 124$\pm$13  & 179$\pm$13   & 6.44$\pm$0.56 & 135$\pm$7   & 180$\pm$7    \\ \hline
Hu$\alpha$                          & 12.37                  & 13.6                  & 0.50$\pm$0.06  & 170$\pm$18  & 475$\pm$42   & 0.77$\pm$0.08 & 162$\pm$17  & 467$\pm$39   \\ \hline
\multirow{3}{*}{[Ne\,II]}           & \multirow{3}{*}{12.81} & \multirow{3}{*}{21.6} & 5.56$\pm$0.32  & -53$\pm$2   & 216$\pm$3    & 9.71$\pm$0.51 & -51$\pm$1   & 215$\pm$3    \\
                                    &                        &                       & 5.21$\pm$0.30  & 171$\pm$1   & 199$\pm$3    & 9.11$\pm$0.48 & 169$\pm$1   & 198$\pm$3    \\
                                    &                        &                       & 5.74$\pm$0.79  & 34$\pm$17   & 1030$\pm$64  & 9.97$\pm$1.30 & 32$\pm$14   & 928$\pm$54   \\ \hline
[Ne\,V]                             & 14.32                  & 97.1                  & 0.29$\pm$0.06  & 183$\pm$24  & 336$\pm$56   & 0.45$\pm$0.08 & 190$\pm$21  & 343$\pm$51   \\ \hline
[Cl\,II]                            & 14.37                  & 13.0                  & 0.56$\pm$0.07  & 41$\pm$15   & 374$\pm$34   & 0.83$\pm$0.09 & 40$\pm$12   & 357$\pm$29   \\\hline
\multirow{3}{*}{[Ne\,III]}          & \multirow{3}{*}{15.56} & \multirow{3}{*}{41.0} & 1.39$\pm$0.08  & -35$\pm$4   & 208$\pm$8    & 2.02$\pm$0.03 & -36$\pm$3   & 204$\pm$6    \\
                                    &                        &                       & 1.23$\pm$0.08  & 183$\pm$3   & 176$\pm$9    & 1.83$\pm$0.03 & 179$\pm$2   & 177$\pm$6    \\
                                    &                        &                       & 2.56$\pm$0.92  & 29$\pm$18   & 607$\pm$68   & 4.33$\pm$0.39 & 29$\pm$11   & 613$\pm$41   \\ \hline
\multirow{2}{*}{\ce{H2}\,0-0\,S(1)} & \multirow{2}{*}{17.04} & \multirow{2}{*}{}     & 9.06$\pm$0.30  & -57$\pm$5   & 196$\pm$5    & 24.2$\pm$0.7  & -44$\pm$3   & 214$\pm$4    \\
                                    &                        &                       & 4.69$\pm$0.31  & 128$\pm$6   & 156$\pm$6    & 9.75$\pm$0.66 & 150$\pm$4   & 145$\pm$4    \\ \hline
[S\,III]                            & 18.71                  & 23.3                  & 7.11$\pm$0.18  & 50$\pm$3    & 341$\pm$6    & 9.55$\pm$0.24 & 50$\pm$3    & 346$\pm$6    \\ \hline
[O\,IV]                             & 25.89                  & 54.9                  & 17.0$\pm$3.1   & -170$\pm$92 & 1524$\pm$219 & 19.3$\pm$3.6  & -178$\pm$97 & 1607$\pm$231 \\ \hline
\end{tabular}
\tablefoot{1): Name of the emission line. (2): Rest wavelength in
    $\mu$m. (3): Ionizing potential, in eV. For each line, we list the best-fit parameters of the individual Gaussian components. (4) \& (7): Component flux, in 10$^{-18}$\,W\,m$^{-2}$. (5) \& (8): Central component velocity relative to the galaxy systemic velocity, in km\,s$^{-1}$. (6) \& (9): Component FWHM in km\,s$^{-1}$. The instrumental FWHM (see Section \ref{sec:miri_data}) has been subtracted in quadrature from the reported values. See Section \ref{sec:miri_analysis} for details. }
\end{table*}

\begin{table*}[h!]
\small
\caption{\label{tab:reg_line} Best-fit parameters for [Ne\,III], [Ne\,II] and \ce{H2}\,0-0\,S(3) lines in region 1 - 9} 
\centering
\begin{tabular}{c|ccc|ccc|ccc}
\hline \hline
Region  &  \multicolumn{3}{c|}{[Ne\,III]}    & \multicolumn{3}{c|}{[Ne\,II]}  & \multicolumn{3}{c}{\ce{H2}~0-0~S(3)} \\ 
        & $f_{\rm obs}$   & $v_{\rm cen}$  & FWHM & $f_{\rm obs}$   & $v_{\rm cen}$  & FWHM & $f_{\rm obs}$   & $v_{\rm cen}$  & FWHM        \\ 
   &  (1)   & (2)   & (3)  & (4) & (5)   & (6) & (7)   & (8) & (9)       \\\hline
\multirow{2}{*}{1}         & 0.43$\pm$0.03 & 170$\pm$3   & 147$\pm$7   & 2.43$\pm$0.24 & 181$\pm$4  & 146$\pm$12 & 1.07$\pm$0.93 & 24$\pm$20  & 277$\pm$10 \\
                           & 0.84$\pm$0.04 & 70$\pm$5    & 350$\pm$5   & 4.44$\pm$0.31 & 65$\pm$8   & 337$\pm$7  & 1.20$\pm$1.00 & 101$\pm$4  & 185$\pm$12 \\\hline
\multirow{3}{*}{2}         & 0.33$\pm$0.02 & -9$\pm$5    & 272$\pm$7   & 1.35$\pm$0.08 & -20$\pm$3  & 234$\pm$7  & 0.89$\pm$0.17 & -91$\pm$3  & 220$\pm$10 \\
                           & 0.23$\pm$0.02 & 191$\pm$2   & 170$\pm$7   & 1.12$\pm$0.07 & 180$\pm$2  & 185$\pm$6  & 1.33$\pm$0.13 & 23$\pm$12  & 352$\pm$7  \\
                           & 0.57$\pm$0.09 & 185$\pm$20  & 507$\pm$18  & 1.44$\pm$0.35 & 106$\pm$15 & 653$\pm$51 &     -          &       -     &    -        \\\hline
\multirow{3}{*}{3}         & 0.03$\pm$0.02 & -65$\pm$5   & 225$\pm$18  & 0.13$\pm$0.01 & -70$\pm$4  & 203$\pm$11 & 1.73$\pm$0.19 & -94$\pm$2  & 168$\pm$9  \\
                           & 0.03$\pm$0.03 & 177$\pm$7   & 260$\pm$16  & 0.14$\pm$0.01 & 155$\pm$4  & 222$\pm$11 & 1.40$\pm$0.33 & -67$\pm$7  & 342$\pm$23 \\
                           & 0.08$\pm$0.05 & 53$\pm$21   & 663$\pm$108 & 0.35$\pm$0.07 & -5$\pm$20  & 599$\pm$64 &     -          &       -     &     -       \\\hline
\multirow{3}{*}{4}         & 0.43$\pm$0.10 & 13$\pm$21   & 235$\pm$43  & 4.23$\pm$0.14 & 42$\pm$5   & 344$\pm$5  & 3.34$\pm$1.53 & 47$\pm$49  & 264$\pm$35 \\
                           & 0.16$\pm$0.09 & 175$\pm$6   & 140$\pm$7   & 1.54$\pm$0.11 & 181$\pm$3  & 134$\pm$8  & 5.01$\pm$1.21 & 166$\pm$4  & 167$\pm$13 \\
                           & 0.87$\pm$0.30 & -24$\pm$11  & 526$\pm$39  &      -         &     -       &     -       &     -          &      -      &      -      \\\hline
\multirow{3}{*}{5}         & 1.16$\pm$0.02 & -35$\pm$5   & 206$\pm$9   & 5.45$\pm$0.27 & -52$\pm$1  & 216$\pm$3  & 4.65$\pm$1.71 & -23$\pm$6  & 231$\pm$5  \\
                           & 1.01$\pm$0.02 & 184$\pm$4   & 169$\pm$12  & 5.08$\pm$0.26 & 171$\pm$1  & 198$\pm$3  & 1.39$\pm$0.23 & 142$\pm$10 & 187$\pm$9  \\
                           & 2.35$\pm$1.08 & 42$\pm$15   & 534$\pm$67  & 5.15$\pm$0.60 & -12$\pm$16 & 985$\pm$53 &     -          &       -     &   -         \\\hline
\multirow{3}{*}{6}         & 0.67$\pm$0.09 & -54$\pm$4   & 172$\pm$8   & 2.68$\pm$0.46 & -73$\pm$4  & 189$\pm$13 & 4.20$\pm$0.14 & -87$\pm$1  & 171$\pm$3  \\
                           & 0.29$\pm$0.04 & 188$\pm$9   & 170$\pm$22  & 4.02$\pm$0.38 & 87$\pm$18  & 343$\pm$16 & 1.29$\pm$0.24 & -47$\pm$6  & 330$\pm$15 \\
                           & 1.63$\pm$0.25 & 113$\pm$12  & 494$\pm$22  &    -           &   -         &   -         &     -          &      -      &     -       \\\hline
\multirow{3}{*}{7}         & 0.06$\pm$0.01 & -104$\pm$10 & 288$\pm$13  & 0.23$\pm$0.02 & -55$\pm$6  & 244$\pm$13 & 2.30$\pm$0.50 & 30$\pm$34  & 314$\pm$24 \\
                           & 0.05$\pm$0.01 & 120$\pm$4   & 220$\pm$13  & 0.21$\pm$0.02 & 148$\pm$4  & 220$\pm$12 & 2.17$\pm$0.52 & 168$\pm$5  & 215$\pm$13 \\
                           & 0.13$\pm$0.03 & -79$\pm$21  & 680$\pm$67  & 0.37$\pm$0.14 & -85$\pm$45 & 688$\pm$84 &      -         &      -      &      -      \\\hline
\multirow{3}{*}{8}         & 0.20$\pm$0.02 & -56$\pm$2   & 186$\pm$5   & 0.94$\pm$0.03 & -64$\pm$2  & 192$\pm$4  & 2.33$\pm$0.21 & -18$\pm$12 & 240$\pm$10 \\
                           & 0.21$\pm$0.02 & 158$\pm$2   & 208$\pm$5   & 1.01$\pm$0.03 & 150$\pm$2  & 212$\pm$4  & 2.03$\pm$0.22 & 163$\pm$10 & 219$\pm$9  \\
                           & 0.49$\pm$0.05 & -137$\pm$22 & 542$\pm$18  & 1.32$\pm$0.18 & -59$\pm$15 & 617$\pm$27 &     -          &      -      &    -        \\\hline
\multirow{3}{*}{9}         & 0.47$\pm$0.09 & -51$\pm$5   & 159$\pm$6   & 3.06$\pm$1.52 & -75$\pm$4  & 183$\pm$12 & 1.93$\pm$0.27 & -34$\pm$3  & 157$\pm$8  \\
                           & 0.24$\pm$0.05 & 158$\pm$15  & 120$\pm$31  & 4.59$\pm$0.48 & 59$\pm$14  & 336$\pm$10 & 2.29$\pm$0.37 & 15$\pm$7   & 249$\pm$6  \\
                           & 1.16$\pm$0.21 & -33$\pm$14  & 473$\pm$16  &     -          &     -       &    -        &   -            &            &    -       \\ \hline
\end{tabular}
\tablefoot{(1), (4), (7): Component flux, in 10$^{-18}$\,W\,m$^{-2}$. (2), (5), (8): Central component velocity relative to the galaxy systemic velocity, in km\,s$^{-1}$. (3), (6), (9): Component FWHM in km\,s$^{-1}$. The instrumental FWHM (see Section \ref{sec:miri_data}) have been subtracted in quadrature from the reported values. These parameters are visualized in Figure \ref{fig:fig6}.}
\end{table*}
\begin{table*}[h!]
\small
\caption{\label{tab:reg_co} Best-fit parameters for CO (1-0), CO (2-1) and CO (3-2) lines} 
\centering
\begin{tabular}{c|ccc|ccc|ccc}
\hline \hline
Region & \multicolumn{3}{c|}{CO (1-0)}              & \multicolumn{3}{c|}{CO (2-1)}             & \multicolumn{3}{c}{CO (3-2)}                 \\
    & $S_\nu \Delta v$          & $v_{\rm cen}$    & FWHM       & 
    $S_\nu \Delta v$          & $v_{\rm cen}$    & FWHM         & $S_\nu \Delta v$          & $v_{\rm cen}$    & FWHM  \\ 
        & (1)         & (2)    &    (3)         & 
          (4)         &  (5)   &    (6)         & (7)          & (8)    & (9) \\\hline
\multirow{2}{*}{1}      & 2.06$\pm$0.27 & 59$\pm$7      & 167$\pm$16 & 2.64$\pm$0.70 & 65$\pm$3      & 125$\pm$16  & 13.9$\pm$0.7  & 61$\pm$3     & 202$\pm$8  \\
                        & -             & -             & -          & 4.61$\pm$1.41 & 60$\pm$5      & 260$\pm$27  & -             & -            & -          \\ \hline
\multirow{2}{*}{2}      & -             & -             & -          & 1.91$\pm$0.53 & $-$21$\pm$34  & 482$\pm$71  & 4.23$\pm$0.73 & $-$85$\pm$10 & 183$\pm$24 \\
                        & -             & -             & -          & 0.79$\pm$0.24 & $-$104$\pm$10 & 143$\pm$32  &               &              &            \\ \hline
3                       & 1.15$\pm$0.30 & $-$125$\pm$18 & 208$\pm$42 & 1.79$\pm$0.17 & $-$95$\pm$5   & 167$\pm$12  & 3.73$\pm$0.57 & $-$87$\pm$8  & 156$\pm$18 \\ \hline
\multirow{2}{*}{4}      & 2.49$\pm$0.27 & 150$\pm$6     & 176$\pm$14 & 6.38$\pm$1.27 & 154$\pm$3     & 149$\pm$11  & 11.9$\pm$3.3  & 160$\pm$4    & 146$\pm$17 \\
                        & -             & -             & -          & 2.00$\pm$1.23 & 25$\pm$99     & 257$\pm$109 & 8.5$\pm$3.3   & 32$\pm$65    & 264$\pm$71 \\ \hline
\multirow{2}{*}{5}      & 7.47$\pm$1.08 & 131$\pm$14    & 193$\pm$16 & 24.8$\pm$0.7  & 140$\pm$2     & 176$\pm$3   & 46.4$\pm$1.8  & 149$\pm$3    & 171$\pm$4  \\
                        & 7.77$\pm$1.08 & $-$44$\pm$17  & 209$\pm$21 & 32.3$\pm$0.7  & $-$44$\pm$3   & 215$\pm$4   & 67.7$\pm$1.9  & $-$41$\pm$4  & 224$\pm$6  \\ \hline
6                       & 2.90$\pm$0.27 & $-$91$\pm$5   & 177$\pm$13 & 7.98$\pm$0.19 & $-$87$\pm$1   & 194$\pm$3   & 15.8$\pm$0.8  & $-$82$\pm$3  & 182$\pm$7  \\ \hline
7                       & 1.22$\pm$0.37 & 172$\pm$29    & 301$\pm$69 & 2.76$\pm$0.20 & 116$\pm$7     & 305$\pm$17  & 9.35$\pm$0.96 & 121$\pm$10   & 300$\pm$23 \\ \hline
\multirow{2}{*}{8}      & 2.07$\pm$0.32 & 171$\pm$16    & 313$\pm$37 & 1.45$\pm$0.41 & 150$\pm$9     & 157$\pm$29  & 13.1$\pm$0.8  & 96$\pm$6     & 285$\pm$14 \\
                        & -             & -             & -          & 5.33$\pm$0.75 & 70$\pm$13     & 358$\pm$19  & -             & -            & -          \\ \hline
9                       & 3.80$\pm$0.27 & $-$26$\pm$4   & 182$\pm$10 & 12.0$\pm$0.2  & $-$27$\pm$1   & 180$\pm$2   & 23.6$\pm$0.6  & $-$25$\pm$2  & 190$\pm$4  \\ \hline
\multirow{2}{*}{C}      & 21.9$\pm$6.7  & 103$\pm$51    & 276$\pm$61 & 44.9$\pm$2.1  & 140$\pm$4     & 182$\pm$5   & 77.0$\pm$5.3  & 151$\pm$5    & 163$\pm$7  \\
                        & 13.1$\pm$7.3  & $-$76$\pm$30  & 202$\pm$38 & 69.1$\pm$2.2  & $-$43$\pm$4   & 216$\pm$6   & 151$\pm$6     & $-$34$\pm$5  & 221$\pm$8  \\  \hline
\end{tabular}
\tablefoot{(1), (4), (7): Component flux, in Jy\,km\,s$^{-1}$. (2), (5), (8): Central component velocity relative to the galaxy systemic velocity, in km\,s$^{-1}$. (3), (6), (9): Component FWHM in km\,s$^{-1}$. These parameters are visualized in Figure \ref{fig:fig7}.}
\end{table*}
\begin{table*}[b!]
\small
\caption{\label{tab:reg_kcwi} Best-fit parameters for H$\beta$ and [O\,III]$\lambda$5007 lines} 
\centering
\begin{tabular}{c|ccc|ccc}
\hline \hline
Region &               & H$\beta$       &             &               & [O\,III]$\lambda$5007   &            \\
      & $f_{\rm obs}$          & $v_{\rm cen}$         & FWHM       & $f_{\rm obs}$          & $v_{\rm cen}$        & FWHM       \\
 & (1) & (2) & (3) & (4) & (5) & (6) \\ \hline

\multirow{2}{*}{exS}    & 1.69$\pm$0.10 & 94$\pm$8   & 409$\pm$19 & 2.61$\pm$0.31 & 75$\pm$10  & 474$\pm$31 \\
                        & -             & -          & -          & 0.70$\pm$0.12 & 109$\pm$5  & 120$\pm$16 \\ \hline
\multirow{2}{*}{SE}     & 1.29$\pm$0.21 & 210$\pm$18 & 806$\pm$58 & 3.28$\pm$0.10 & 159$\pm$6  & 704$\pm$15 \\
                        & 0.81$\pm$0.09 & 146$\pm$6  & 297$\pm$22 & 0.09$\pm$0.03 & 181$\pm$8  & 60$\pm$22  \\ \hline
\multirow{2}{*}{SW}     & 0.58$\pm$0.09 & 203$\pm$8  & 111$\pm$17 & 0.79$\pm$0.11 & 172$\pm$13 & 176$\pm$24 \\
                        & 0.34$\pm$0.07 & 311$\pm$6  & 48$\pm$11  & 0.29$\pm$0.08 & 287$\pm$4  & 46$\pm$14  \\ \hline
\multirow{3}{*}{C}      & 3.46$\pm$0.80 & 44$\pm$28  & 244$\pm$24 & 2.51$\pm$0.24 & 14$\pm$10  & 204$\pm$16 \\
                        & 2.93$\pm$0.87 & 200$\pm$20 & 216$\pm$17 & 1.67$\pm$0.22 & 170$\pm$7  & 142$\pm$10 \\
                        & 1.75$\pm$0.30 & 137$\pm$21 & 874$\pm$65 & 3.91$\pm$0.37 & 101$\pm$8  & 637$\pm$26 \\ \hline
\multirow{3}{*}{NW}     & 0.40$\pm$0.10 & -75$\pm$16 & 301$\pm$57 & 4.12$\pm$0.09 & -30$\pm$5  & 688$\pm$11 \\
                        & 0.34$\pm$0.04 & 187$\pm$4  & 111$\pm$11 & 0.35$\pm$0.03 & 154$\pm$3  & 66$\pm$7   \\
                        & 1.46$\pm$0.24 & -75$\pm$22 & 970$\pm$76 & -             & -          & -          \\ \hline
\multirow{2}{*}{NE}     & 0.64$\pm$0.08 & 119$\pm$16 & 209$\pm$21 & 0.53$\pm$0.13 & 19$\pm$31  & 214$\pm$48 \\
                        & 0.22$\pm$0.06 & 224$\pm$4  & 70$\pm$17  & 0.50$\pm$0.11 & 176$\pm$9  & 124$\pm$17 \\
exN                     & 2.47$\pm$0.15 & 119$\pm$8  & 393$\pm$18 & 5.10$\pm$0.16 & 115$\pm$4  & 402$\pm$10 \\ \hline
\end{tabular}
\tablefoot{(1)(4): Component flux, in 10$^{-16}$\,erg\,s$^{-1}$\,cm$^{-2}$. (2)(5): Central component velocity relative to the galaxy systemic velocity, in km\,s$^{-1}$. (3)(6): Component FWHM in km\,s$^{-1}$. These parameters are visualized in Figure \ref{fig:fig9}.}
\end{table*}

\begin{table*}[h!]
\small
\caption{\label{tab:reg_kcwi_other} Flux measurements of optical emission lines detected with KCWI} 
\centering
\begin{tabular}{clllllll}
\hline \hline
Region & [O\,II]$\lambda3726$           & [O\,II]$\lambda3729$           & H$\alpha$         & [N\,II]$\lambda$6584            & [O\,I]$\lambda6300$            & [S\,II]$\lambda$6716              & [S\,II]$\lambda$6731        \\ \hline 
exS                     & 5.20$\pm$1.52  & 6.17$\pm$1.38  & 10.3$\pm$1.0 & 8.66$\pm$0.96  & 4.01$\pm$0.45  & 1.61$\pm$0.12  & 1.11$\pm$0.08  \\
SE                      & 7.75$\pm$2.36  & 2.98$\pm$7.92  & 11.0$\pm$0.2 & 21.0$\pm$0.3 & 3.46$\pm$0.19  & 5.30$\pm$0.05  & 4.06$\pm$0.05  \\
SW                      & 3.35$\pm$0.39  & 3.97$\pm$0.37  & 6.21$\pm$0.10  & 4.47$\pm$0.10  & 1.93$\pm$0.18  & 2.95$\pm$0.05  & 2.03$\pm$0.04  \\
C                       & 10.3$\pm$1.0 & 15.0$\pm$0.8 & 46.3$\pm$0.6 & 42.7$\pm$0.6 & 10.7$\pm$0.3 & 18.1$\pm$0.1 & 13.0$\pm$0.1 \\
NW                      & 6.77$\pm$17.37 & 9.85$\pm$4.38  & 7.81$\pm$2.02  & 10.6$\pm$0.3 & 1.34$\pm$0.11  & 2.45$\pm$0.08  & 1.80$\pm$0.06  \\
NE                      & 2.25$\pm$0.22  & 2.39$\pm$0.21  & 3.09$\pm$0.50  & 2.51$\pm$0.35  & 1.32$\pm$0.16  & 1.75$\pm$0.16  & 1.21$\pm$0.11  \\
exN                     & 7.92$\pm$1.53  & 8.97$\pm$1.35  & 19.7$\pm$1.7 & 6.75$\pm$0.86  & 1.59$\pm$0.21  & 1.46$\pm$0.14  & 1.58$\pm$0.14 \\ \hline
\end{tabular}
\tablefoot{Values are in units of 10$^{-16}$\,erg\,s$^{-1}$\,cm$^{-2}$.}
\end{table*}
\section{Kinematic modelling of the central molecular gas disk with $^{3\mathrm{D}}$Barolo }\label{ap:bbarolo}
We estimate the contribution from the rotating molecular disk to the total CO (J = 2 - 1) flux measured within the central 3\,kpc covered by region C, by performing tilted-ring modeling with $^{3\mathrm{D}}$Barolo \citep{3dbarolo}. \cite{agostino26} already performed kinematic modelling of the nuclear molecular disk with $^{3\mathrm{D}}$Barolo using CO (J = 2 - 1) observations with an angular resolution of $\sim 0\farcs02$ (19\,pc), which is more than 10\,times higher than the CO (J = 2 - 1) observation used this work that focuses on the extended emission. Here we assume that the massive nuclear molecular gas disk modeled by \cite{agostino26} on $\sim$10-pc scale also dictates the rotational motion on kpc-scale, and hence we model the inner kpc ($\sim 1''$) of the low-resolution CO (J = 2 - 1) emission by fixing the inclination and PA to values derived by \cite{agostino26} from high-resolution observations, which are 60$^{\circ}$ and 256$^{\circ}$, respectively. The rotational velocity and dispersion are allowed to vary freely. The modelling results are shown below in Figure \ref{fig:bbarolo} in red contours over the PV diagrams extracted along the major ($\phi = 256^{\circ}$) and minor axis ($\phi = 345^{\circ}$), shown in grey scales. The model accounts for $\sim$\,60\% of the total CO (J = 2 - 1 ) flux, as shown in the rightmost panel. Region C, indicated with vertical dashed cyan lines in the PV diagrams, covers the bulk of the cold molecular gas outflow that is most visible along the minor axis, while excluding the tidal feature seen towards NE that is most visible along the major axis. We note that the modelling performed here is only meant as an independent assessment of the disk contribution, rather than an accurate representation of the underlying gas kinematics, given that the observed gas motions within the central kpc may still be impacted by the outflow. 
\begin{figure}[h!]
    \centering
    \includegraphics[width=0.9\linewidth]{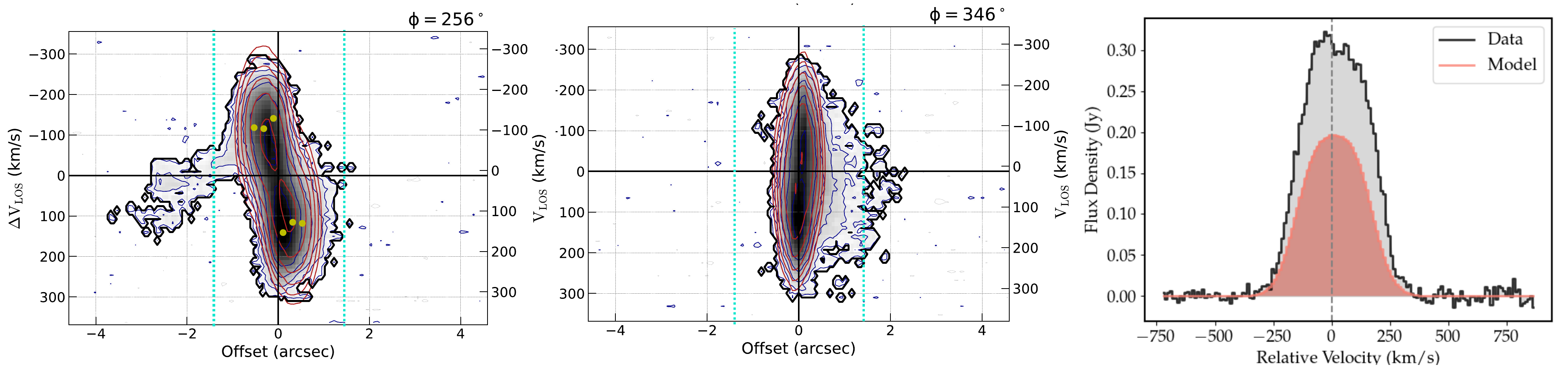}
    \caption{Modelling results of the central kpc of the CO (J = 2 - 1) emission from $^{3\mathrm{D}}$Barolo. The left and middle panels show the PV diagrams of the fitted model of the nuclear rotating molecular disk (in red contours), along the major ($\phi = 256^{\circ}$) and minor axis ($\phi = 345^{\circ}$) previously determined by \cite{agostino26}. The observed emission are shown in grey scales, and area covered by region C is indicated with vertical dashed cyan lines. The right panel compares the observed and modeled CO ( J = 2 - 1) line profile extracted from region C, shown in grey and red, respectively. The modeled nuclear disk contributes 60\% of the total flux contained within region C.} 
    \label{fig:bbarolo}
\end{figure}

\section{Channel maps for the multi-phase gas tracers}\label{sec:chanmaps}
\begin{figure}[h!]
    \centering
    \includegraphics[width=0.7\linewidth]{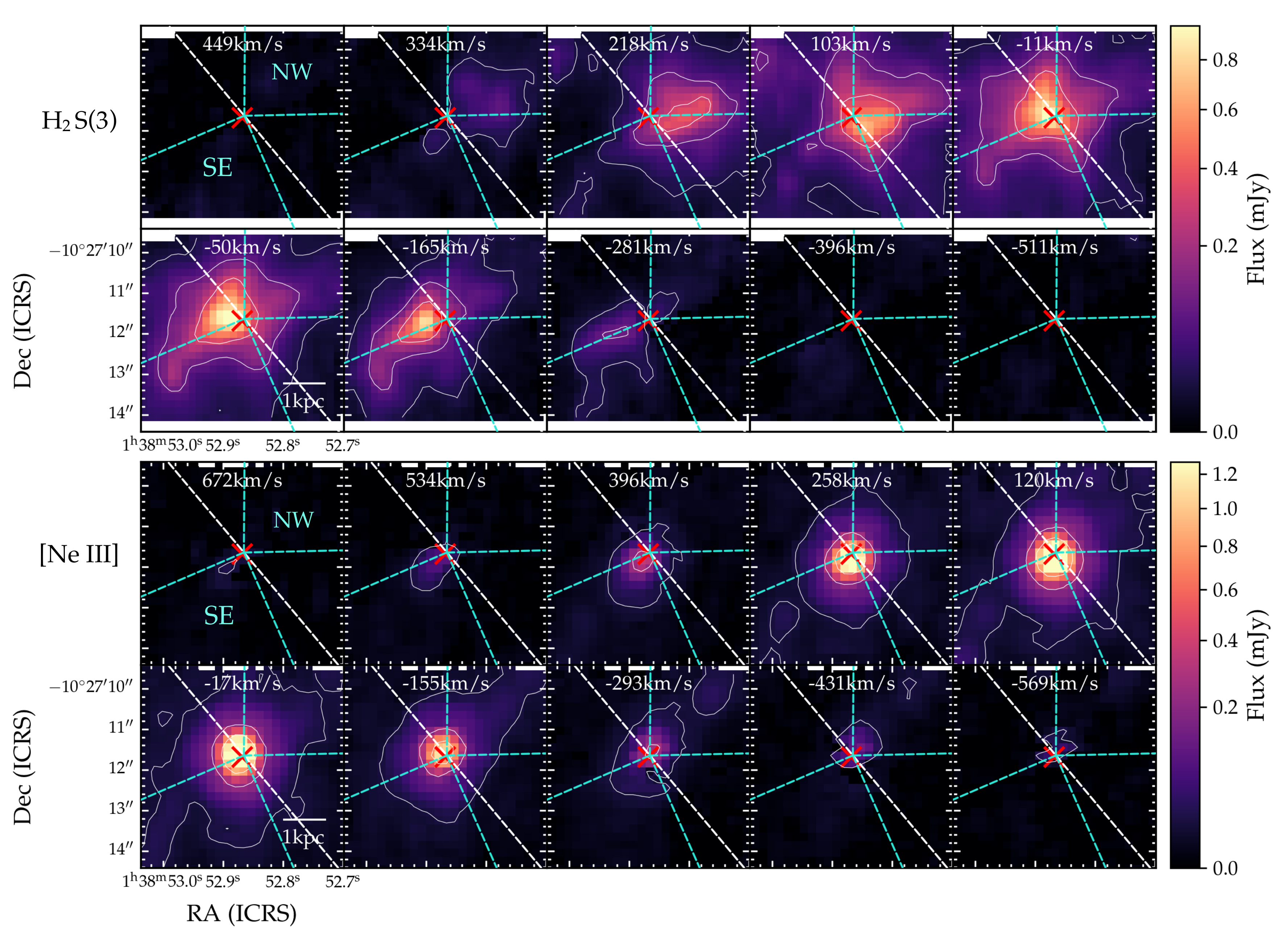}
    \caption{Channel maps for (top) \ce{H2}\,0-0\,S(3) and (bottom) [Ne\,III]$15.6\mu$m lines observed with \textit{JWST/MIRI-MRS}. Red cross mark the position of the nucleus based on the ALMA 334\,GHz continuum image. White and cyan dash lines represent the kinematic major axis of the stellar disk, and the outlines of the NW and SE outflow cones, respectively. White contours show levels of 0.015, 0.09 , 0.3\,mJy, and 0.016, 0.24 , 0.8\,mJy, for the top and bottom panels respectively.  
    \label{fig:miri_chan}}
\end{figure}

\begin{figure}[h!]
    \centering
    \includegraphics[width=0.75\linewidth]{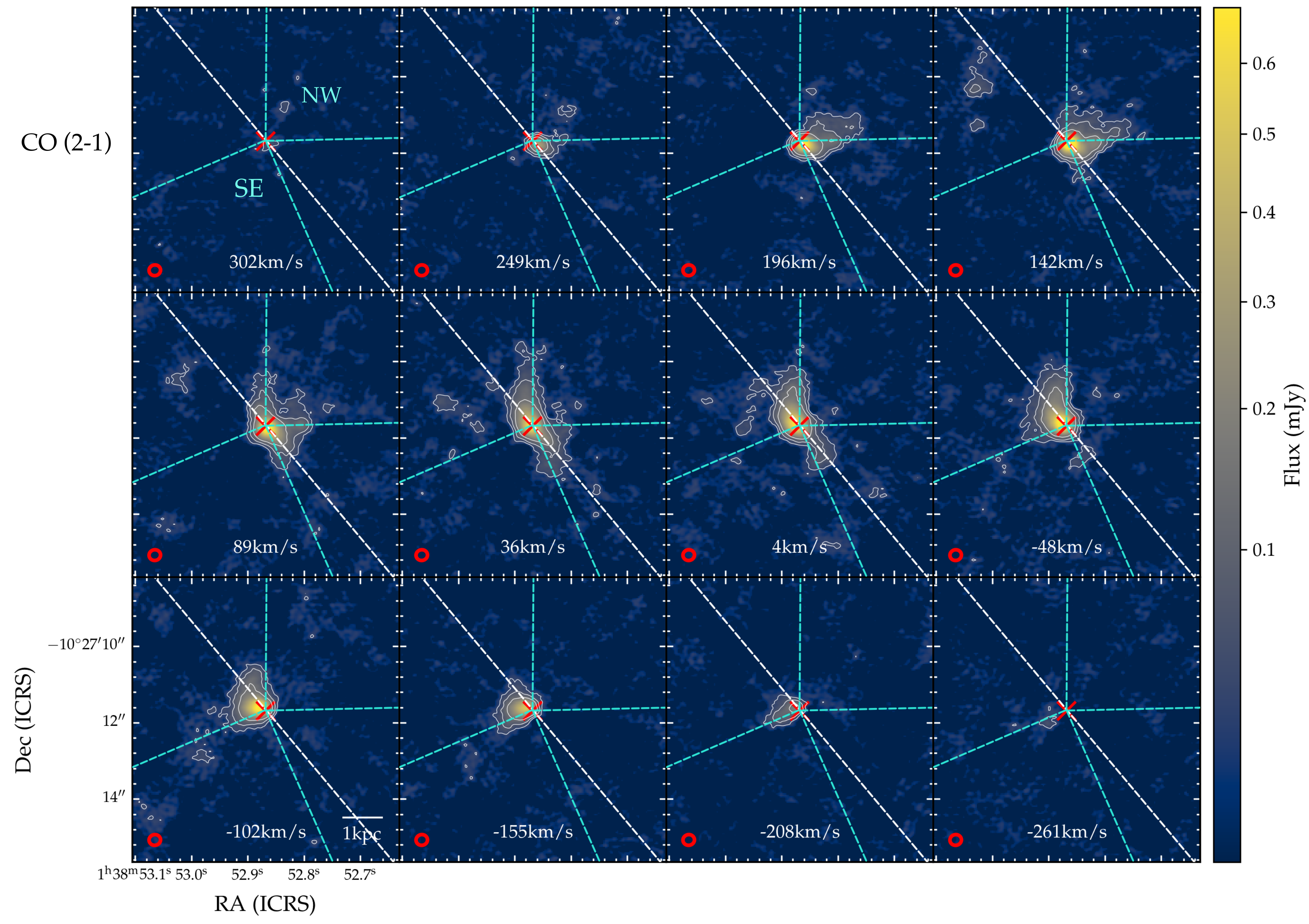}
    \caption{Channel maps for the CO(J = 2 - 1) line observed with ALMA. Red cross, white and cyan dashed lines are the same as in Figure \ref{fig:miri_chan}. White contours show levels of 0.04, 0.07, 0.14\,mJy. } 
    \label{fig:co21_chan}
\end{figure}

\begin{figure}[h!]
    \centering
    \includegraphics[width=0.75\linewidth]{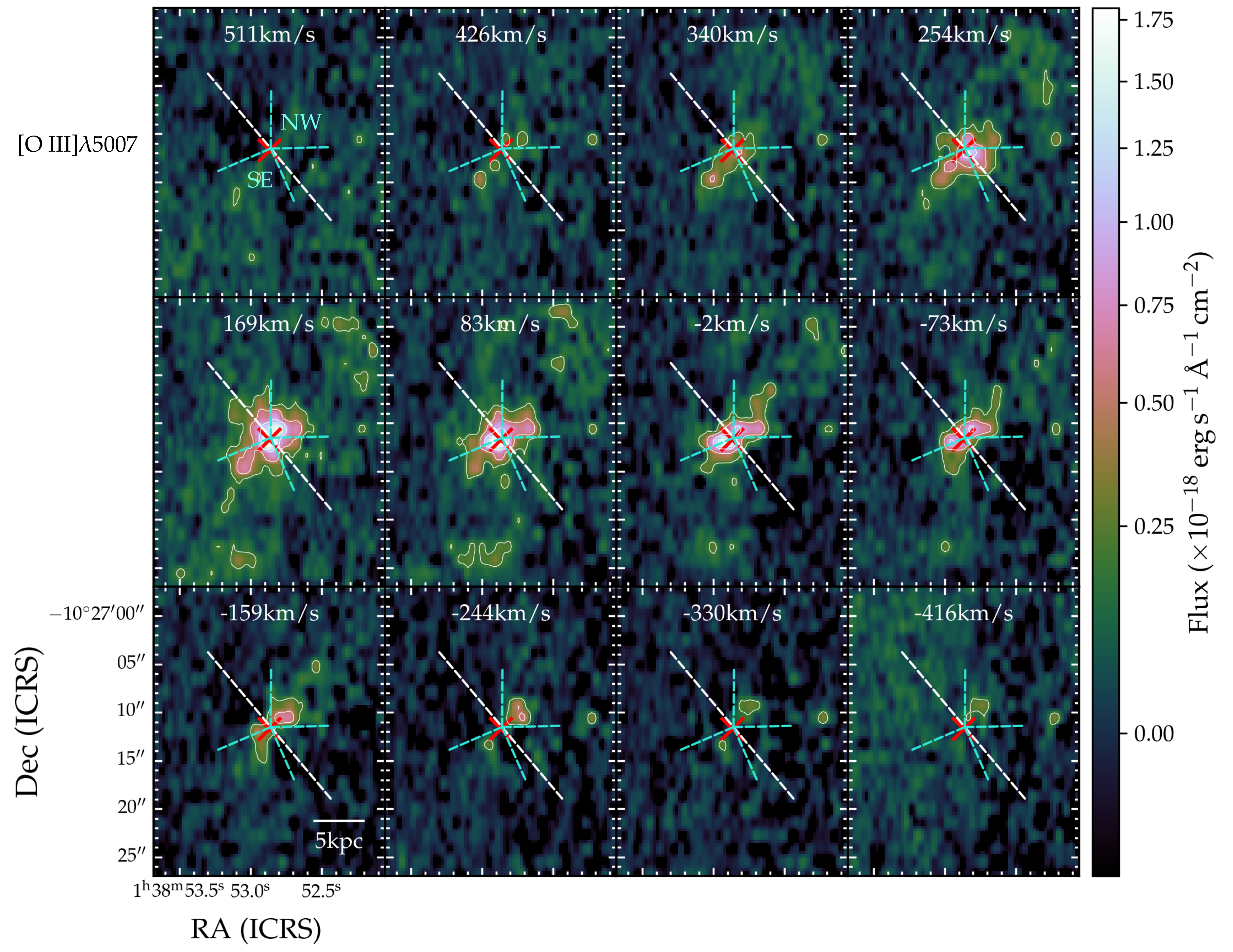}
    \caption{Channel maps for the [O\,III]$\lambda5007$ line observed with Keck/KCWI. Red cross, white and cyan dashed lines are the same as in Figure \ref{fig:miri_chan}. White contours show levels of 0.2, 0.5, 1.0\,$\times 10^{-18}$\,erg\,s$^{-1}$\,$\AA^{-1}$\,cm$^{-2}$. } 
    \label{fig:oiii_chan}
\end{figure}

To better visualize the kinematic features present in IRAS\,F01364, we present here channel maps for [Ne\,III]\,15.6$\mu$m, \ce{H2}\,0-0\,S(3), CO\,($J=2-1$) and [O\,III]\,$\lambda5007$ lines, respectively in Figure \ref{fig:miri_chan}, \ref{fig:co21_chan} and \ref{fig:oiii_chan}. We only show channels over the velocity ranges where emission is seen, which is approximately $-500 - +350$\,km/s for \ce{H2}\,0-0\,S(3), $-600 - +700$\,km/s for [Ne\,III]\,15.6$\mu$m, $-300 - +300$\,km/s for CO(2-1) and $-450 - +500$\,km/s for [O\,III]\,$\lambda5007$. The channels displayed were selected to center around the velocity channel closest to 0\,km/s with spacings of 3, 3, 5 and 6 channels for [Ne\,III]\,15.6$\mu$m, \ce{H2}\,0-0\,S(3), CO\,($J=2-1$) and [O\,III]\,$\lambda5007$, respectively, to account for the different instrument spectral resolutions. In all panels, we additional show the direction of the kinematic major axis of the stellar disk (i.e., Figure \ref{fig:fig8}), represented in white dashed lines, and the outlines of the NW and SE outflow cones, in dashed cyan lines. The latter were defined based on the FWHM linewidth-enhanced regions identified in [O\,III]$\lambda 5007$ (i.e., Figure \ref{fig:fig9}). 


\end{appendix}
\end{document}